\documentclass[12pt]{report}
\usepackage{graphicx, amsmath, bm, amssymb}
\usepackage{cite}
\usepackage{epstopdf}
\usepackage{enumerate}
\usepackage{appendix}
\usepackage{lineno}
\usepackage{setspace}
\usepackage{xfrac}
\usepackage{footmisc}
\usepackage{wasysym}
\usepackage{physics}
\usepackage{longtable}

\usepackage[usenames]{color}
\definecolor{navyblue}{rgb}{0,0.08,0.45}
\usepackage[colorlinks=true, urlcolor=navyblue, linkcolor=navyblue, citecolor=navyblue]{hyperref}

\usepackage[mathscr,scaled=1.15]{urwchancal}
\DeclareFontFamily{OT1}{pzc}{}
\DeclareFontShape{OT1}{pzc}{m}{it}%
{<-> s * [1.15] pzcmi7t}{}
\DeclareMathAlphabet{\mathpzc}{OT1}{pzc}{m}{it}

\begin{document}


\begin{flushright}
{\small JLAB-PHY-23-3758 \\ \vspace{1pt}
}
\end{flushright}
\begin{flushright}
{\small SLAC--PUB--17724 \\ \vspace{1pt}
}
\end{flushright}
\begin{flushright}
{\small NJU-INP 071/23 \\ \vspace{1pt}
}
\end{flushright}

\vspace{40pt}

\begin{center}{\huge \bf QCD Running Couplings\\[0.5ex] and Effective Charges}\end{center}

\vspace{80pt}

\centerline{Alexandre~Deur}

\vspace{5pt}

\centerline {\it Thomas Jefferson National Accelerator Facility, Newport News, VA 23606, USA}

\vspace{15pt}

\centerline{Stanley J.~Brodsky}

\vspace{5pt}

\centerline {\it SLAC National Accelerator Laboratory, Stanford University, Stanford, CA 94309, USA}

\vspace{15pt}

\centerline{Craig D.~Roberts}

\vspace{5pt}

\centerline {\it School of Physics, Nanjing University, Nanjing, Jiangsu 210093, China}
\centerline {\it Institute for Nonperturbative Physics, Nanjing University, Nanjing, Jiangsu 210093, China}

\vspace{15pt}

{\small \centerline{2023 January 20}}

\vspace{60pt}

{\small
\centerline{\href{mailto:deurpam@jlab.org}{\tt deurpam@jlab.org}, \, \href{mailto:sjbth@slac.stanford.edu}{\tt sjbth@slac.stanford.edu},\, 
\href{mailto:cdroberts@nju.edu.cn}{\tt cdroberts@nju.edu.cn}
}}

 \vspace{40pt}

 \centerline{ \it (Review article commissioned by Progress in Particle and Nuclear Physics)}

\newpage

{\centerline{ \bf \large Abstract}

\vspace{10pt}

We discuss our present knowledge of $\alpha_s$, the fundamental running coupling or effective charge of Quantum Chromodynamics (QCD). 
A precise understanding of the running of $\alpha_s(Q^2) $ at high momentum transfer, $Q$, is necessary for any perturbative QCD calculation. 
Equally important, the behavior of $\alpha_s$ 
at low $Q^2$ in the nonperturbative QCD domain is critical for understanding strong interaction phenomena, including the emergence of mass and quark 
confinement. 
The behavior of $\alpha_s(Q^2)$ at all momentum transfers also provides a connection between perturbative and nonperturbative QCD phenomena, such as hadron spectroscopy and dynamics. 
We first sketch the origin of the QCD coupling, the reason why its magnitude depends on the scale at which hadronic phenomena are probed, and the resulting consequences for QCD phenomenology. 
We then summarize latest measurements in both the perturbative and nonperturbative domains. 
New theory developments include the derivation of the universal nonperturbative 
behavior of $\alpha_s(Q^2)$ from both the Dyson-Schwinger equations and light-front 
holography. 
We also describe theory advances for the calculation of gluon and quark Schwinger functions in the nonperturbative domain and the relation of these quantities to $\alpha_s$. 
We conclude by highlighting how the nonperturbative knowledge of $\alpha_s$ is now providing a parameter-free determination of hadron spectroscopy and structure, a central and long-sought goal of QCD studies.}

\tableofcontents

\chapter{Preamble}
The precise knowledge of the strong coupling $\alpha_s$ is essential to studies of quantum chromodynamics (QCD)\footnote{A list of abbreviations used in this document is provided in an appendix, which begins on page~\pageref{PageAbbreviations}.}
-- the gauge theory of the strong interaction of gluons and quarks that underlies hadronic and nuclear physics -- because it sets the magnitude of the force and triggers the onset of many complex and essential phenomena, such as the emergence of mass, asymptotic freedom, and confinement. 
Another crucial role of $\alpha_s$ is as the expansion parameter for first-principles perturbative QCD (pQCD) calculations. 
The functional dependence of $\alpha_s(Q^2)$ varies with the value of $Q^2$, the energy-momentum scale at  which hadronic phenomena are probed. 
The precise quantitative knowledge of $\alpha_s(Q^2)$ is necessary both at high $Q^2$ (\textit{i.e}., short distance scales, also called the``ultraviolet'' UV regime) and at low $Q^2$ (long distances, also called the ``infrared'' IR regime): at high $Q^2$, the accurate determination of $\alpha_s$ is needed to meet the precision required by hadron scattering experiments,  testing the standard model and searching for
its extensions \cite{high-precision-QCD, Anastasiou:2016cez, Heinrich:2020ybq}. 
For instance, the uncertainty on $\alpha_s$ directly contributes to the total uncertainty of the theoretical prediction for Higgs production in hadronic collisions. 
Overall, in order to ensure that it does not dominate other uncertainties, $\alpha_s$ in the UV is needed to be known at 
the sub-percent level \cite{Salam:2017qdl}. 
As of 2023, this goal is only marginally met ($\Delta \alpha_s/\alpha_s = 0.85\%$ \cite{Workman:2022ynf}), revealing that studies of $\alpha_s$ in the UV must actively continue. 
It is also interesting to note that, today, even this sub-percent goal remains far from the precision of 
other fundamental 
force couplings, {\it viz}. $\Delta \alpha/\alpha=1.5\times 10^{-10}$ (quantum electrodynamics -- QED), $\Delta G_F/G_F=5.1\times 10^{-7}$ (weak interaction) and $\Delta G_N/G_N=2.2\times 10^{-5}$ (gravitation) \cite{Workman:2022ynf}. 

At low $Q^2$, where $\alpha_s(Q^2)$ is even less known, its uncertainties affect the determination of the hadron mass spectrum and other hadronic static properties, the calculation of the production of heavy quarks near threshold, and the behavior of hadronic  distribution amplitudes, structure functions, and form factors. 
In addition, the behavior of $\alpha_s(Q^2)$ at low $Q^2$ directly connects with recent efforts to elucidate and quantify the various mechanisms responsible for the proton mass \cite{Roberts:2021nhw, Binosi:2022djx, Papavassiliou:2022wrb, Ding:2022ows, Roberts:2022rxm, Ferreira:2023fva}.
This should not be surprising, given that the values of the proton mass and size  are related to the scales which are typically used to characterize $\alpha_s$, such as the mass scale  $\Lambda_s$ which determines its rate of running, and  in some models, the string tension $\sigma$ \cite{Deur:2016tte}. 

In this article, we will review our present empirical and theoretical knowledge of $\alpha_s$, as well as presenting
recent advances. 
We first briefly explain the origin and meaning of $\alpha_s(Q^2)$ and why it depends on $Q^2$, 
avoiding technicalities in order to 
remain accessible to nonspecialists. 
A formal recapitulation of that material is presented in Ch.\,\ref{ChShortDistance} in order to define the notations and definitions linked to $\alpha_s$. 
We then review the latest measurements at the high energy-momentum scale. 
The second major theme of this review is presented in Ch.\,\ref{low Q regime}, which discusses the latest phenomenological and theoretical developments at low energy-momentum scales. 
We close in Ch.\,\ref{epilogue} with a perspective on how the improved knowledge of $\alpha_s$ and its running have brought new fundamental understandings of hadronic properties and phenomena, and we highlight how recent nonperturbative information about $\alpha_s$ has provided new insights into the determination of hadron masses and structure. 
This determination, which has no free parameters, apart from quark current masses, realizes the central and long-sought goal of predicting strong interaction phenomena directly from QCD.


We will not discuss in this article the running couplings of QCD-related theories, 
such as gauge theories with unphysical numbers of colors (\textit{e.g}., SU(2)), flavors or space dimensions;
nor gauge theories at nonzero temperature.  
Neither will we report work relating to $\alpha_s$ in the IR timelike domain\footnote{In the purely perturbative case, the  timelike and spacelike behaviors of $\alpha_s$ are identical.}.
Notwithstanding these omissions, which have been made to satisfy the need for a focused discussion, these topics are areas of active research.

Numerical tools are available for accurately computing $\alpha_s$, \textit{e.g}., a Mathematica package \cite{Chetyrkin:2000yt, Herren:2017osy, Hoang:2021fhn}, and several complementary $\alpha_s$ reviews exist -- see 
Refs.\,\cite{Deur:2016tte, Altarelli:2013bpa, Prosperi:2006hx, Dissertori:2015tfa, Pich:2018lmu, Proceedings:2019pra, Aoki:2021kgd, dEnterria:2022hzv}. 
A snapshot of some standard contemporary perspectives is provided by the Particle Data Group (PDG) \cite{Workman:2022ynf} and Refs.\,\cite{dEnterria:2016cww, dEnterria:2018cye} offer an explanation of recent updates of the PDG determination.
Most current reviews cover the status of $\alpha_s$ at high $Q^2$, but some also discuss its low $Q^2$ behavior, \textit{e.g}., Refs.\,\cite{Deur:2016tte, Prosperi:2006hx}. 
This review updates and expands upon Ref.~\cite{Deur:2016tte}, to which we will refer for details of several nonperturbative frameworks used to compute $\alpha_s$ at low $Q^2$, the history of these calculations, and of some relevant measurements.

\chapter{Pedestrian view of $\alpha_s$}
\label{introduction}
Classically, the strength exerted by a force is characterized by a universal coefficient -- the coupling constant, which quantifies the force between two static bodies of unit ``charge'', \textit{i.e}., the electric charge for QED, the color charge for QCD, the weak isospin for the weak force, or the mass for gravity.\footnote{Since there is no presently known ``unit mass'', the gravitational constant, $G$, has mass dimension negative-two when defined conventionally.} 
Consequently, the coupling is defined as being proportional to the elementary charge squared, \textit{e.g}., 
$\alpha \equiv e^2/4\pi$ where $e$ is the elementary electric charge, or $\alpha_s \equiv g^2/4\pi$ where $g$ is the elementary gauge field coupling in QCD. 
For theories in which the superposition principle holds, with static sources and massless force carriers, the coupling relates the force to the ``charges'' of the two sources divided by $r^2$, where $r$ is their separation.
This $1/r^2$ dependence was interpreted first by Faraday as the dilution of the force flux as it diffuses uniformly through space. 
In quantum field theory (QFT), $1/r^2$ is the coordinate-space expression for the propagator of the force carrier (gauge boson) at leading-order in perturbation theory: in momentum space, the analogous propagator is proportional to $1/q^2$, where $q$ the boson 4-momentum.  
(Hereafter we write $Q^2=-q^2$ because most of the processes we consider involve spacelike momenta, $Q^2>0$.)
The interpretation of a propagator as expressing a particle's probability to travel from one point to another \cite[Ch.\,2.5]{IZ80}, together with the isotropic emission of the bosons by a source and their free propagation, is consistent with Faraday's picture.

For sources interacting weakly, the one-boson exchange representation of interactions is a good first approximation. 
However, when interactions become strong (with ``strong" to be defined below), higher orders in perturbation theory become noticeable and the $1/r^2$ law no longer stands. 
In such cases, it makes good physics sense to fold the extra $r$-dependence into the coupling, which thereby becomes $r$, or equivalently $q^2$, dependent.
This reveals that the running of the coupling is a typical quantum phenomenon, a manifestation of 
the contribution of higher order interaction diagrams to the magnitude of the force.

\begin{figure}[t!]
\centering
\includegraphics[width=13.0cm]{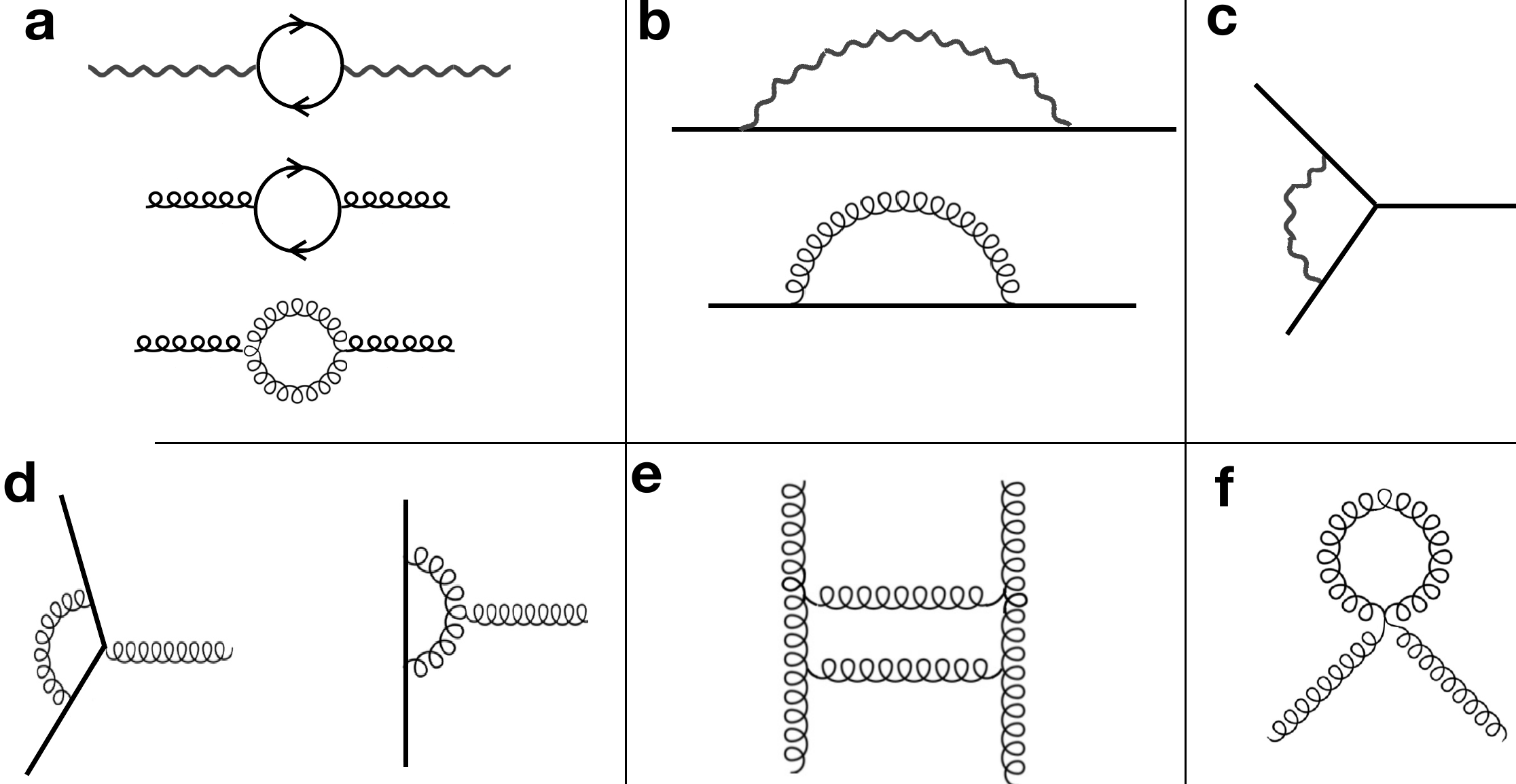}
\caption{\label{quantum_effects}\small  
Short distance phenomena for QED (panels a, b and c) and QCD (all panels).
Panel (a): QED and QCD vacuum polarizations;
Panel (b): fermion self-energy (virtual photon or gluon corrections for the propagating electron or quark);
Panel (c) QED vertex correction;
Panel (d) QCD vertex corrections;
Panel (e): ladder, or H, graph;
Panel (f): tadpole graph.}
\end{figure}

It is clear from this sketch that choosing to absorb the extra $r$-dependence into the coupling
is a matter of convenience and convention. 
The convention is not always followed, however. 
For instance, the traditional approach to publishing experimental results from lepton scattering experiments is to perform QED radiative corrections on the data \cite{Mo:1968cg}, so that quantities within the one-photon exchange approximation are provided, such as the Born cross-section rather than the physical all-order observables. 
These radiative corrections include vacuum polarization corrections -- see Fig.\,\ref{quantum_effects}(a), which is the quantum effect responsible for the running of QED's $\alpha$ \cite{GellMann:1954fq}.
In that case, the usual large-distance value, $\alpha \approx 1/137$, is used 
-- rather than the running value of $\alpha$ at the specific $Q^2$ of the experiment -- in the calculations of the factors involved in forming the observables, \textit{e.g}., the Born cross-section or radiative corrections.
Hence, in lepton-hadron scattering experiments, the short distance quantum effects responsible for the running of couplings are treated differently for QED and QCD: for QCD, they are folded into the definition of $\alpha_s$; while for QED, they are corrected for by experimentalists and $\alpha$ remains constant.  
On the other hand, in electron-muon elastic scattering, one uses $\alpha(Q^2)$ in the one-photon exchange amplitude, which sums all vacuum polarization corrections to all orders.
Furthermore, when one includes the ladder and crossed ladder two-photon exchange contributions, 
a new renormalization scale enters, reflecting the reduced virtuality of the exchanged photons.






That it has been a matter of convention whether or not to absorb specific extra $r$-dependence into the coupling is especially important for understanding why there are seemingly disagreements in the literature regarding the behavior of $\alpha_s$ in the nonperturbative domain of QCD.\footnote{
Such differences of choices on the definition of the coupling also exists in the perturbative domain; however, since the relations between couplings calculated in different scheme are precisely known, the  running on $\alpha_s$ in the pQCD domain is unambiguous.}
The disagreements typically reflect the different choices of conventions and definitions for $\alpha_s$. The effect of these different choices becomes acute at large distance, as we will discuss in Section~\ref{low Q regime}. 
The question is thus not of a disagreement, but of which is the most appropriate choice of definition and convention; and the answer here may depend on the practitioner.

While in QED, the extra $r$-dependence comes only from the vacuum polarization -- Fig.\,\ref{quantum_effects}a-top graph, in some common treatments of QCD, $\alpha_s$ receives contributions from the vacuum polarization -- Fig.\,\ref{quantum_effects}a-lower graphs,
quark self-energy -- Fig.~\ref{quantum_effects}b, 
vertex corrections -- Fig.~\ref{quantum_effects}d, 
and gluon loop corrections to the elementary three-gluon and four-gluon couplings;
but not tadpole graphs -- Fig.~\ref{quantum_effects}f. 
Moreover, the quark self-energy -- Fig.~\ref{quantum_effects}b-bottom graph -- does not contribute in the oft-used Landau gauge; and gluon emissions for which the gluon reconnects to a different parton do not contribute to the perturbative running of $\alpha_s$, if they are UV finite.
%
In contrast, using a combination of pinch technique (PT) \cite{Cornwall:1981zr, Cornwall:1989gv, Pilaftsis:1996fh, Binosi:2009qm, Cornwall:2010upa} and background field method (BFM) \cite{Abbott:1980hw, Abbott:1981ke}, one can rearrange and recombine diagrams in order to arrive at a unique QCD running coupling that is determined solely by a modified form of gluon vacuum polarisation \cite{Binosi:2016nme, Cui:2019dwv}.
Evidently, the manner by which the relevant amplitude is separated into different graphs is also a matter of convention; and once again, different practitioners may choose distinct resummation schemes, each of which is perceived by its proponent to have special merits.

\begin{figure}[t]
\centering
\includegraphics[width=15.0cm]{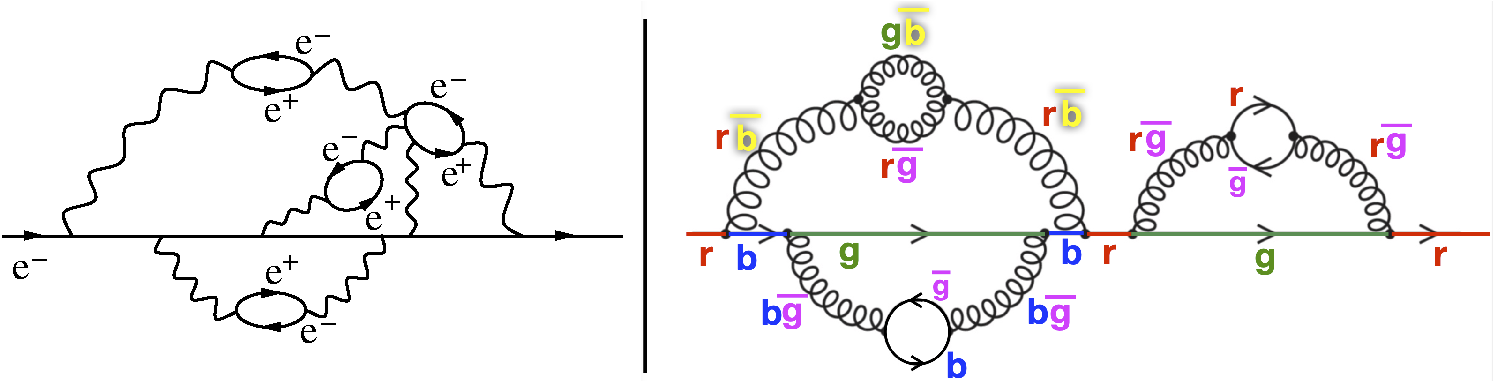}
\caption{\label{Flo:screening} \small  
Left: screening of the QED charge by vacuum polarization. 
The positrons are preferably closer to the propagating negative charge than the electrons.
Right: anti-screening of the QCD charge. 
Contrary to QED, whose force carrier is electrically neutral, the gluons are color-charged, which leads to a spatial dilution of the propagating color charge. 
The evolution of the propagating quark color is illustrated in the right figure. 
The initially red quark changes to either a blue or green quark when its red color is carried away by the emitted gluon.  
The anti-blue color is symbolized as yellow and anti-green as magenta.
The spread of the initial red color charge suppresses any small distance interaction with anti-red or (green+blue) charges, \textit{i.e}., it makes $\alpha_s$ weaker at shorter scales.
}
\end{figure}

Now that we understand that microscopic effects alter the $1/r^2$ behavior of the force,
we can go further and check how the force departs from $1/r^2$. 
The left panel of Fig.~\ref{Flo:screening} illustrates the QED case: electron+positron pairs
are created around the initial charge, here an electron. 
The positrons, because of their 
opposite charge, will tend to be nearer to the initial charge than the created electrons.
Therefore, the total charge inside a sphere of radius $r$, which by Gauss' law effectively sets the
magnitude of the coupling, is smaller than the original (bare) charge of the initial electron. 
We conclude that the QED running coupling decreases as $r$ increases (charge screening), tending to its macroscopic value $\alpha \approx 1/137$, and increases at small distances.

For QCD, the fact that gluons are color-charged fundamentally alters the above charge-screening process. An emitted gluon carries away the color of the initial source -- see the right panel of Fig.\,\ref{Flo:screening}. 
Hence, the initial charge becomes spatially diluted: in the example of Fig.\,\ref{Flo:screening}, the initially red quark is green most of its time and thus does not couple to the high-resolution anti-red gluons that would otherwise interact with it. 
Consequently, a charge ``anti-screening'' occurs, opposite to QED. 
In details, gluon loop anti-screening dominates over the QED-like quark+antiquark loop screening. 
The color charge dilution is further enhanced by vertex correction while the quark self-energy either does not contribute or screens the charge, depending on gauge choice. 
Overall, $\alpha_s$ decreases at smaller distances. 

The coupling runs logarithmically because QCD has no intrinsic scale, quarks being point-like and typically nearly massless. Indeed, with $Q^2$ the only available scale, dimensionality imposes
\begin{equation}
\frac{d\alpha_s}{dQ^2}= \frac{a_2 \alpha_s^2 +\cdots a_n  \alpha_s^n}{Q^2} 
\end{equation}
with the dimensionless $a_i$ determined by the theory\footnote{There is no $a_0$ or $a_1$ term since positive $a_0$ or $a_1$ would imply, respectively, an increasing or a constant $\alpha_s$ at short distances, contradicting the anti-screening argument. If they are negative, they would lead to an unphysical negative coupling.}. 
For short distances, where $\alpha_s \ll 1$, the $a_2$ term dominates and 
\begin{equation}
\alpha_s(Q^2) = \frac{-1}{a_2 \ln(\sfrac{Q^2}{C})} 
\label{eq:simple_coupling}
\end{equation}
with $C$ an integration constant. 
Evidently, for a meaningful coupling in the UV, one must have $a_2 < 0$. 
As distance increases, $\alpha_s$ increases, too, and additional $\ln^n(Q^2)$ terms arise.
This offers a phenomenological insight into  the ``asymptotic freedom" phenomenon at large $Q^2$ and into the ``strong QCD'' regime at low momentum transfer. 
The logarithmic increase of $\alpha_s$ at large distances should be expected to stop owing to finite hadron size, since the confinement of colors within the hadron suppresses the parton wave functions when their wavelengths reach the typical hadronic size. 
The quantum effects responsible for the evolution of $\alpha_s$ are suppressed and $\alpha_s$  becomes constant at hadronic scales \cite{Brodsky:2008be, Binosi:2014aea, Binosi:2016xxu, Binosi:2016wcx, Gao:2017uox}.

Hitherto, the discussion has been phenomenological. 
More formal arguments can 
explain the running of couplings as originating from renormalization procedures. 
In renormalizable QFT, the strength of the interaction described by the Lagrangian
is set by the coupling constant. 
This coupling becomes scale dependent after the regularization and renormalization of UV-divergent integrals. 
In the renormalization process, the coupling acquires a dependence on the {\it arbitrary} choice of the UV cut-off value. 
This unphysical dependence is eliminated by allowing the Lagrangian's couplings and masses to become scale-dependent and by normalizing them to the values measured at a given scale. 
This ``renormalization"  procedure entails a running of the couplings which, in effect, become ``effective couplings''. 

For instance, in QED, the process of renormalizing the virtual photon propagator implies replacing  the coupling parameter (bare coupling) appearing in the Lagrangian by the running coupling $\alpha(Q^2) = \alpha(0)/\big( 1- \Pi(Q^2)\big)$, where $\Pi(Q^2)$ sums all vacuum polarization loops. 
The logarithmic UV divergence stemming from vacuum polarization loop integration is typically regulated with the Pauli--Villars method \cite{Pauli:1949zm} in which a massive photon is employed. 
The divergence is then eliminated by normalizing the coupling to a known (measured) value at a chosen value of $Q_0$.
The UV divergences from the fermion propagator self-energy and the vertex renormalization factor cancel 
each other (Ward identity) \cite{Ward:1950xp}. 

A more formal and general explanation for the running of couplings in QFT is that predictions of phenomena must be independent of which renormalization scheme (RS) is used, since its choice is arbitrary. 
This invariance forms a symmetry group, the renormalization group (RG), and  group theory techniques can then be used to calculate the behavior of the coupling. 
Since couplings are not typically observable\footnote{Exceptions exist, \textit{e.g}., process-dependent effective charges -- Sec.\,\ref{EffectiveCharge}, and the coupling entering the QCD potential -- Sec.\,\ref{EffectivePotential}.},
they need not be independent of RS \cite{Celmaster:1979km} or gauge choice. 
Using $\alpha_s$ as an example, an observable $R$ is expanded in pQCD as
%
$R(Q^2)=\sum_{n}r_{n}\left[\alpha_s(Q^2)\right]^{n}$.
%
While overall $R$ should be RS-independent\footnote{In practice however, the perturbative series is calculated at finite order and the approximant of $R$ displays a residual RS-dependence.}, the $r_n$ and $\alpha_s$ may depend on the RS choice. 

This freedom is denied to $r_0$ and $r_1$, as may be seen as follows.
Since $R \xrightarrow[Q^2 \to \infty]{} r_0$, owing to asymptotic freedom, $r_0$ must be RS-independent; accordingly, the residual RS-dependence of $R$ must be minimal in the UV, while it can become important in the IR. 
Moreover, the first order perturbative expression of $\alpha_s$ -- Eq.\,\eqref{Eq.one loop} below -- is RS-independent. 
This is because, whilst $\alpha_s$ is not directly observable, the phenomena responsible for the running of the coupling are physical processes. 
Therefore, any two couplings $\alpha_s$ and $\alpha^\prime_s$, calculated perturbatively in two distinct RSs, must satisfy
\begin{equation}
\alpha'_s=\alpha_s+\mbox{v}_2\alpha^2_s+\mbox{v}_3\alpha^3_s+\cdots\,.
\label{eq:alpha_rel_2RS}
\end{equation}
Consequently, like $r_0$, $r_1$ is RS-independent, as shown by expanding the observable in a different scheme, $R(Q^2)=\sum_{n}r'_{n}\left[\alpha^\prime_s(Q^2)\right]^{n}$ and inserting $\alpha^\prime_s$ in terms of its $\alpha_s$ expansion.
Formally, the reason for the UV-weakening of the RS-dependence is because the different forms of the renormalization group equation (RGE) obtained using distinct schemes differ only in mass terms. 
Therefore, in the UV limit, where QCD running masses are negligible, they take the same form, {\it viz}.\ the RS-dependence vanishes. 

For QED, the RS-dependence of $\alpha$ is small and the coupling is often taken as an observable. Furthermore, a low energy theorem based on Compton scattering shows that $\alpha$ is nearly constant and becomes measurable in the IR, making it an observable in this limit~\cite{GellMann:1954fq}. 

In contrast for pQCD, $\alpha_s$ has a marked RS-dependence and, {\it prima facie}, cannot be determined in the IR because quarks and gluons are confined and interactions take place between hadronic degrees of freedom -- see, however, Sec.\,\ref{EffectiveCharge}.  
Therefore, $\alpha_s$ is often taken as an intermediate quantity without definite physical meaning.
Indeed, if a scheme for computing the running coupling enables $\alpha_s$ to vary by an arbitrary factor, depending on RS choice, then this coupling cannot characterize the physical strength of the strong force. 
Nevertheless, so long as studies are confined to the deep UV regime, the phenomenological description given previously is still approximately relevant, owing to the fact that such unphysical features are minimal in the UV. 
In the IR, however, it is commonly thought that any attempt at a phenomenological understanding is doomed to failure. 
However, we will see in Sec.\,\ref{EffectiveCharge} that $\alpha_s$ can be restored to the status of an observable and thus defined at long distances while maintaining its physical and intuitive meaning. 

Thus far, our discussion pertains to the weak-coupling UV regime of QCD. 
The pQCD prediction at longer distances is that $\alpha_s$ becomes infinite as $Q^2\to \Lambda_s^2$.
This so-called Landau pole is a general feature of perturbatively calculated running couplings.
It is also found in QED at very high energy. 
These divergences signal the breakdown of the perturbative approach to estimating observables and the theory's $\beta$-function (see Eq.~(\ref{eq:Callan-Symanzik2}) below for its definition and meaning) when the coupling becomes too large. 
(\textit{N.B}.\ The individual $\beta$-function expansion coefficients, $\beta_n$, are nevertheless reliable because they are obtained via an expansion in $\hbar$ rather than the coupling.)

Sometimes, the Landau pole is erroneously interpreted as a physical feature of the theory. 
In fact, in the early QCD days, it was imagined to be the origin of color confinement. 
This fallacious solution of the confinement problem was called ``IR slavery'' \cite[Sec.\,3.1]{Marciano:1977su}. 
In that context, it is useful to recall that the coupling need not become infinite to produce confinement, \textit{e.g}.:
QCD in (1+1)-dimensions is a confining theory with a finite coupling; 
Gribov's supercritical binding model, in which confinement occurs for $\alpha_s/\pi \lesssim 0.137$ \cite{Gribov:1999ui};
and, related to the former, the violation of reflection positivity by elementary QCD two-point Schwinger functions \cite{Binosi:2019ecz, Boito:2022rad}, which is known to occur when $\alpha_s(Q^2)< \infty$ $\forall Q^2 \geq 0$.

Another reason for considering the pole to be unphysical is that it is located on the positive
real axis of the complex $Q^{2}$-plane; and a simple-pole singularity at $Q^2=\Lambda_s^2$ corresponds to a causality violating tachyon-related process. 
In contrast, poles located on the negative real $Q^2$-axis are permitted because they may be associated with production of on-shell particles.
(If one considers the analytic structure of $\alpha_s$ in the entire complex-$Q^2$ plane, it is conceivable that singularities may occur off the positive real semi-axis \cite{Contreras:2020hme}.)

As already mentioned, instead of diverging at the Landau pole, the scale-dependence of a physical $\alpha_s$ should disappear past an IR scale since the wavelengths of the secondary colored particles 
that dilute the color charge are suppressed at the physical hadron size \cite{Brodsky:2008be, Binosi:2014aea, Binosi:2016xxu, Binosi:2016wcx, Gao:2017uox}. 
The IR scale-independence of $\alpha_s$ is known as the ``freezing of $\alpha_s$'', or QCD's ``conformal window",
and it has been established using a combination of continuum and lattice techniques \cite{Cui:2019dwv}.
This fact permits one to employ formalisms based on gauge-gravity duality \cite{Maldacena:1997re} to develop models with relevance to nonperturbative QCD. 
Such applications to $\alpha_s$ will be discussed in Sec.\,\ref{HLFQCD} and their expressions in QCD
phenomenology will provide the perspective given in Sec.\,\ref{epilogue}.

In closing this section it is worth noting that whilst the argument based on the wavelength cut-off at confinement scale {\it a priori} seems incontrovertible, one may encounter physically motivated calculations yielding forms of $\alpha_s$ that do not freeze in the IR, \textit{e.g}., that based on the static long-range potential, Eq.~\eqref{eq:Q-Q stat pot. Mom space.}, or those based on the dispersive approach (Section~\ref{AnalyticApproaches}). 
This is because such definitions include processes other than those shown in Fig.\,\ref{quantum_effects}. The extra processes need not have a well-defined wavelength, in which cases the argument does not apply and $\alpha_s$ may not freeze. 
For example, one cannot define wavelengths for the instantaneous long-range potential that is present in the definition of the coupling via Eq.~\eqref{eq:Q-Q stat pot. Mom space.} and causes  $\lim_{Q^2\to \infty} \alpha_s(Q^2) \to \infty$.  
It is a matter of debate whether one should add, to the IR running of $\alpha_s$, processes unconnected with those causing its UV running or if one should preserve a connection to the theoretically better understood pQCD coupling. 
Ultimately, we judge that it is a definition's usefulness in terms of enhancing QCD's predictability that should arbitrate on the ``best'' coupling definition.

\chapter{Short distance: the perturbative regime\label{high Q regime}}
\label{ChShortDistance}
The scale-dependence of $\alpha_s$ at short distances is well understood because perturbation theory is applicable at the associated scales.  
Nevertheless, considerable efforts in experiment and lattice gauge theory (LGT) are devoted to obtaining the absolute magnitude of $\alpha_s$ or, equivalently in the perturbative regime, its rate of evolution, determined by $\Lambda_s$, since these quantities are nonperturbative.\footnote{ \label{ftnt:alpha-lambda}
The equivalence between the information provided by $\alpha_s$ in the UV and $\Lambda_s$ holds only for a given $\alpha_s$ perturbative evolution equation, \textit{i.e}., at given perturbation order and given form of the evolution equation. 
The latter may differ even at the same perturbative order, compare, \textit{e.g}., Eq.\,\eqref{eq:alpha_s} truncated at next-to-next-to-leading order (NNLO) and Eq.\,\eqref{eq:GGK beta2}. 
In other words, because the exact evolution equation is not known at NNLO or beyond, a given $\Lambda_s$ may lead to different $\alpha_s$ unless the evolution equation is also specified.} 
Furthermore, confronting accurate determinations of $\alpha_s$ extracted from 
different observables and obtained at different scales is a fundamental internal consistency test of QCD and of the Standard Model. 
In fact, in the first comprehensive review of various extractions of $\alpha_s$, all at $Q \ll M_{\rm Z}$, G. Altarelli compiled a world average of $\alpha_s (\sim M_{\rm Z}) = 0.11\pm0.01$ (1989) and stated that ``{\it Establishing that this prediction is experimentally true would be a very quantitative and accurate test of QCD}'' \cite{Altarelli:1989ue}.
This prediction was verified, see, \textit{e.g}., the current world average (2023) of $\alpha_s (M_{\rm Z}) = 0.1179\pm0.0009$ \cite{Workman:2022ynf}.
Finally, with $\alpha_s$ remaining relatively large even for the momenta characterizing high-energy experiments, theoretical methods are still actively developed in order to minimize RS dependence and other arbitrary factors arising from the truncation of the pQCD series \cite{Wang:2023ttk, Boito:2016pwf, Jamin:2016ihy, Vafaee:2017nze, Boito:2020hvu}. 

Evidently, the study of $\alpha_s$ at short distance remains an active field of research. 
We discuss this regime first because many practitioners will consider it familiar; furthermore, it sets the stage for our analysis of the lesser known subject of $\alpha_s$ in the IR. 
For a standard review, frequently updated, see the PDG \cite[Sec.\,9.4]{Workman:2022ynf}. 
Comments on the PDG method of combining the world experimental and LGT determinations of the strong coupling, together with a partial review of $\alpha_s$, may be found elsewhere \cite{Salam:2017qdl}.

\section{Short-distance behavior of $\alpha_s$}
\label{alpha_s pQCD}

The QCD Lagrangian density is: 
\begin{equation}
\mathcal{L}_{QCD}=\sum_f \overline{\psi}_i^{(f)}\big(i\gamma_{\mu}D^{\mu}_{ij} -m_f\delta_{ij}\big)\psi_j^{(f)}- 
\frac{1}{4}F^{\mu\nu}_a F_{\mu\nu}^a.  \label{eq:QCD Lagrangian}
\end{equation}
In the matter part of the Lagrangian, 
$\psi^{(f)}$ is the field of a quark of flavor $f$ and mass $m_f$;  
and $D^\mu_{ij} \equiv \partial^\mu\delta_{ij} + i\sqrt{4\pi \overline\alpha_s}t^a_{ij} A_a^\mu$ is the covariant derivative, with 
$\overline\alpha_s \equiv \overline{g}^2/4\pi$ being the bare coupling constant;
$\{t^a_{ij}\}$ are the generators of SU(3) in the fundamental representation, where
$a=1,\ldots,8$ are color indices and $A_{\mu}^{a}$ are the associated gluon fields.
In the Lagrangian force part, 
$F^{\mu\nu}_a \equiv \partial^{\mu}A^{\nu}_{a}-\partial^{\nu}A^{\mu}_{a}+\sqrt{4\pi \overline\alpha_s}f_{abc}A^{\mu}_{b}A^{\nu}_{c}$ is the gluon field strength tensor, wherein 
$f^{abc}$ are the SU(3) structure constants. 
We will generally neglect $m_f$ in comparison with $Q$.%
\footnote{For low $Q$ reactions and/or large $m_f$, effective degrees of freedom (dof) and effective couplings are often used in lieu of partonic dof and $\alpha_s$. 
For example for heavy (nonrelativistic) quarks, the quarks remain dof but the gluon field may sometimes be described  as a QCD string, with the string tension, $\sigma$, serving as an effective coupling -- see Sec.\,\ref{EffectivePotential}.}
Furthermore, terms associated with gauge fixing do not materially affect this discussion and are suppressed in writing Eq.\,\eqref{eq:QCD Lagrangian}.

A key feature of Eq.\,(\ref{eq:QCD Lagrangian}) is that, when treated classically for negligible $m_f$, $\mathcal{L}_{QCD}$ defines a conformally invariant theory, \textit{i.e}., it has no mass or energy scale. 
Nevertheless, two scales must appear in the calculation of any processes. 
One is the process' 4-momentum flow, $Q$, and the other, $\mu$, emerges from the regularization and renormalization procedure.  
The emergence of a scale in a classically conformal theory is called ``dimensional transmutation" \cite{Coleman:1973sx}, and $\mu$ is called the ``regularization scale" or, sometimes,  ``subtraction point". 
The interpretation of $\mu$ depends upon the choice of regularization method and RS.%
\footnote{Some examples follow. 
In the cut-off regularization procedure, $\mu$ is the UV cut-off value, while it is the (large) regulating mass in the Pauli-Villars regularization scheme. 
In LGT, it is the inverse lattice spacing size.
In the momentum subtraction (MOM) RS, $\mu$ corresponds to the momentum at which the renormalization is done, while in QED's on-shell RS, it is the mass of the relevant lepton. 
In the minimal subtraction, ${\rm MS}$, and modified-${\rm MS}$, $\overline{\rm MS}$, RSs, a mass scale must appear since $\alpha_s$ is dimensionless only in 4D space and regularization is achieved by working in a $4-2\varepsilon$ space.}
This, and the fact that the choice of value for $\mu$ is arbitrary means that any observable, $R$, must be independent of $\mu$. 

Consider, therefore, the following series for an observable: 
\begin{equation}
R(Q^2, {\bm y})=\sum_{n=0} r_{n}(Q^2/\mu^2, {\bm y})\overline\alpha_s^n \,,
\end{equation}
where ${\bm y}$ are kinematic variables chosen to be dimensionless since they describe a system without intrinsic physical scale; here, $R$ is dimensionless; and the $r_n$ are calculated perturbatively. 
Once diverging loops appear in their corresponding Feynman graphs, which happens for $n\geq2$, the $r_n$ will depend on $\mu^2$, or rather $Q^2/\mu^2$ since there are no intrinsic physical scales.
As explained in Ch.\,\ref{introduction}, one conventionally transfers all the scale dependence of the process into the coupling, apart for that owing to the probing boson propagator, $\propto 1/Q^2$. 
Therefore, 
\begin{equation}    
R(Q^2, {\bm y})=\sum_{n=0} r_{n}({\bm y, \alpha_s})\alpha_s^n(Q^2/\mu^2)\,,
\end{equation}
with $\alpha_s(Q^2/\mu^2)$ being a running coupling.

Since $R$ is $\mu^2$-independent, then%
\footnote{\label{fn:gauge-dependence}
For simplicity here, we assume there is no partial derivative with respect to the gauge parameter, $\xi$, either because it is fixed, \textit{e.g}., $\xi=0$ as in the Landau gauge, 
or that the $r_n$ (\textit{i.e}., $\alpha_s$) are $\xi$-independent as, \textit{e.g}., for the ${\rm MS}$ and $\overline{\rm MS}$ RSs. 
Note that since neither $r_n$ nor $\alpha_s$ are observables, gauge invariance is not mandatory. 
Also, there is no partial derivative with respect to the the mass because we consider $m$ to be negligible in comparison with all other scales.}
\begin{equation}
 \frac{d}{d\mu^2}R=
 \left(\frac{\partial}{\partial \mu^2}+ 
 \frac{\partial \alpha_s}{\partial\mu^2} \frac{\partial}{\partial\alpha_s}+
 \frac{\partial Q^2}{\partial\mu^2} \frac{\partial}{\partial{Q^2}} +
 \sum_i \frac{\partial {y_i}}{\partial\mu^2} \frac{\partial}{\partial{y_i}}
  \right)R=0\,.
 \label{eq:Callan-Symanzik1}
\end{equation}
Multiplying Eq.\,(\ref{eq:Callan-Symanzik1}) by $\mu^2$ and noting that the last two terms are null, 
because $\bm y$ and $Q^2$ are physical quantities (kinematic variables), one obtains the RGE \cite{Petermann:1953wpa, GellMann:1954fq, Callan:1970yg, Symanzik:1970rt, Symanzik:1971vw}: 
%
\begin{equation}
\left(-\frac{\partial}{\partial \tau}+\beta(\alpha_s)\frac{\partial}{\partial\alpha_s}\right)R=0\,,
\label{eq:Callan-Symanzik2}
\end{equation}
where $\beta(\alpha_s) \equiv \mu^2\frac{\partial\alpha_s}{\partial\mu^2}$ and we introduced $\tau=\ln\left(Q^2/\mu^2\right)$ because $\alpha_s$ can only depend on $Q^2/\mu^2$ since $\mathcal{L}_{QCD}$ defines a scale invariant classical theory.
As discussed in Ch.\,\ref{introduction}, the breaking of classical conformal invariance in the theory defined by $\mathcal{L}_{QCD}$, owing to dimensional transmutation, is interpreted as the result of short-distance quantum effects.  
It constitutes a typical example of a ``quantum anomaly''.

The distinct origin of the coupling's running is at the origin of the BFM, where the gluon field is separated into a classical background component and a quantum component that is responsible for the running \cite{Abbott:1980hw, Abbott:1981ke}. 
This is one of the approaches employed to handle the forbidding number of perturbative graphs involved in the calculation of the coupling at high orders -- see, \textit{e.g}., Ref.\,\cite{Herzog:2017ohr}.
 
Equation~(\ref{eq:Callan-Symanzik2}) introduces the QCD $\beta$-function, which determines the scale dependence of $\alpha_s$. 
This RGE is completely general, \textit{viz}.\ valid nonperturbatively; but in the UV, it is typically developed as a perturbative series:
\begin{equation}
\beta\left(\alpha_s\right) := \mu^2\frac{\partial \alpha_s}{\partial \mu^2} 
=-\frac{\alpha_s^2}{4\pi}\sum_{n=0}\left(\frac{\alpha_s}{4\pi}\right)^{n}\beta_{n},\label{eq:alpha_s beta series}
\end{equation}
where the expansion coefficients, $\beta_{0,1,2,3}$, are available in a variety of RSs, \textit{e.g}., 
MOM (Landau gauge) \cite{Chetyrkin:2000dq, Boucaud:2008gn, vonSmekal:2009ae}, 
V (static potential) \cite{Fischler:1977yf, Peter:1996ig, Peter:1997me, Schroder:1998vy, Melles:2000dq, Smirnov:2009fh, Anzai:2009tm, Kataev:2015yha}, 
MS and $\overline{\rm MS}$ \cite{Tarasov:1980au, Larin:1993tp, vanRitbergen:1997va};
and $\beta_4$ has been calculated in $\overline{\rm MS}$ \cite{Baikov:2016tgj, Luthe:2016ima, Luthe:2017ttc, Luthe:2017ttg, Chetyrkin:2017bjc}.
(The $\overline{\rm MS}$ values are listed in Eqs.\,\eqref{eq:beta4 list} below.)
By definition, and as seen in Eq.\,\eqref{eq:alpha_s beta series}, the $\beta_n$ are independent of $\alpha_s$. 
They are expansions in $\hbar$ and depend only on the number of active quark flavors, $n_f$, \textit{i.e}., quarks with mass $m_f\ll \mu$, so that they may participate in the loops causing $\alpha_s$'s scale dependence -- see details in Sec.\,\ref{Quark thresholds}. 

For negligible $m_f$, the one-loop, $\beta_0$~\cite{Gross:1973id, Politzer:1973fx}, 
and two-loop, $\beta_1$~\cite{Caswell:1974gg, Jones:1974mm, Egorian:1978zx}, coefficients are RS-independent.\label{b0,b1 RS-indep} 
Again, this is because the RS-dependence of $\alpha_s$ vanishes in the UV -- see Eq.\,\eqref{eq:alpha_rel_2RS},
and may readily be verified.  Using Eq.\,\eqref{eq:alpha_s beta series}, the  $\beta$-function of $\alpha^\prime_s$ appearing in Eq.\,\eqref{eq:alpha_rel_2RS} can be expressed as
\begin{eqnarray}
\beta'\left(\alpha'_s\right) &\equiv& \mu^2\frac{\partial \alpha'_s}{\partial \mu^2} 
=-\beta'_0\frac{\alpha{'}_s^2}{4\pi}-\beta'_1\frac{\alpha{'}_s^{3}}{(4\pi)^2}-\beta'_2\frac{\alpha{'}_s^{4}}{(4\pi) ^3}+\mathcal{O}(\alpha{'}_s ^{5})
=\beta\left(\alpha_s\right) / \frac{\partial \alpha_s}{\partial\alpha'_s} \nonumber \\ 
&=&-\beta_0\frac{\alpha{'}_s^2}{4\pi}-\beta_1\frac{\alpha{'}_s^{3}}{(4\pi)^2}-\big[-(4\pi)^2(3\mbox{v}_2^2+\mbox{v}_3)\beta_0+4\pi\mbox{v}_2\beta_1+ \beta_2\big]\frac{\alpha{'}_s^{4}}{(4\pi)^3}+\mathcal{O}(\alpha{'}_s ^{5}),  \nonumber
\end{eqnarray}
where the $\alpha_s$ in $\beta(\alpha_s)$ was developed in its $\alpha'_s$ series. 
Consequently, 
\begin{equation}
\beta_0= \beta^\prime_0\,, \quad \beta_1= \beta^\prime_1,
\end{equation}
but 
\begin{equation}
\beta_2 \neq \beta^\prime_2=-(4\pi)^2(3\mbox{v}_2^2+\mbox{v}_3)\beta_0+4\pi\mbox{v}_2\beta_1+ \beta_2\,.
\end{equation}

For the physical number of active quarks, $\beta(\alpha_s) < 0$ which, with Eq.\,\eqref{eq:alpha_s beta series}, implies that  $\alpha_s \xrightarrow[\mu \to \infty]~$~0, \textit{i.e}., asymptotic freedom.%
\footnote{The critical number of light-quark flavors is $n_f^c = 9 \pm 1$ \cite{LSD:2009yru, Hayakawa:2010yn, Cheng:2013eu, Aoki:2013xza, Binosi:2016xxu}.}
This limit constitutes a UV {\it Gaussian fixed-point} \cite{Zinn-Justin:2010} for $\alpha_s$. 
In the second part of this review, we will see that $\alpha_s$ also presents a fixed-point in the IR. 

\section{First-order calculation of $\alpha_s(Q^2)$ \label{alpha_s 1-loop}}
\subsection{Direct calculation}
To better understand the origin of the various contributions to the running of  $\alpha_s$, it is useful to follow the first-order (1-loop) calculation. 
We use $\overline{\rm MS}$ and set $m_q=0$. 

\begin{figure}[t]
\centering
\includegraphics[width=10cm]{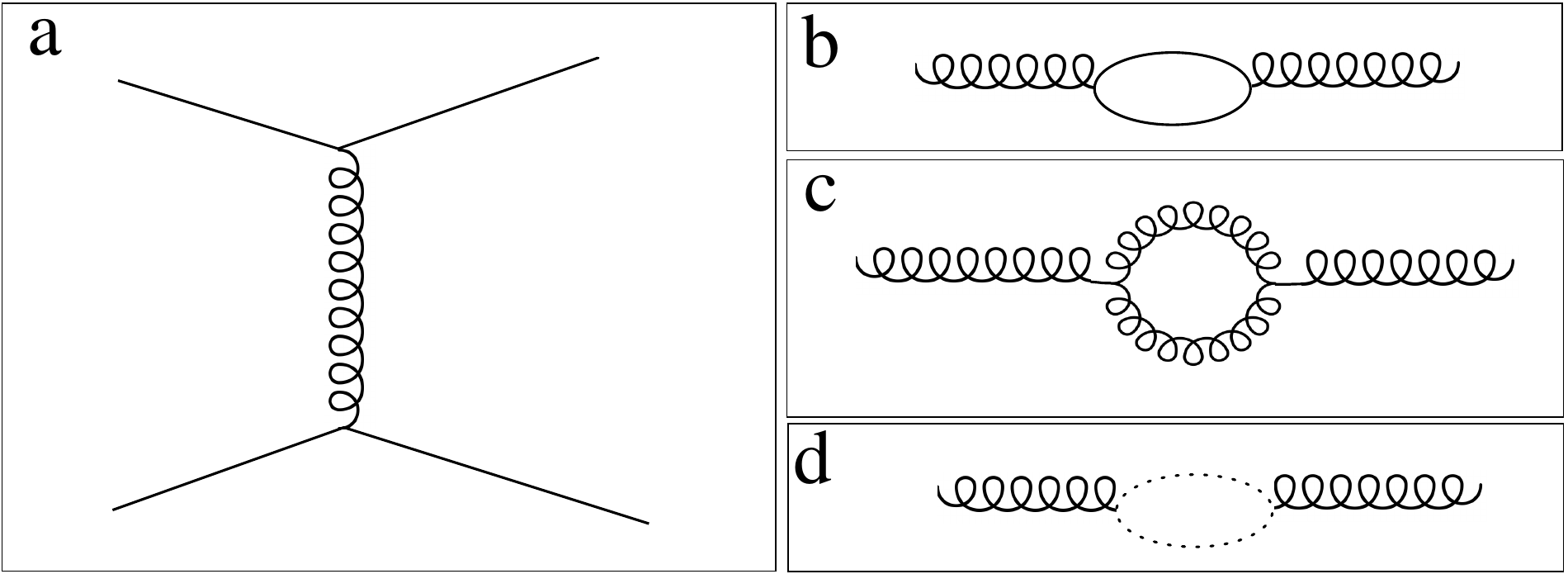}
\caption{\label{Flo:1st order corrections} \small Panel (a): lowest order quark--quark scattering (no quantum loops). 
Panels (b), (c) and (d): respectively, quark, gluon and ghost
loop on the gluon propagator.}
\end{figure}
An interaction at the classical level is shown by the tree diagram of Fig.\,\ref{Flo:1st order corrections}a. 
Quantum effects, which are responsible for the running of the effective coupling, appear as loops attached to the propagators or vertices of the tree diagram. 

We first compute the quark loop influence on the gluon propagator -- see Fig.~\ref{Flo:1st order corrections}b.
QCD's Feynman rules yield:
\begin{equation}
D_q^{\mu\nu}\left(q\right)=-\frac{\overline{\alpha_s}}{3\pi}\left(q^{\mu}q^{\nu}-\eta^{\mu\nu}q^2\right)\ln\left(\frac{Q^2}{\mu^2}\right)n_f\frac{\delta_{ab}}2,
\label{eq:quark loop}
\end{equation}
with $\eta^{\mu\nu}$ being the metric tensor and $a$, $b$, gluon color indices.
Equation~\eqref{eq:quark loop} is similar to the one giving the running of the QED coupling. 

Gluon loops -- Fig.~\ref{Flo:1st order corrections}c -- supply an additional  contribution to the  gluon propagator:
\begin{equation}
D_{g}^{\mu\nu}\left(q\right)=\frac{\overline{\alpha_s}}{4\pi}N_c\delta_{ab}\left[\frac{11}{6}q^{\mu}q^{\nu}-\frac{19}{12}\eta^{\mu\nu}q^2
+\frac{1-\xi}2\left(q^{\mu}q^{\nu}-\eta^{\mu\nu}q^2\right)\right]\ln\left(\frac{Q^2}{\mu^2}\right),\label{eq:gluon loop}
\end{equation}
where $\xi$ is a gauge fixing term ($\xi = 0$ is Landau gauge). 
In contrast to the quark loop case, $D_{g}^{\mu\nu}(q) \not\propto \left(q^{\mu}q^{\nu}-\eta^{\mu\nu}q^2\right)$, \textit{viz}.\ gluon loops introduce a longitudinal component to the gluon propagator resulting in a violation of current conservation: $D_{g}^{\mu\nu}q_{\mu}\neq0$.
This can be addressed by using a gauge in which gluons are always transverse, \textit{e.g}., axial gauge $A_3^a=0$ or light-cone gauge $A^{+}_a\equiv A_0^a+A_3^a=0$. 

More commonly, the quantum analogue of gauge invariance (BRST symmetry \cite{Becchi:1975nq, Tyutin:1975qk}, \cite[Ch.\,II]{Pascual:1984zb}, for Becchi-Rouet-Stora-Tyutin) is ensured by introducing Faddeev-Popov ghost fields \cite{Faddeev:1967fc} whose characteristics ensure that their loop contribution -- Fig.~\ref{Flo:1st order corrections}d -- cancels the gluon propagator's longitudinal component.  
These ghosts are scalar Grassmann fields (so obey Fermi--Dirac statistics despite being spin-0 fields), whose one-loop contribution is
\begin{equation}
D_{gh}^{\mu\nu}\left(q\right)=-\frac{\overline{\alpha_s}}{4\pi}N_c\delta_{ab}\left[\frac{1}{6}q^{\mu}q^{\nu}+\frac{1}{12}\eta^{\mu\nu}q^2\right]\ln\left(\frac{Q^2}{\mu^2}\right).
\label{eq:ghosts}
\end{equation}
Plainly, $(D_{g}^{\mu\nu}+D_{gh}^{\mu\nu})$ is purely transverse. 

QCD's fermion self-energy (Fig.~\ref{quantum_effects}b) and vertex (Fig.~\ref{quantum_effects}d)  corrections also contribute to the running coupling, in contrast to QED, where they cancel against each other. 
Using the $\overline{\rm MS}$ RS, the self-energy and vertex corrections are, respectively, 
\begin{equation}
G_q\left(p\right)=\frac{\not \! p}{p^2}\delta_{ab}\left[1-\xi\frac{\overline{\alpha_s}}{4\pi}\frac{N_c^2-1}{2N_c}\ln\left(\frac{-p^2}{\mu^2}\right)\right]\,
\label{eq:quark_self_E}
\end{equation}
and
\begin{equation}
\Gamma_{\mu}^{\alpha\beta;a}\left(q\right)=-i\sqrt{4\pi\overline{\alpha_s}}\frac{\lambda_{\alpha\beta}^{a}}2\gamma_{\mu}\left[1-\frac{\overline{\alpha_s}}{4\pi}\ln\left(\frac{Q^2}{\mu^2}\right)\left\{ \xi\frac{N_c^2-1}{2N_c}+N_c\left(1-\frac{1-\xi}{4}\right)\right\} \right],
\label{eq:vertex_cor}
\end{equation}
where $\lambda_{\alpha\beta}^{a}$ are the Gell-Mann matrices, with $\alpha, \beta$ the color indices of the two quark lines.
There is no tadpole contribution (Fig.~\ref{quantum_effects}f) to the running coupling in the $\overline{\rm MS}$ RS.

In the $\overline{\rm MS}$ RS, the coupling is independent of the gauge parameter-- see footnote~\ref{fn:gauge-dependence}. 
Thus, to sum the amplitudes of the five graphs that use the propagators, Eqs.\,\eqref{eq:quark loop}-\eqref{eq:quark_self_E} and vertex, Eq.\,\eqref{eq:vertex_cor}, no generality is lost by choosing a convenient value for $\xi$, \textit{e.g}., the Landau gauge $\xi=0$. 
This yields the next-to-leading order (NLO) quark--quark interaction amplitude:
\begin{equation}
\mathcal{M}=\mathcal{M}_{Born}\left[1+\frac{\overline{\alpha_s}}{4\pi}\left\{ \frac{2n_f}{3}-\frac{13N_c}6-\frac{3N_c}2\right\}\ln\left(\frac{Q^2}{\mu^2}\right) \right]\,,
\end{equation}
where the $\mathcal{M}_{Born}$ factor is the tree diagram amplitude. 
The ``$2n_f/3$'' term originates from the quark loop, 
the ``$-13N_c/6$'' term from the gluon and ghost loops, and 
the ``$-3N_c/2$'' term from the vertex correction. 
Plainly, the quark self-energy correction, Eq.\,\eqref{eq:quark_self_E}, does not contribute to the one-loop $\beta$-function in Landau gauge.

Absorbing the quantum corrections into the coupling constant, $\overline{\alpha_s}$, defines the effective coupling $\alpha_s(Q^2)$:
\begin{equation}
\alpha_s\left(Q^2\right)=\alpha_s\left(\mu^2\right)\left[1+
\frac{\alpha_s\left(\mu^2\right)}{4\pi}\frac{2n_f-11N_c}{3}\ln\left(\frac{Q^2}{\mu^2}\right) \right].
\end{equation}
For small $\alpha_s(\mu^2)\ln(\sfrac{Q^2}{\mu^2})$, this yields:
\begin{equation}
\frac{4\pi}{\alpha_s\left(Q^2\right)}=\frac{4\pi}{\alpha_s\left(\mu^2\right)}+\frac{11N_c-2n_f}{3}\ln\left(\frac{Q^2}{\mu^2}\right).
\label{eq:alpha_s_1loop_1}
\end{equation}
Differentiating with respect to $Q^2$ gives:
\begin{equation}
\frac{d\alpha_s\left(Q^2\right)}{\alpha_s^2\left(Q^2\right)}=-\frac{1}{4\pi}\frac{11N_c-2n_f}{3}\frac{dQ^2}{Q^2},
\end{equation}
{\it viz}, from the definition of $\beta$ and its series expansion, Eq.~(\ref{eq:alpha_s beta series}):
\begin{equation}
\beta_0=\frac{11N_c - 2n_f}{3}.\label{eq:beta0}
\end{equation}
It is interesting to note that for $N_c=0$, $\beta_0=-\tfrac{2}{3}n_f$, \textit{i.e}., 
one recovers the 1-loop Abelian running of QED with $n_f$ fermions.
In fact, with careful attention to $N_c$ rescaling of the SU($N_c$) Casimir factor, $C_F$, and also $n_f$, one can assign a finite $N_c\to 0$ limit to colored quantities that yields a theory with Abelian character in this limit \cite{Brodsky:1997jk, Brodsky:1998kb, Brodsky:2015eba}.~\label{NC-->limit}

\subsection{Calculation using renormalization constants \label{alpha_s from renorm Z_i}}
In the previous subsection, $\alpha_s$ was obtained from amplitudes already regularized and renormalized. It is also interesting to derive $\alpha_s$ using renormalization constants because it illuminates the connection between the running and renormalization.
Furthermore, this method is often used to compute $\alpha_s$ -- \textit{e.g}., Refs.\,\cite{Celmaster:1979km, Muta:1998vi}), including on the IR domain. 

The QCD Lagrangian, Eq.~(\ref{eq:QCD Lagrangian}), can be separated into gluon and quark parts, 
$\mathcal{L}_{QCD}=\mathcal{L}_g+\mathcal{L}_q$, where
\begin{equation}
\mathcal{L}_g = - \frac{1}{4}F^{\mu\nu}_a F_{\mu\nu}^a, \quad
\mathcal{L}_q=\sum_f \overline{\psi}_i^{(f)}\big(i\gamma_{\mu}D^{\mu}_{ij} -m_f\delta_{ij}\big)\psi_j^{(f)}\,.
\end{equation}
After gauge fixing, introducing the Faddeev-Popov ghost fields, two additional terms appear:
\begin{equation}
\mathcal{L}_{gf} = -\frac{1}{2\xi}\big(\partial^\mu A^a_\mu \big)^2; \qquad 
\mathcal{L}_{fp}= i\big(\partial^\mu {\bar \chi^a})D^{ab}_\mu \chi^b, 
\end{equation}
being, respectively, a gauge fixing term and a ghost field term. Here, $\chi^{a}$ is the ghost field 
and $D^{ab}_\mu = -\partial_\mu \delta^{ab}+g f^{abc} A_\mu^c$.
Combining all contributions, one obtains the complete Lagrangian density:
\begin{equation}
\mathcal{L}_{QCD}=\mathcal{L}_g+\mathcal{L}_q+\mathcal{L}_{gf}+\mathcal{L}_{fp}.
\label{QCD FP Lagrangian}
\end{equation}

$\mathcal{L}_{QCD}$ can be rewritten as a renormalized Lagrangian using renormalization constants and counterterms \cite[Ch.\,III]{Pascual:1984zb}. 
The renormalization constants depend on the regularization scale, $\mu_{\rm R}$ -- implicit in the definitions of the bare fields, coupling and masses, and the renormalization point, $\mu$.  They are defined thus:
\begin{equation}
\label{eq:Z_renorm_cst}
\begin{array}{ll}
Z_\alpha(\mu^2,\mu_{\rm R}^2) = \overline{\alpha_s}(\mu_{\rm R}^2)/\hat\alpha_s(\mu^2); &
%
Z_m(\mu^2,\mu_{\rm R}^2) = m(\mu_{\rm R}^2)/\hat{m}(\mu^2); \\ 
Z_2(\mu^2,\mu_{\rm R}^2) = \big(\psi(\mu_{\rm R}^2) / \hat\psi(\mu^2)\big)^2; & 
Z_3(\mu^2,\mu_{\rm R}^2)=\big(A_\mu^a(\mu_{\rm R}^2)/\hat{A_\mu^a}(\mu^2)\big)^2; \\
\tilde{Z}_3(\mu^2,\mu_{\rm R}^2)=\big(\chi_i(\mu_{\rm R}^2) / \hat{\chi^a}(\mu^2)\big)^2 . &
\end{array} 
\end{equation}
Here we have used a circumflex to highlight the $\mu^2$-dependent renormalized quantities.  Subsequently,  it is typically suppressed for notational simplicity.

In dimensional regularization (MS and $\overline{\rm MS}$ RSs), UV divergences are expressed in 
$1/\varepsilon$ poles that appear when calculating Feynman graph integrals in $4-2\varepsilon$ dimensions. 
Further, as a consequence of working in $D=4-2\varepsilon$ spacetime dimensions, the coupling acquires a mass dimension.
(This may be seen, \textit{e.g}., by constructing the $D$-dimensional action associated with Eq.\,\eqref{eq:QCD Lagrangian}.)
Introducing $\mu_0$, an arbitrary reference scale, the product $\alpha_s^{(\varepsilon)}(\mu^2)=\alpha_s(\mu^2)/\mu_0^{2\varepsilon}$ is dimensionless.
%
%
Consequently, when using the MS or $\overline{\rm MS}$ RS, the first entry in Eq.\,\eqref{eq:Z_renorm_cst} becomes:
\begin{equation}
Z_\alpha(\mu^2,\mu_0^2)=\bigg(\frac{\mu_0}{\mu}\bigg)^{2\epsilon} ~ \frac{\overline{\alpha_s}^{(\varepsilon)}(\mu_0^2)}{\alpha_s^{(\varepsilon)}(\mu^2)}.
\label{eq:alpha_s renorm}
\end{equation}
$Z_{\alpha}$ thus defines the scale dependence of $\alpha_s^{(\varepsilon)}$, respecting $\alpha_s^{(\varepsilon \to 0)} \to \alpha_s$; and  
$\alpha_s(Q^2)=Z_{\alpha}(Q^2,\mu^2)\alpha_s(\mu^2)$, so 
$Z_\alpha(Q^2,\mu^2) = Z_\alpha(Q^2,\mu_0^2)/Z_\alpha(\mu^2,\mu_0^2)$.

In fact, the collection of renormalization constants $\{Z_{\alpha}\}$ constitute a group, with 
an identity element, $Z_{\alpha}(Q^2,Q^2)=1$;
each element possessing an inverse, 
$Z_{\alpha}(Q^2,\mu^2)^{-1}=1/Z_{\alpha}(\mu^2,Q^2)$;
and a composition law,
$Z_{\alpha}(Q^2,\mu^2)=Z_{\alpha}(Q^2,\mu_0^2)Z_{\alpha}(\mu_0^2,\mu^2)$.
At this point, requiring observables to be invariant under $Q\to Q+\delta Q$ leads to the RGE; and using Eq.\,\eqref{eq:alpha_s beta series}, one obtains the leading-order result
\begin{equation}
Z_{\alpha}(Q^2,\mu_0^2)=1-\frac{1}{\varepsilon} \frac{1}{4\pi}
 \beta_0 \,\alpha_s^{(\varepsilon)}(Q^2) 
\label{eq:z_alpha_beta0}.
\end{equation}
The $1/\varepsilon$ factor, necessary to cancel the UV-divergence, is evident.

Rewriting $\mathcal{L}_{QCD}$ in terms of renormalized fields, masses, coupling, and gauge parameter, one obtains \cite[Ch.\,III]{Pascual:1984zb}:
\begin{equation}
\mathcal{L}_{QCD}=\hat{\mathcal{L}}_g+\hat{\mathcal{L}}_q+\hat{\mathcal{L}}_{gf}+\hat{\mathcal{L}}_{fp} -
(Z_3-1)\mathcal{L}_g + (Z_2-1)\mathcal{L}_q+(\tilde{Z}_3-1)\mathcal{L}_{fp}+\mbox{mixing}
\end{equation}
where the $\hat{\mathcal{L}}$ are the same as their corresponding $\mathcal{L}$, except written using renormalized quantities; and the counterterms are expressed in the $(Z_i-1)\mathcal{L}$ and mixing terms.
All counterterms are functions of $1/\varepsilon$, deliberately constructed to cancel the $1/\varepsilon$ poles that represent the regularized UV divergences. 

It is convenient to employ an equivalent set of renormalization constants:
\begin{equation}
Z_1=Z_\alpha^{\sfrac{1}{2}} Z_3^{\sfrac{3}{2}}; \quad
\tilde{Z}_1=Z_\alpha^{\sfrac{1}{2}} Z_3^{\sfrac{1}{2}} \tilde{Z_3}; \quad
Z_{1q}=Z_\alpha^{\sfrac{1}{2}} Z_2 Z_3^{\sfrac{1}{2}}; \quad
Z_4=Z_\alpha Z_3^2.
\label{Eq:alternate_Zs}
\end{equation}
They correspond, respectively, to the renormalization constants for the  3-gluon, ghost-gluon, quark-gluon, and 4-gluons vertices.
The omnipresent $Z_\alpha$ indicates that quantum effects can be incorporated into the bare couplings of four different interaction vertices in $\mathcal{L}_{QCD}$. 
That the same running results from the distinct vertices is, {\it a priori}, not assured, but
this is guaranteed, at every order in perturbation theory, by the Slavnov-Taylor identities (STIs) \cite{Slavnov:1972fg, Taylor:1971ff} which owe to the BRST symmetry of QCD, {\it viz}.
%
$\sfrac{Z_1}{Z_3}=\sfrac{\tilde{Z}_1}{\tilde{Z}_3}=\sfrac{Z_{1q}}{Z_2}=\sfrac{Z_4}{Z_1}.$

Thanks to the STIs, $Z_\alpha$, and therefore $\alpha_s(Q^2)$ {\it via} Eq.\,\eqref{eq:alpha_s renorm}, can be calculated in different ways, in particular: 
\begin{equation}
Z_\alpha=\frac{\tilde{Z}_1^2}{\tilde{Z}_3^2 Z_3},
\label{eq:zg}
\end{equation}
where ${\tilde{Z}_1}$, $\tilde{Z}_3$ and $Z_3$ are obtained by computing the renormalized ghost-gluon vertex, ghost self-energy and gluon self-energy, respectively. 
This yields, at first (1-loop) order:
\begin{subequations}
\label{eq:z1z3z3}
\begin{eqnarray}
\tilde{Z}_1(Q^2,\mu_0^2)&=&
1- \frac{1}{\varepsilon}\alpha_s^{(\varepsilon)}(Q^2) \frac{N_c\xi}{8\pi};  \\
\tilde{Z}_3(Q^2,\mu_0^2)&=& 1- 
\frac{1}{\varepsilon}\alpha_s^{(\varepsilon)}(Q^2) \frac{N_c(3-\xi)}{16\pi};   \\
Z_3(Q^2,\mu_0^2)&=& 1- 
\frac{1}{\varepsilon}
\alpha_s^{(\varepsilon)}(Q^2) \frac{\tfrac{N_c}{2}(\tfrac{13}{3}-\xi)-\tfrac{2}{3}n_f}{4\pi}.
\end{eqnarray}
\end{subequations}
Inserting Eqs.\,\eqref{eq:z1z3z3} into Eq.\,\eqref{eq:zg}, one finds at first order
\begin{equation}
Z_\alpha= 1- \alpha_s^{(\varepsilon)}(Q^2)\frac{11N_c-2n_f}{12\pi\epsilon},
\label{eq:zg2}
\end{equation}
where the gauge parameter, $\xi$, has cancelled, as expected for a $\overline{\rm MS}$ derivation. 

The leading $\beta$-function coefficient, $\beta_0$, is readily obtained by comparing Eqs.\,\eqref{eq:z_alpha_beta0} and (\ref{eq:zg2}). 
Alternatively, it can be computed from $\alpha_s^{(\varepsilon)}(Q^2)$ as obtained with Eqs.\,\eqref{eq:alpha_s renorm}, (\ref{eq:zg2}). 
Differentiating the resulting $\alpha_s^{(\varepsilon)}(Q^2)$ with respect to $Q^2$, taking the $\varepsilon \to 0$ limit, and recalling the definition of $\beta(\alpha_s)$ -- Eq.\,\eqref{eq:alpha_s beta series}, then:
\begin{equation}
\beta(\alpha_s)=-\frac{\alpha_s^2}{4\pi}\frac{11N_c-2n_f}{3} \;
\Rightarrow \; \beta_0 = (11N_c - 2n_f)/3\,.
\end{equation}

It is readily established that the other equalities in Eq.\,\eqref{Eq:alternate_Zs} provide the same result -- see, \textit{e.g}., the derivation in \cite{Deur:2016tte} using the gluon self-energy, quark self-energy and 3-gluon vertex renormalization factors, $Z_3$, $Z_2$, $Z_1$, respectively.

\subsection{Solution of the 1-loop RGE}
The solution of Eq.\,\eqref{eq:alpha_s beta series} is known exactly and analytically only at 1-loop ($\beta_0$) order. 
It is given by Eq.\,\eqref{eq:alpha_s_1loop_1}, which exemplifies the generic fact that pQCD calculations can only provide scale-evolution  rather than absolute quantities (for rare exceptions to that rule, see Ref.\,\cite{Deur:2018roz}).  
This is  even more clearly seen by introducing the renormalisation group independent QCD scale parameter $\Lambda_s$, defined at $\beta_0$-order as:
\begin{equation}
\Lambda_s^2 \equiv\mu^2e^{-\sfrac{4\pi}{\beta_0\alpha_s(\mu^2)}}.
\label{eq:lambda-1loop}
\end{equation}
Together with Eq.\,\eqref{eq:alpha_s_1loop_1}, this yields:
\begin{equation} 
\label{Eq.one loop}
\alpha_s(Q^2)=\frac{4\pi}{\beta_0 \ln\left(Q^2/ \Lambda_s^2\right)}.
\end{equation}

\section{Higher order results for $\beta_n$ and $\alpha_s(Q^2)$ \label{alpha_s n-loops}}
The analogue of Eq.\,\eqref{eq:alpha_s_1loop_1} at two loops ($\beta_1$ order) is also exact and analytical:
\begin{equation}
\frac{4\pi}{\alpha_s \big( Q^2 \big)}-\frac{ \beta_1}{\beta_0} \ln\bigg(\frac{4\pi}{\alpha_s\big(Q^2\big)}+\frac{ \beta_1}{\beta_0}\bigg)= \frac{4\pi}{\alpha_s\big(\mu^2\big)}-\frac{ \beta_1}{\beta_0} \ln\bigg(\frac{4\pi}{\alpha_s\big(\mu^2\big)}+\frac{ \beta_1}{\beta_0}\bigg)+\beta_0 \ln\bigg(\frac{Q^2}{\mu^2}\bigg);
\end{equation}
but this equation does not yield an exact, analytic solution for $\alpha_s$. An exact, nonanalytic solution is \cite{Gardi:1998qr}:
\begin{equation}
\alpha_s(Q^2)=-\frac{4\pi\beta_0}{ \beta_1}\frac{1}{1+W_{-1}\left(z\right)}\,,
\quad z = -\frac{1}{e}\frac{\beta_0}{\beta_1}\left(\frac{\Lambda_s^2}{Q^2}\right)^{\beta_0^2/ \beta_1}\,,
\label{eq:GGK beta1}
\end{equation}
where $e = \exp(1)$ and $W_{-1}\left(z\right)$ is the lower branch of the real-valued solution of $z=W(z)e^{W(z)}$, with $W(z)$ being the Lambert function.
An approximation to $W_{-1}$ is given in Ref.\,\cite{BARRY200095}.
At $\beta_1$-order, the scale parameter is approximately:
\begin{equation}
 \Lambda_s^2 \approx Q^2\left(\frac{4\pi}{\beta_0\alpha_s\left(Q^2\right)}+\frac{ \beta_1}{4\pi\beta_0}\right)^{\tfrac{ \beta_1}{4\pi\beta_0}}e^{-\tfrac{4\pi}{\beta_0\alpha_s\left(Q^2\right)}},
 \label{eq:lambda-2loop}
\end{equation}
as can be inferred from Eq.\,\eqref{eq:alpha_s} truncated at order $\beta_{1}^1$. 
Its exact expression, valid nonperturbatively, is available -- see, \textit{e.g}., Ref.\,\cite{Aoki:2021kgd}.

No exact solution at the 3-loop ($ \beta_2$) order is known. The following approximation is derived in Ref.\,\cite{Gardi:1998qr}:
\begin{equation}
\alpha_s(Q^2)=-\frac{4\pi\beta_0}{ \beta_1}\frac{1}{1-\frac{ \beta_2\beta_0}{ \beta_1^2}+W_{-1}\left(z\right)}\,,\quad
z = -\frac{1}{e}\frac{\beta_0}{ \beta_1}\left(\frac{\Lambda_s^2}{Q^2}\right)^{\sfrac{\beta_0^2}{\beta_1}}e^{\sfrac{ \beta_2\beta_0}{ \beta_1^2}}\,.
\label{eq:GGK beta2}
\end{equation}

A more commonly employed approximation is that obtained using an iterative method \cite{Chetyrkin:1997sg},
which can systematically be used to higher orders. 
It is presently known at five-loop ($\beta_4$) order in the $\overline{\rm MS}$ RS \cite{Kniehl:2006bg}, and written with $t=\ln Q^2/\Lambda_s^2$:
{\allowdisplaybreaks
\begin{align}
\alpha_s^{\overline{\rm MS}}& (Q^2) \nonumber \\
& =\frac{4\pi}{\beta_0t} \left\{ 1-\frac{ \beta_1}{\beta_0^2}\frac{\ln(t)}{t} 
+\frac{ \beta_1^2}{\beta_0^{4}t^2}
\bigg(\ln^2(t)-\ln(t) -1+\frac{\beta_2\beta_0}{ \beta_1^2}\bigg) \right. \nonumber \\
& \quad + \frac{ \beta_1^{3}}{\beta_0^{6}t^{3}}\biggl(-\ln^3(t)+\frac{5}2\ln^2(t)  
 +2 \ln(t)-\frac{1}2-3\frac{\beta_2\beta_0}{ \beta_1^2}\ln(t)+\frac{ \beta_3\beta_0^2}{2 \beta_1^{3}}\biggr)
 \nonumber \\
& \quad + \frac {\beta_1^4} {\beta_0^8 t^4} \bigg(\ln^4(t) -\frac{13} {3} \ln^3(t) -\frac{3} 2  \ln^2(t)  +4 \ln(t) +\frac{7}{6} +  \frac {3\beta_2 \beta_0}{\beta_1^2} \left( 2 \, \ln^2(t) - \ln(t) -1\right) 
\nonumber \\
& \qquad \left.  -  \frac{\beta_3\beta_0^2}{\beta_1^3}\left[2 \, \ln(t) + \frac{1}{6}\right] +\frac{5\beta_2^2 \beta_0^2}{3\beta_1^4}+\frac{\beta_4 \beta_0^3}{3 \beta_0^4} \biggr) \right\}\,.
\label{eq:alpha_s}
\end{align}
In the $\overline{\rm MS}$ RS (and other RSs for $n\leq1$), the $\beta_n$-values are
\begin{subequations}
\label{eq:beta4 list}
\begin{align}
\beta_0 & = 11-\frac2{3}n_f,  \label{eq:beta0 list} \\
\beta_1 & =102-\frac{38}{3}n_f,  \\
\beta_2 & = \frac{2857}2-\frac{5033}{18}n_f+\frac{325}{54}n_f^2,  \label{eq:beta2 list} \\
\beta_3 & =  \frac{149753}{6}+3564 {\zeta_3} -\bigg(\frac{1078361}{162}+\frac{6508}{27}\zeta_3 \bigg)n_f \nonumber \\
& \qquad +\bigg(\frac{50065}{162}+\frac{6472}{81} \, \zeta_3 \bigg)n_f^2+\frac{1093}{729}n_f^{3},  \\
\beta_4 & = \frac{8157455}{16} + \frac{621885}2\zeta_3 - \frac{88209}2\zeta_4 -288090 \zeta_5 \nonumber \\
& \qquad +\bigg(-\frac{336460813}{1944} -\frac{4811164}{81}\zeta_3 + \frac{33935}{6}\zeta_4 +\frac{1358995}{27}\zeta_5 \bigg) n_f  \nonumber \\
& \qquad + \bigg(\frac{25960913}{1944} + \frac{698531}{81}\zeta_3
-\frac{10526}{9}\zeta_4 - \frac{381760}{81}\zeta_5 \bigg)n_f^2   \nonumber \\
& \qquad + \bigg(-\frac{630559}{5832} -\frac{48722}{243}\zeta_3 
+ \frac{1618}{27}\zeta_4 + \frac{460}{9}\zeta_5 \bigg)n_f^3 \nonumber \\
& \qquad +\bigg(\frac{1205}{2916} -\frac{152}{81}\zeta_3 \bigg)n_f^4,   
\end{align}
\end{subequations}
where $\zeta_n$ is the Riemann zeta function. 
}

The QCD values of $\beta_2$ and $\beta_3$ in the MOM (Landau gauge) and V RSs are collected in Ref.\,\cite[Sec.\,3.2]{Deur:2016tte}.  
More generally, expressions for $\beta_{0,1,2,3}$ appropriate to all interactions in the Standard Model (including QCD) have recently been derived in the $\overline{\rm MS}$ RS \cite{Davies:2019onf, Bednyakov:2021qxa}. 
Direct relations between $\alpha_s$ calculated in $\overline{\rm MS}$ and results in various implementations of the MOM RS, up to 3 loops, are provided in Refs.\,\cite{Chetyrkin:2000fd, Chetyrkin:2000dq}.  

\begin{figure}[t]
\centering
\includegraphics[width=0.6\textwidth]{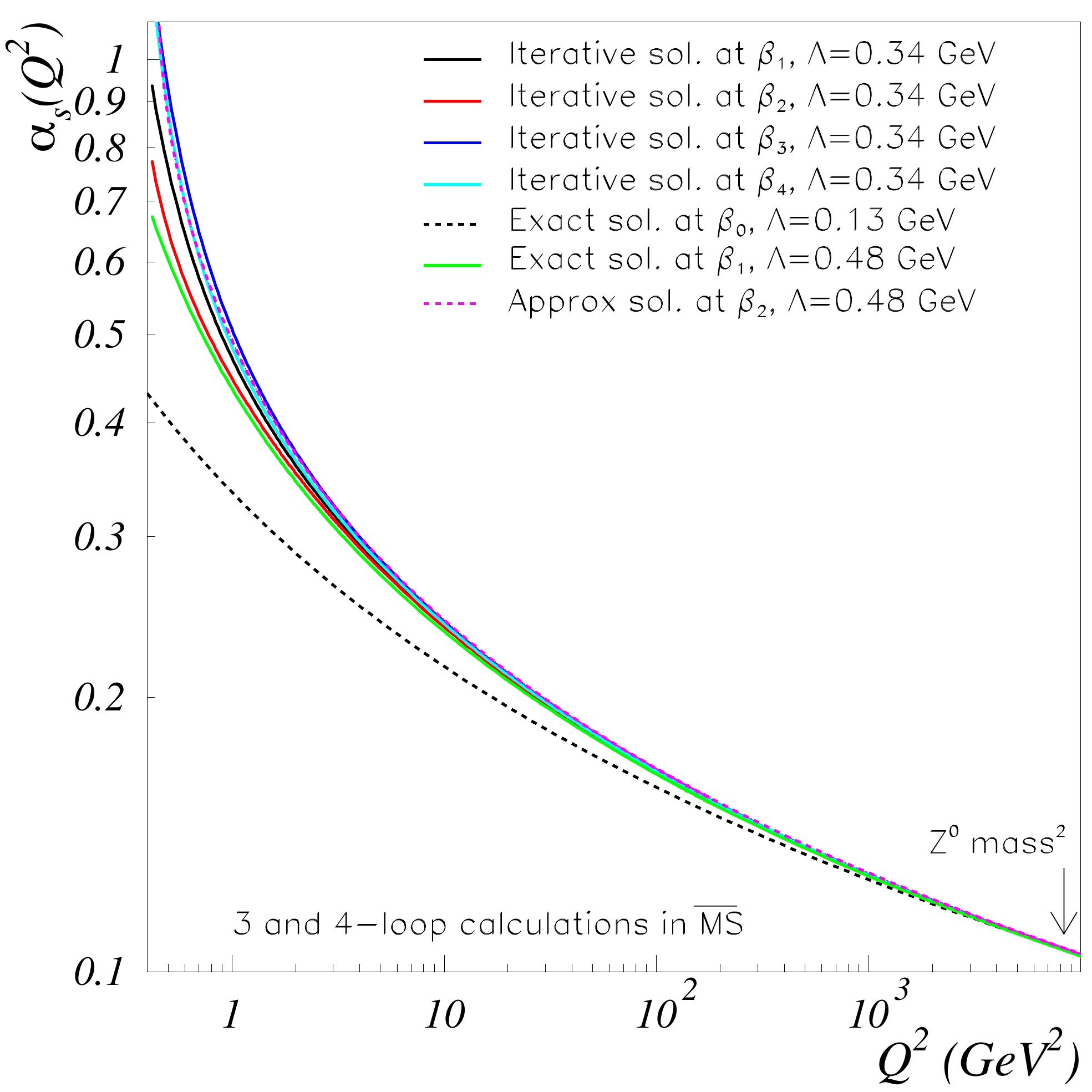}
\caption{\label{Flo:alpha_s from pQCD} \small 
The QCD coupling $\alpha_s(Q^2)$ computed perturbatively at different loop orders.
The exact solutions used Eq.\,\eqref{Eq.one loop}, valid at $\beta_{0}$ (1-loop), and 
Eq.\,\eqref{eq:GGK beta1}, valid up to $\beta_{1}$ (2-loop), and are shown by the dashed 
black and solid green lines, respectively. 
The approximate solution from Eq.\,\eqref{eq:GGK beta2} for $\beta_{2}$ (3-loop) is shown by the magenta dashed line.
The solid black, red, blue and cyan lines use the iterative method, Eq.\,\eqref{eq:alpha_s}, 
at orders $\beta_{1}$, $\beta_{2}$, $\beta_{3}$ (4-loop) and $\beta_{4}$ (5-loop), respectively. 
We use $n_{f}=3$ and $\Lambda_s=0.34$ GeV for the iterative method and 0.13\,GeV or 0.48\,GeV for the others (see legend) in order for the calculations to match near the $Z^0$ mass squared $M^2_Z$, the conventional scale at which the coupling UV value is usually computed (arrow).
The $\overline{\rm MS}$ scheme is used for the $\beta_{2}$ and $\beta_{3}$ computations.}
\end{figure}

Comparisons between exact and approximate estimates of $\alpha_s$ at orders up to $ \beta_4$ are provided in Fig.~\ref{Flo:alpha_s from pQCD}.  
The relative behavior of all curves and the similarity between the solid cyan curve (iterative method at 5-loop) and the dashed magenta curve (3-loop approximate solution using the Lambert function, Eq.\,\eqref{eq:GGK beta2}) suggests that the approximation to $\alpha_s$ provided by the latter should be favored over the $\beta_{2,3,4}$ iterative results.
It is also worth noting that the different estimation methods require distinct values of $\Lambda_s$ in order to reach agreement at $Q^2=M_{Z}^2$.  Thus, in addition to being RS dependent, the value of $\Lambda_s$ also depends on the method used for its computation within a given scheme.  This emphasizes again that $\Lambda_s$ is not an observable.

Once the $\beta_i$ are determined to the desired order, $\Lambda_s$ remains the only unknown parameter
in Eq.\,\eqref{eq:alpha_s}. 
Since that formula is a monotonic function of $Q^2$ then, at any value of $Q^2$ on the perturbative domain within which it is valid, one can use Eq.\,\eqref{eq:alpha_s} to unambiguously determine the value of $\alpha_s(Q^2)$ from either its value at a given scale, chosen by convention, often $Q^2 = M_{\rm Z}^2$, or from the value of $\Lambda_s$.
Thus the perturbative result for $\alpha_s$ is completely determined by the value of a given observable at one particular scale.  
As noted above, this provides a stringent test on the internal consistency of QCD: results obtained from two distinct observables at different scales can be compared by evolving each to a common scale using Eq.\,\eqref{eq:alpha_s}.  
Modern comparisons \cite[Sec.\,9.4]{Workman:2022ynf} establish the universality of the perturbative running coupling and thereby the RGE, Eq.\,\eqref{eq:alpha_s beta series}, and the BRST symmetry of QCD, at least perturbatively. 

\section{$\Lambda_s$: the QCD scale parameter}
The renormalization group invariant mass-scale $\Lambda_s$ originally appears as an integration constant -- see Eq.\,\eqref{eq:simple_coupling}. 
It controls the rate at which $\alpha_s(Q^2)$ evolves with changes in $Q^2$ on any domain within which perturbation theory is valid.
On the other hand, $\Lambda_s$ also marks the scale whereat $\alpha_s \to \infty$ and perturbation theory has certainly failed.
In the absence of quark current-masses, $\Lambda_s$ is the only scale present in pQCD and this explains why it is called the ``(p)QCD scale parameter''.

\subsection{$\Lambda_s$ as the evolution rate parameter \label{lambda_s}}
As revealed by Eqs.\,\eqref{Eq.one loop}, \eqref{eq:alpha_s}, $\Lambda_s$ determines the pace at which the perturbative running coupling evolves: in any pQCD expression, a larger value of $\Lambda_s$ places the Landau pole at a larger value of $Q^2$; consequently, the coupling appears to fall more quickly with increasing $Q^2$ on any domain upon which the original expression is valid.
However, this rate is not necessarily physical because $\alpha_s$ need not be an observable. 
In fact, past 2-loop order in pQCD, the value of $\Lambda_s$ becomes RS-dependent, therefore arbitrary, with some RSs implying faster evolution than others. 

The determination of $\Lambda_s$ (or $\alpha_s$ at some conventional scale) demands either an actual measurement or nonperturbative calculations.
Since the measurement is interpreted using a pQCD series, which depends on the RS,
the value of $\Lambda_s$ thereby obtained is RS-dependent. 
 The same is true if $\Lambda_s$ is extracted from a nonperturbative LGT calculation since the latter is matched in the UV to a perturbative quantity expressed in a specific RS.
Table~\ref{Flo:Table of Lambda} provides examples of values for commonly used RSs.
\begin{table}[t]
\centering
\begin{tabular}{|c|c|c|c|c|c|c|}
\hline 
\rule{0ex}{2.5ex}RS/eff. charge & ${\overline{\rm MS}}$  & MS & MOM & V or R & $g_1$ & $\tau$\tabularnewline
\hline 
 $\Lambda_s$ (GeV) & 0.34 & 0.30 & 0.62 & 0.48 & 0.92 & 1.10 \tabularnewline
\hline
\end{tabular}
\caption{
\label{Flo:Table of Lambda} \small Approximate values of  $\Lambda_s$ for 
commonly used renormalization schemes, determined in the momentum range where $n_f=3$.}
\end{table}

The relation between different $\Lambda_s$ values is obtained at first order by taking the ratio of Eq.\,\eqref{eq:lambda-1loop} expressed in the two RSs.
If one works at leading- or next-to-leading-order in the coupling, then $\alpha_s^\prime = \alpha_s$ -- see Eq.\,\eqref{eq:alpha_rel_2RS}; hence, $\Lambda_s^\prime=\Lambda_s$.
More generally, one considers the 2-loop order expression for $\Lambda_s$, Eq.\,\eqref{eq:lambda-2loop}, and exploits the fact that $\Lambda_s$ is scale-independent \cite{Celmaster:1979km}, to derive the following all-orders result:
\begin{equation}
\Lambda^\prime_s=\Lambda_se^{\frac{2\pi\mathrm{v}_2}{\beta_0}},
\label{eq:lambda relation}
\end{equation}
with $\mathrm{v}_2$ the (scale independent) leading-order difference between the couplings 
in the two RSs -- see Eq.\,\eqref{eq:alpha_rel_2RS}. 
For instance, the $\overline{\rm MS}$ and MOM scale parameters are related
by 
\begin{equation}
\Lambda_{\rm MOM}=\Lambda_{\overline{\rm MS}} {\rm e}^{\frac{507 - n_f 40}{792-n_f 32 /3}},
\end{equation}
%
yielding, for $n_f=3$, $\Lambda_{\rm MOM}=1.817\Lambda_{\overline{\rm MS}}$. 

We close this subsection by mentioning that one could define $\Lambda_s$ so that it is RS-independent \cite{Sonoda:1994cu}; or, once a RS is chosen, relate $\Lambda_s$ to a physical scale -- more on the latter possibility below.


\subsection{$\Lambda_s$ as the confinement scale \label{Landau pole}}
The Landau pole designates the divergence of the perturbative expression of $\alpha_s(Q^2)$, Eq.\,\eqref{eq:alpha_s}, which occurs at $Q^2=\Lambda_s^2$.  
The existence of Landau poles was first recognized in QED \cite{Landau:1959fi, Landau:1965nlt}. 
However, its occurs at $\Lambda\sim10^{30-40}$ GeV, well above the Planck scale \cite{Gockeler:1997dn}; thus, is of no practical concern because new physics would have emerged before that scale is reached, rendering standard QED irrelevant.

As noted above, the divergence is an unphysical feature of the perturbative treatment of quantum gauge field theories.
It signals the breakdown of all perturbative approximants that use series in $\alpha_s$, at $Q^2$-values well above $\Lambda_s^2$, \textit{viz}.\ all observables and the $\beta$-function, but not the individual $\beta_n$ terms, which are obtained via an expansion in $\hbar$. 
Indeed, if the pole were genuine, its location on the positive real axis of the complex $Q^2$-plane would produce imaginary mass particles, {\it viz}.\ tachyons.
Therefore, the existence of the pole need not have any relationship with quark confinement; there is no proof that it does \cite[Sec.3.1.2]{Marciano:1977su} and numerous reasons to judge otherwise.
Nevertheless, $\Lambda_s$ is often referred to as the ``confinement scale'', simply because it defines a scale that is well below that for which pQCD fails.
The fact that $\Lambda_s$ suggests a scale at which $\alpha_s$ becomes large; thus, where the effective charge may become sufficient to spur a confinement mechanism, is all that may be understood: there is no known mathematical or physical connection between $\Lambda_s$ and confinement.

Furthermore, confinement is a physical phenomenon, whereas the value of $\Lambda_s$ is conventional: 
it depends on the arbitrary choice of RS (Table~\ref{Flo:Table of Lambda});
on the order at which the $\beta$-series, Eq.\,\eqref{eq:alpha_s beta series}, is truncated;
and, at more than 2-loop, on the approximation used to solve Eq.\,\eqref{eq:alpha_s beta series}. 
This limitation understood, one can still explicitly and precisely relate, within a particular theoretical framework, $\Lambda_s$ to a physical scale, such as that characterising the hadron mass spectrum, as done, \textit{e.g}., using continuum \cite{Cui:2019dwv} and LGT \cite{Aoki:2021kgd} Schwinger function methods or HLFQCD \cite{Deur:2014qfa}.

\section{Quark thresholds  \label{Quark thresholds}}
Hitherto, we have neglected quark current masses in discussing the pQCD running coupling.
In fact, the mass-dependent terms \cite{Chetyrkin:1996hm} that would be present in Eqs.\,\eqref{eq:Callan-Symanzik1} or \eqref{eq:Callan-Symanzik2} do have a negligible effect. 
Nevertheless, even if these terms are omitted, the Higgs-generated quark current masses still have an indirect impact on the $Q^2$-dependence of the effective charge because they force one to distinguish between different numbers of quark flavors: the $\beta_n$ in Eqs.\,\eqref{eq:beta4 list} change with $n_f$.

\begin{figure}[t]
\centering
\includegraphics[width=9cm]{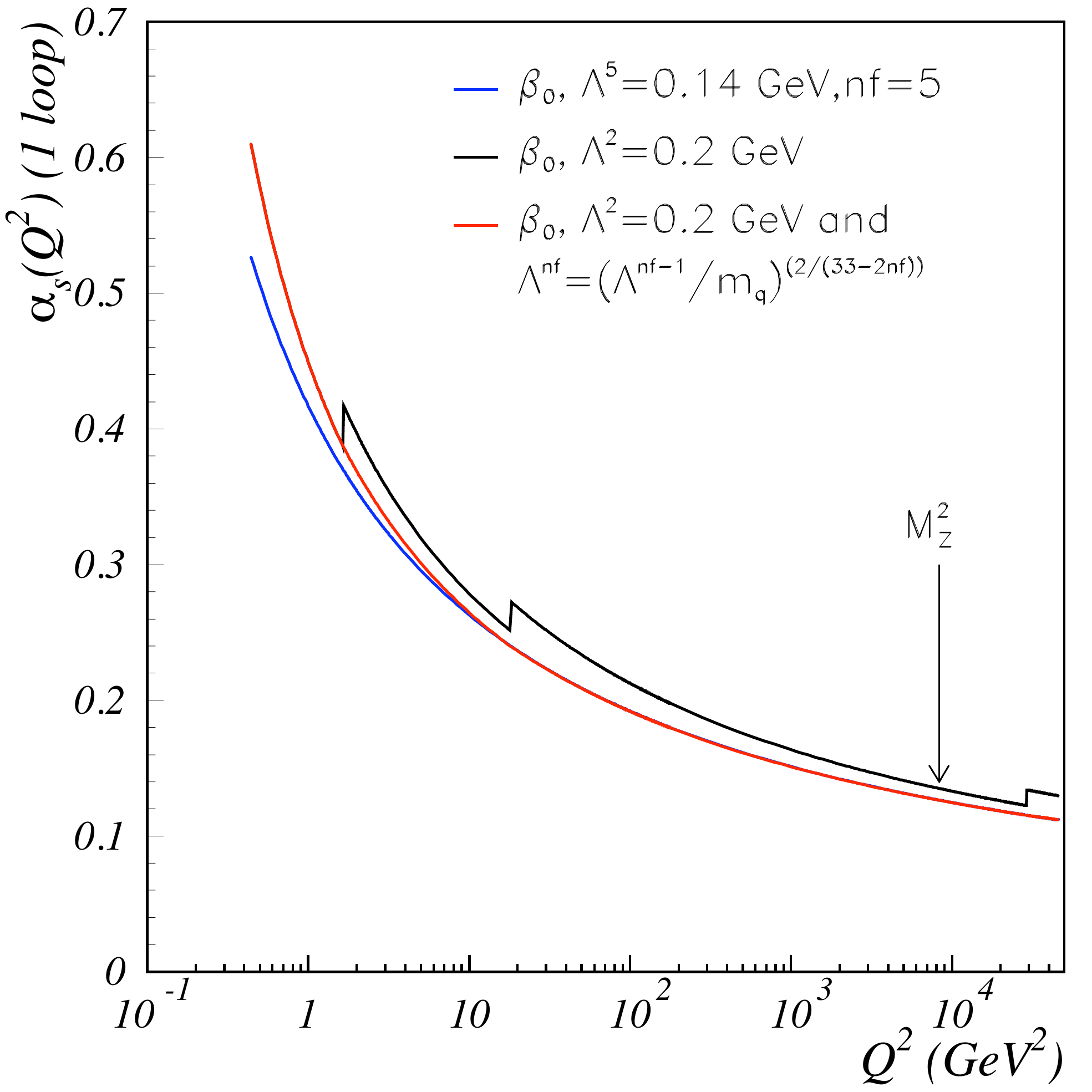}
\caption{\label{Flo:quark_thr} \small 
The running of $\alpha_s(Q^2)$ at 1-loop ($\beta_{0}$) for $n_f=5$ (blue line), with $\Lambda_s^{n_f=5}$
set so that $\alpha_s(M_{\rm Z}^2)$ matches the phenomenologically determined coupling ($Q^2$ location shown by the arrow). 
The black line shows the effect on Eq.\,\eqref{eq:beta0 list} of suddenly enabling new flavors. 
The result of the threshold smoothing perturbative procedure is shown by the red line for which $\Lambda_s^{n_f=2}$ was adjusted so that $\alpha_s(M_{\rm Z}^2)$ matches the phenomenological coupling. 
(Figure from \cite{Deur:2016tte})}
\end{figure}

It is typically assumed that a quark of flavor $f$ becomes ``active'', \textit{i.e}., begins to play a role in physical processes, once the momenta involved in the reaction exceed its mass threshold: $Q^2 \gg m_f^2$. 
This loose condition is implemented such that $\alpha_s(Q^2)$ varies smoothly across a quark threshold, instead of generating the discrete steps depicted in Fig.\,\ref{Flo:quark_thr}.
Consequently, $\Lambda_s$ comes to be dependent on $n_f$.  
The condition $\alpha_s^{n_f-1}(m_q^2)=\alpha_s^{n_f}(m_q^2)$ imposes, at leading order and for the perturbative domain: 
\begin{equation}
\Lambda_s^{n_f}=\Lambda_s^{n_f-1}\left(\frac{\Lambda_s^{n_f-1}}{m_q}\right)^{2/(33-2n_f)}.
\label{eq:lambda(nf) LO}
\end{equation}
The  formula at $\beta_1$ is provided in Ref.\,\cite{Larin:1994va}; that at $\beta_2$ in \cite{Melnikov:2000qh}; and at $\beta_3$ in Ref.\,\cite{Chetyrkin:1997sg}. 
Other smoothing procedures exist where, \textit{e.g}., $\Lambda_s(n_f)$ remains analytic \cite{Aoki:2021kgd, Binger:2003by}.

\section{Determinations of \mbox{$\alpha_s$} at short distances}
\label{state of the art}
We now discuss the methods available to access $\alpha_s$ and list the determinations that have become available in the past quinquennium.  Earlier estimations are discussed elsewhere \cite{Deur:2016tte}.

The standard global compilation of $\alpha_s(M_{\rm Z}^2)$ determinations is provided by the PDG \cite[Eq.\,(9.25)]{Workman:2022ynf}: 
\begin{equation}
\label{PDGalpha}
\alpha_s(M_{\rm Z}^2)=0.1179 \pm 0.0009 \quad
(\mbox{$\overline{\rm MS}$\,RS}). 
\end{equation}
The criteria applied in choosing the results to be included in the global average include, {\it inter alia}, that $\alpha_s$ is extracted from perturbative approximants expressed at least to NNLO, \textit{i.e}., $\alpha_s^3$.
Discussions of the high precision phenomenological extraction of $\alpha_s(M_{\rm Z}^2)$ can be found in the $\alpha_s$-2019 workshop proceedings \cite{Proceedings:2019pra}.
A compilation of LGT estimations of $\alpha_s(M_{\rm Z}^2)$ is provided in \cite{Aoki:2021kgd}.

The current range over which QCD's effective charge has been measured is 
$4\times 10^{-4}<Q^2/{\rm GeV}^2<4\times 10^{6}$ -- see, \textit{e.g}., Refs.\,\cite[Fig.\,9.4]{Workman:2022ynf}, \cite{Deur:2022msf}.
Notably, however, it is commonly judged that only data on $Q^2\gtrsim 4\,$GeV$^2$ can safely be evolved with pQCD equations. 
Above this scale, the breadth of range offers a basic test of pQCD and the Standard Model {\it via} the running of the coupling, Eq.\,\eqref{eq:alpha_s}.

The coupling $\alpha_{s}$ can be extracted experimentally, from measurements involving hadronic reactions \cite[Fig.\,9.4]{Workman:2022ynf};
determined numerically via LGT computations\linebreak \cite[Sec.\,9.4.7]{Workman:2022ynf};
calculated using continuum Schwinger function methods \cite{Cui:2019dwv}; 
and estimated using nonperturbative modeling of the strong force, \textit{e.g}., using AdS/QCD \cite{Brodsky:2014yha}. 
In all cases, a renormalization scheme must be chosen: the $\overline {\rm MS}$ scheme is widely used.
As already explained, it is critical to know the value of $\alpha_s$  in the UV (or equivalently, $\Lambda_s$, with the caveat pointed out in footnote~\ref{ftnt:alpha-lambda}) at the sub-percent level
so that $\alpha_s$ is not a dominant source of uncertainty in precision QCD studies and searches for physics beyond the Standard Model. 
It is therefore important to identify observables and techniques that enable accurate extraction and good control and estimation of uncertainties.  

There are three classes of uncertainties that contribute to the extraction of $\alpha_s$ from experimental data: 
(A) experimental uncertainty;
(B) effects of the pQCD series truncation; and
(C) nonperturbative contributions, which may not be well-known. 
The last two also significantly affect systematic calculations of $\alpha_s$.
.%
\paragraph{Series truncation.\label{BLM/PMC}} 
Regarding (B), the uncertainty on the series truncation is often estimated by varying the
renormalization scale, $\mu$, typically between $\mu/2$ and $2\mu$. 
This is an arbitrary procedure.
Alternatives exist, \textit{e.g}., schemes that optimize perturbative approximants and thus the accuracy of $\alpha_s$ extracted from the corresponding observables.
One such approach is the principle of maximum conformality (PMC) \cite{Wang:2023ttk, Brodsky:2011ta, Brodsky:2012rj, Brodsky:2011ig, Mojaza:2012mf, Brodsky:2013vpa}. 
It extends the Brodsky-Lepage-Mackenzie (BLM) procedure \cite{Brodsky:1982gc} where optimal perturbative predictions are obtained by summing all $\beta$-terms into the running coupling. 
The concept, generalized to the nonperturbative domain, 
will be discussed in Sec.\,\ref{EffectiveCharge}.

Another approach designed to optimize perturbative series is the definition, proposed in Ref.\,\cite{Boito:2016pwf}, of a strong coupling, $\hat{\alpha}_s$, which has its $\beta$-function explicitly RS invariant. 
The RS dependence of $\hat{\alpha}_s$ itself is captured by a single parameter $C$ ($C$-scheme). 
Variations of $C$ are equivalent to changing the value of $\Lambda_s$, \textit{i.e}., to choices of RS.
Therefore, in contrast to the standard definition, where the higher orders of $\beta_i$, 
Eqs.\,\eqref{eq:beta4 list}, and the $\alpha_s$ series, \textit{e.g}., Eq.\,\eqref{eq:alpha_s}, are scheme-dependent, their expressions are universal in the $C$-scheme, with their numerical values depending only on $C$. 
Perturbative series in term of $\hat{\alpha}_s$ are then optimized by an appropriate choice of $C$.   
Compared with the BLM/PMC procedure, the main difference is that the perturbative series is optimized {\it via} scheme dependence rather than optimizing the renormalization scale.

\begin{figure}[t]
\centering
\includegraphics[scale=0.45]{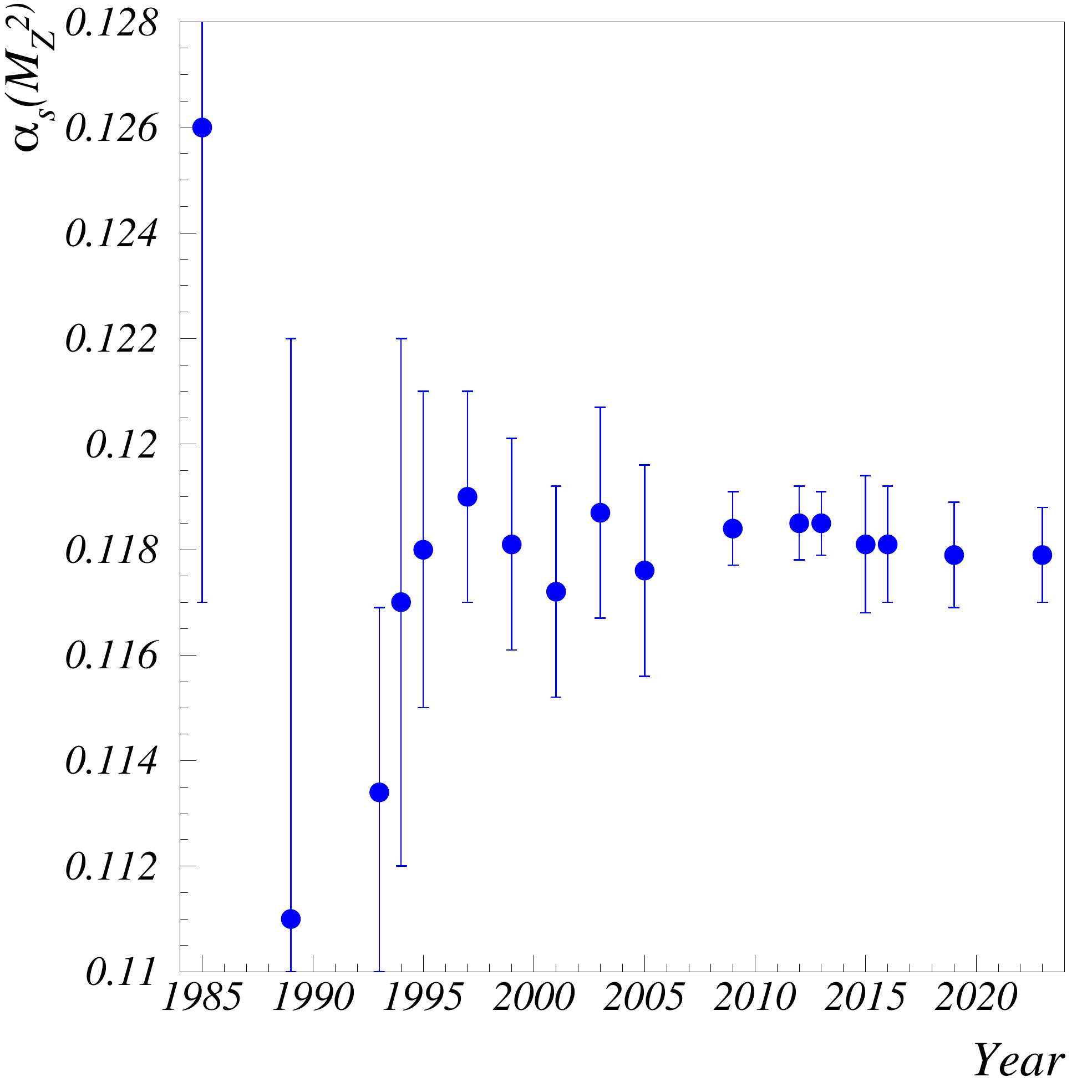}
\caption{\label{Flo:history} \small
Evolution (1985-2021) of the average world data for $\alpha_{s}(M_{\rm Z}^{2})$, as determined by the PDG.  The current result is $\alpha_s(M_{\rm Z}^2)=0.1179 \pm 0.0009$ ($\overline{\rm MS}$ RS) \cite[Eq.\,(9.25)]{Workman:2022ynf}.}
\end{figure}

\paragraph{Evolution of the determinations of $\alpha_{s}(M_{\rm Z}^{2})$.}
Progress in the determinations of $\alpha_{s}(M_{\rm Z}^{2})$, per the PDG \cite{Workman:2022ynf}, can be seen in Fig.\,\ref{Flo:history}. 
The significant improvement after 1989 owes to the appearance of CERN's LEP ($e^+ e^-$ collisions) and DESY's HERA ($e^{+/-} p$ collisions) facilities, as well as theoretical developments including, {\it inter alia}, the availability of NNLO perturbative series, which enable the extraction of $\alpha_s$ to the few\,\% level (\textit{cf}.\ NLO determinations typically yield 10\% accuracy).
The next jump, in the mid-2000, owes to improvements in LGT determinations, which now dominate the global averages. 
Notice, however, that the assessment of the global uncertainty does not necessarily decrease with time because new precision measurements may be in tension with a previous global average or as systematic uncertainties are re-estimated, becoming larger than initially assessed. 
This is what happened for the 2015 global average after the uncertainties on LGT results were judged to be larger than previously estimated. 
A short history of the determination of $\alpha_s$ is provided elsewhere \cite{Dissertori:2015tfa}.

%

We now discuss some methods used to precisely access $\alpha_s$. 

\subsection{Deep Inelastic Scattering and global PDF fits}
\label{DIS determinations}
Deep inelastic scattering  (DIS) is the elementary process of lepton--quark scattering,
$\ell q \to \ell^\prime  q^\prime$.
DIS data allow access to $\alpha_s$ {\it via} violations of Bjorken scaling \cite{Bjorken:1969ja, Feynman:1969wa}, {\it viz}.\ the near scale-independence at large $Q^2$ of nucleon inclusive structure functions, $F_{2}(x,Q^{2})$,  $g_{1}(x,Q^{2})$ and $F_{3}(x,Q^{2})$, where $x=\sfrac{Q^2}{2M\nu}$ is the Bjorken scaling variable and $\nu$ is the energy transferred from the lepton to the quark. 
Moments ($x$-weighted integrals of these structure functions) can also be employed \cite{Deur:2016tte}, including using Bernstein polynomials to weight the integrand over the measured domain, which ameliorates complications encountered at low and large $x$ \cite{Yndurain:1977wz, vanNeerven:1991nn, Zijlstra:1992qd, Larin:1993vu, Larin:1996wd, Santiago:1999pr, Santiago:2001mh}.

Scaling violations arise at LO from gluon emission by the struck quark (gluon bremsstrahlung), pair creation from the emitted gluon, and photon--gluon fusion. 
Other processes start contributing at NLO: quark self-energy and quark--photon vertex corrections -- see Fig.\,\ref{quantum_effects}.
The resulting $Q^2$-dependence is formalized by the Dokshitzer--Gribov--Lipatov--Altarelli--Parisi (DGLAP) equations \cite{Dokshitzer:1977sg, Gribov:1971zn, Lipatov:1974qm, Altarelli:1977zs}. 

The extraction of $\alpha_s$ is based on fits of one or several structure functions using the 
DGLAP formalism and parameterizations of the nonperturbative quark and gluon $x$-distributions (PDFs: parton distribution functions).\footnote{
Structure function moments that are involved in sum rules constitute an exception. In sum rules, the PDFs are integrated into measured quantities, thereby removing the need for model-dependent inputs. An example is discussed in Sec.\,\ref{EffectiveCharge}}
The PDFs contain some of the free parameters of the global fit because $\alpha_s$ cannot be reliably extracted from structure functions -- or, more generally, any physical process that depends on PDFs -- without simultaneously fitting the PDFs themselves \cite{Forte:2020pyp}. 
The advantage of global fits including several observables is that they maximize statistics and partially average out systematic uncertainties since they are largely uncorrelated. 
The caveat is that it is more complicated than fitting individual observables, especially managing the correlated theoretical uncertainties, which adds an additional source of systematic uncertainty.\\

In addition to DIS, jet production is another reaction providing $\alpha_s$ in a similar way.
For instance, the rate for jets originating from the $\gamma^\ast q \to q g$ reaction is
proportional to $\alpha_s$ since, in the UV, $\alpha_s$ essentially gives the probability of gluon emission. 
Considering $e^+ e^-$ collisions, the LO process is the two-jet event arising from $e^+ e^- \to q \bar{q}$.
Three-jet events stem from $e^+ e^- \to q \bar{q} g$ since a ``hard gluon'' (a high-energy gluon radiated at sufficiently large angle from its emitter) must hadronize, generating an additional jet. 
This occurs at a rate approximately $\alpha_s$ lower than two-jets. 
Thus, $\alpha_s$ may be drawn directly from the ratio of two to three jet event rates.
Three-jet events (one hard gluon emission) are thus a good observable for obtaining $\alpha_s$.
The mechanisms of gluon jet production are the same as those underlying the DGLAP formalism, \textit{viz}.\ gluon bremsstrahlung and photon--gluon fusion. 

An advantage of fitting jet data is that the $Q^2$-dependence of $\alpha_s$ is directly and independently determined -- see, \textit{e.g}., Ref.\,\cite{H1:2017bml}. 
A caveat is the ambiguity associated with defining jets.
The most common definitions employ the Durham \cite{Catani:1991hj} or Cambridge \cite{Dokshitzer:1997in} algorithms.   
Both use a parameter $y_{\rm cut}$ that establishes whether the emission is energetic enough to be a jet. 

Beside inclusive and jet production cross sections, other processes for precision measurements of 
$\alpha_s$ in $ep$ collision are event-shape -- see Sec.\,\ref{ep-em collisions determinations}, jet-substructure observables, and heavy flavor production cross-sections. 
Crucial experimental developments in the past quinquennium are significantly increased high-precision 
data from the full LHC Run-1 and availability of the final H1 jet DIS data from $ep$ collision at HERA-II \cite{H1:2015ubc}. 
Of equal importance is the theoretical advance in determining the approximants of the involved observables to NNLO  (i.e., $\alpha_s^3$, which enables global fits to be performed at this order \cite{Currie:2016ytq}.
Thanks to these developments, state-of-the-art global fits now include, in addition the traditional DIS and jet production data, hadron collider data. 
The latter often dominate the determinations of the PDFs and $\alpha_s$.

The kinematic domain whereupon a pQCD DGLAP analysis can be applied is chosen such that:
(A) higher-twist (HT) corrections%
\footnote{$1/Q^n$ corrections arising from multiparton processes and therefore involving poorly known nonperturbative multiparton-distributions.}
are negligible or small enough to be reliably modelled by the introduction of phenomenological $1/Q^n$ terms; 
(B) there are no noticeable contributions from nucleon resonances, whose structure would also add nonperturbative contributions, not described by $1/Q^n$ terms; and
(C) low-$x$ resummation effects are avoided. 
The data are usually restricted to kinematics with at least $Q^2 > 1\,$GeV$^2$ and invariant mass $W>1.7\,$GeV. 
For jet analysis, beside the $Q^2$ criterion, the mean transverse momentum, $p_T$, of the two highest energy jets must also be large, normally $p_T > 10\,$GeV, an order of magnitude above the typical mass of the created hadrons. 
This criterion is strongly correlated with the  $y_{\rm cut}$ requirement mentioned above.

An existing caveat on the method is that part of the high statistics data resides at relatively low $Q^2$, where nonperturbative corrections add a significant uncertainty. 
This issue is expected to become less important following collection of future DIS data at, \textit{e.g}., EIC \cite{Accardi:2012qut} and other future facilities \cite{LHeC:2020van, Anderle:2021wcy, Behnke:2013xla}. 
Such data will enable an increase in the minimum $Q^2$ over which the fits are performed.

A major uncertainty in the extraction of $\alpha_s$ from DIS is the selection of PDF parameterization functions, which are typically chosen at a practitioner's discretion and without reference to Standard Model constraints \cite{Cui:2021gzg, Cui:2021mom, Cui:2022bxn, Chang:2022jri, Lu:2022cjx}.
The gluon PDF, $g(x,Q^2)$, is particularly challenging because DIS is only indirectly sensitive to this distribution. 
Such uncertainty leads to parameter degeneracy when one attempts to simultaneously constrain both $\alpha_s$ and $g(x,Q^2)$.  
In the global assessment of $\alpha_s$ from DIS literature~\cite{Workman:2022ynf, dEnterria:2016cww}, the dominant uncertainty when combining DIS and global PDF fits comes from systematic differences in fits to the same data. 

Several DIS determinations of $\alpha_s$ have become available in the past quinquennium.
\begin{description}
\item[CTEQ-Jefferson Lab CJ15.] This global fit \cite{Accardi:2016qay} features lower than typical $Q^2$ and $W$ cuts in order to achieve higher $x$ coverage. 
It also includes neutron data from deuterium. 
Consequently, kinematical and dynamical HT corrections and nuclear corrections are considered. 
The kinematic range of the data selected for the fit is $Q^2  > 1.7\,$GeV$^2$  and $W > 1.7\,$GeV. 
The fit yields $\alpha_s(M_{\rm Z}^2)=0.1183\pm0.0002$ (NLO -- $\alpha_s^2$, and $\overline{\rm MS}$). 
\item[ABMP16.] This fit \cite{Alekhin:2017kpj}, which supersedes earlier analyses by the same group, is obtained from a global NNLO fit of DIS, Drell-Yan reactions, and hadro-production of single-top and top-quark pairs. 
It yields $\alpha_s(M_{\rm Z}^2)=0.1147\pm0008$ (NNLO, $\overline{\rm MS}$), or $\alpha_s(M_{\rm Z}^2)=0.1191\pm0.0011$ at NLO \cite{Alekhin:2018pai}.
The selection of the kinematic ranges over which the fit is performed depends on the process type, but they may be characterized as $Q^2> 1\,$GeV$^2$, $x<0.75$, $W>1.8\,$GeV, $p_T>20\,$GeV. 
The fit includes estimates of HT corrections. 
\item[H1PDF2017.] This determination of $\alpha_s(Q^2)$ and PDFs is obtained by fitting structure functions and jet cross-sections using neutral-current DIS data from HERA H1. 
It yields  \cite{H1:2017bml} $\alpha_s(M_{\rm Z}^2)=0.1147\pm0.0025$ (NNLO, $\overline{\rm MS}$). 
The scale uncertainty largely dominates the error, while the PDF model dependence is reported to be small. 
The evolution of $\alpha_s(Q^2)$ was determined over the range $7 \lesssim Q/{\rm GeV} \lesssim 90$ and agrees with the pQCD expectation.
The selected data range depends on the observable and may be identified with $Q^2> 10\,$GeV$^2$, $0.003<x<0.5$, $p_T>4.5\,$GeV. 
\item[NNPDF.] This collaboration updated its global PDF determinations, with NNPDF4.1 \cite{Ball:2022hsh} replacing NNPDF3.1 \cite{Ball:2018iqk}.
Whilst $\alpha_s$ is fixed in NNPDF4.1, it was a fit parameter in arriving at NNPDF3.1. 
That fit included LHC data in addition to DIS, with the former dominating the result. 
The analysis yields \cite{Ball:2022hsh, Ball:2018iqk} $\alpha_s(M_{\rm Z}^2) = 0.1185\pm0.0005$(exp)$\pm0.0001$(method)$\pm0.0011$(th) (NNLO, $\overline{\rm MS}$).
The uncertainties are dominated by theory (th), which stems mostly from truncation of the perturbative series of the various observable approximants.
\item[CTEQ-TEA] The most recent fits from this collaboration are available in Ref.\,\cite[CT18]{Hou:2019efy}. 
CT18 produces $\alpha_s(M_{\rm Z}^2) = 0.1164\pm0.0026$ (NNLO, $\overline{\rm MS}$). 
The fit includes fixed target lepton-$p$ DIS, Tevatron $p\bar p$, HERA $ep$, and LHC $pp$ data.
\item[MSHT20] This global fit to data from DIS and other hard processes yields \cite{Cridge:2021qfd}
$\alpha_s(M_{\rm Z}^2)=0.1174\pm0.0013$ (NNLO, $\overline{\rm MS}$).
\end{description}
The average of these results is $\alpha_s(M_{\rm Z}^2)=0.1174 \pm 0.0007$, which is consistent with the PDG value $\alpha_s(M_{\rm Z}^2)=0.1179 \pm 0.0009$ ($\overline{\rm MS}$ RS) \cite[Eq.\,(9.25)]{Workman:2022ynf}.
It is worth remarking here that Ref.\,\cite[CJ15]{Accardi:2016qay} reports an uncertainty on its $\alpha_s(M_{\rm Z}^2)$ result that is 5-times smaller than any other determination.  This is not explained therein but may indicate an incomplete accounting of the fit's error budget.

\subsection{Decay widths and event shapes in $e^+e^-$ annihilation}
\label{ep-em collisions determinations}
Beside $e^+e^-$ jet rate ratios, decay widths are other inclusive observables that permit access to $\alpha_{s}(M_{\rm Z}^2)$. 
At LO, the colliding $e^{+}e^{-}$ pair annihilates into a virtual photon or a $Z^0$. 
Their decay rate into a $q \bar q$ pair depends on $\alpha_{s}$, mostly through gluon bremsstrahlung on the $q$-$\bar q$ lines. 
Available observables at the $Z^0$-pole are the $Z^0$ and $W^\pm$ hadronic decay widths $\Gamma_{Z}^{\rm had}$ and $\Gamma_{W}^{\rm had}$, respectively -- see Ref.\,\cite{Pich:2020gzz} for a recent review; 
the $R_{Z}$ ratio of $\Gamma_{Z}^{\rm had}$ over the leptonic decay width; 
the ratio of the hadron production to lepton pair-production rates;
and hadronic and leptonic cross-sections.%
\footnote{The leptonic cross-section contains $\Gamma_{Z}^{\rm had}$ and is thus dependent on $\alpha_{s}$.}
All these observables are sensitive to $\alpha_{s}$ through $\Gamma_{Z}$ which is dominated by  hadronic decay.  

As mentioned, $\Gamma_{Z}^{\rm had}$ depends on $\alpha_s$, mostly through gluon bremsstrahlung, but at 
NLO, quark self-energy corrections arise. 
The pQCD corrections are typically computed at NNLO and are available to 
N$^{3}$LO for $R_{Z}$ and $R_{\tau}$ \cite{Baikov:2008jh, Baikov:2010je}.  ($R_{\tau}$ is the $\tau$-lepton analogue of $R_{Z}$.)
In addition to $\Gamma_{Z}^{\rm had}$, extraction of $\alpha_s$ from 
$\Gamma_{W}^{\rm had}$ and the branching ratio $\mathcal{B}_{W}^{\rm had}$ have reached N$^3$LO and NNLO, respectively. 
Their analysis from both $e^+e^-$ and hadron collision data yields 
$\alpha_s(M_{\rm Z}^2 ) = 0.117 \pm 0.0042$ \cite{dEnterria:2016rbf},
with the error largely dominated by the experimental data uncertainty.
An update of the associated analysis, which accounts for theoretical and experimental progress, yields 
$\alpha_s(M_{\rm Z}^2 ) = 0.1203 \pm 0.0028$ \cite{dEnterria:2020cpv}.
Another global fit, performed at  N$^3$LO and 
next-to-next-to-leading logs%
\footnote{Soft and collinear resummations are necessary for an event shape observable, $y$, because of the enhancement in its approximant's perturbative series of each $\alpha_s^n$ term by factors $(\ln y)^m$, where $m\leq n+1$ and $m=n+1$ is the leading $\log$. 
These factors are important for small $y$ values. 
They degrade the convergence of the pQCD fixed order $\alpha_s$ series thereby reducing their kinematical range of applicability.}  
(NNLL), using the data on two- and three-jet rates from $e^+ e^-$ collisions at LEP and PETRA, results in 
$\alpha_s(M_{\rm Z}^2 ) = 0.11881 \pm 0.00063$(exp)$\pm 0.00116$(th) \cite{Verbytskyi:2019zhh} (N$^3$LO, NNLL, $\overline{\rm MS}$).
Finally, Ref.\,\cite{Boito:2019pqp} recently developed the use of moments of the $R_{c\bar{c}}$ ratio for the hadronic $e^+e^- \to c\bar{c} + X$ cross-section to extract 
$\alpha_{s}(M_{\rm Z}^{2})$, obtaining $0.1168\pm0.0019$ (NLO).
Here the average is $\bar \alpha_s(M_{\rm Z}^2 ) = 0.1181 \pm 0.0013$.

One can also extract $\alpha_s$ from event shape observables \cite{Dasgupta:2003iq, OPAL:2004wof, Gehrmann:2008kh}
, particularly, the thrust, $T$, C-parameter, total and wide jet broadenings, $B_T$ and $B_W$, heavy jet mass $M_H$, two-to-three jet transition parameter $y_{23}$, jet cone energy fraction, oblateness, $O$, and energy-energy correlations $EEC$. 
Their approximants are known to NNLO 
and can be resummed to at least NNLL accuracy. 

To explain how these observables are sensitive to $\alpha_s$, we use the most familiar, \textit{i.e}., event shape: 
\begin{equation}
T= \max \bigg[ \frac{\sum_i |p_i \cdot n|}{\sum_i |p_i|} \bigg],
\end{equation}
where $p_{i}$ are the momenta of produced particles or jets, and $n$ is the unit vector along the {\it thrust axis}, defined as the direction maximizing $T$. 
In the infinite energy limit of back-to-back produced $q\bar{q}$, $T\to1$. 
At finite energy, this distribution leaks to lower values of $T$ owing to gluon emissions from the quark lines. 
Therefore, the shape of the $T$ distribution away from $T=1$ allows access to $\alpha_s$.
Likewise, other event shape observables are also extremized in back-to-back $q\bar{q}$ production 
in the Born limit. 

Emissions of gluons transform the singular distribution into something measurable and directly sensitive (\textit{i.e}., at LO) to $\alpha_s$.
Compared to the inclusive quantities previously discussed, event shape variables are more sensitive to nonperturbative QCD corrections \cite{Dokshitzer:1995qm}. 
They arise from the hadronization of the $q$, $\bar{q}$ or hard gluons and are especially important in the case of back-to-back events, when $T\to 1$.
 
Ref.\,\cite{Kardos:2018kqj} performed a global analysis of $EEC$ in $e^+ e^-$ collisions from LEP, PEP, PETRA, SLC and TRISTAN \cite{CELLO:1982rca, JADE:1984taa, Fernandez:1984db, PLUTO:1985yzc, TASSO:1987mcs, Wood:1987uf, TOPAZ:1989yod, OPAL:1991uui, L3:1992btq, DELPHI:1992qrr, OPAL:1993pnw, SLD:1994idb}, with nonperturbative hadronization corrections treated by Monte Carlo simulation. It yields 
$\alpha_s(M_{\rm Z}^2 ) = 0.11750 \pm 0.00287$ 
(NNLO+NNLL, $\overline{\rm MS}$), with the uncertainty on the renormalization scale largely dominating the total error. 
The $T$ and C-parameter data from $e^-e^+$ collisions at LEP, PEP, PETRA and TRISTAN have also been studied, with the result \cite{Kardos:2020igb}
$\alpha_s(M_{\rm Z}^2 ) = 0.12911 \pm 0.00177$(exp)$\pm 0.0123$(scale)
(NNLO with N$^3$LO estimate, at least NNLL, $\overline{\rm MS}$). 
An alternate analysis, performed with an analytic model rather than treating the hadronization corrections by Monte Carlo methods, yields 
$\alpha_s(M_{\rm Z}^2 ) = 0.121 \pm^{0.001}_{0.003}$ (NNLL+NNLO, $\overline{\rm MS}$) \cite{Tulipant:2017ybb}.
Ref.\,\cite{Haller:2018nnx} used an updated global electroweak fit to analyze three hadronic $Z$-decay observables from LEP data \cite{ALEPH:2005ab}. 
Keeping only $\alpha_s$ as a free parameter, assigning all other standard model parameters to their otherwise determined values, yields 
$\alpha_s(M_{\rm Z}^2 ) = 0.1194 \pm 0.0029$.
The four results listed in this paragraph produce 
$\bar \alpha_s(M_{\rm Z}^2 ) = 0.1218 \pm 0.0032$, which exceeds the PDG value, Eq.\,\eqref{PDGalpha},  by $1.3\sigma$.

With the accuracy now reaching NNLO, or even N$^3$LO and NNLL, the dominant uncertainty is often the variation of the renormalization scale, $\mu$. 
The standard procedure, which varies it between $\mu/2$ and $2\mu$ to assess truncation uncertainty, is arbitrary and the large uncertainty that ensues, unnecessary: 
as mentioned in Sec.\,\ref{BLM/PMC}, methods exist which eliminate the scheme and renormalization scale 
ambiguities of perturbative calculations, \textit{e.g}., BLM/PMC \cite{Wang:2023ttk}.
The thrust, $T$, measured in $e^-e^+$ colliders by the ALEPH (LEP), AMY (TRISTAN), DELPHI (LEP), HRS (PEP), JADE (PETRA), L3 (LEP), MARKII (PEP-II), OPAL (LEP), SLD (SLC), and TASSO (PETRA)
experiments was analyzed using the PMC \cite{Wang:2019ljl, Wang:2019isi, Wu:2019mky, Wang:2021tak}. 
The analysis, which shows that $\mu$ is not a fixed constant in processes like $e^+ e^- \to$ jets, but depends in detail on $T$, yields a running of $\alpha_s(Q^2)$ over $4\leq Q/{\rm GeV} \leq 16$ that agrees well with that obtained at NNLO from the world average and a value 
$\alpha_s(M_{\rm Z}^2 ) = 0.1185 \pm0.0012$ (NNLO).
A similar analysis of the C-parameter yields $\alpha_s(M_{\rm Z}^2 ) = 0.1193^{+0.0021}_{-0.019}$ (NNLO).

\subsection{Hadron collisions}
\label{p-p collisions determinations}
Until recently, determinations of $\alpha_{s}$ from hadron collisions were limited to NLO, but several NNLO analyses are now available, \textit{e.g}., that of $t \bar{t}$ production \cite{Czakon:2013goa}. 
The CMS collaboration at LHC has provided several extractions of $\alpha_{s}$ from the inclusive cross section for top-quark pair $t\overline{t}$ \cite{Khachatryan:2014waa} 
($\alpha_{s}(M_{\rm Z}^{2})=0.1185_{-0.0042}^{+0.0063}$, NNLO) 
and inclusive production cross sections for $W^\pm$ and $Z_0$  
($\alpha_{s}(M_{\rm Z}^{2})=0.1175^{+0.0025}_{-0.0028}$, NNLO) \cite{CMS:2019oeb}.
A simultaneous extraction of $\alpha_s$ and the top quark mass using $t \bar{t}$ cross-section measurement from the CMS 2016 data provided 
$\alpha_s(M_{\rm Z}^2) = 0.1135^{+0.0021}_{-0.0017}$ (NLO) \cite{CMS:2019esx}.
A similar analysis, using both ATLAS and CMS data and performed to NNLO, yields 
$\alpha_s(M_{\rm Z}^2) = 0.1159^{+0.0013}_{-0.0014}$ (NNLO) \cite{Cooper-Sarkar:2020twv}.
CMS also analyzed inclusive jet cross-sections \cite{CMS:2016lna}
($\alpha_s(M_{\rm Z}^2)=0.1164 \pm^{0.0060}_{0.0043}$, NLO) 
and dijet cross-sections \cite{CMS:2017jfq}
($\alpha_{s}(M_{\rm Z}^{2})=0.1199 \pm 0.0015$(exp)$_{-0.0020}^{+0.0031}$(th), NLO).
%
An analysis following a CMS method developed to extract $\alpha_{s}$ from the cross-section for $t\bar{t}$ production, including both LHC (CMS+ATLAS) and Tevatron data, yields
$\alpha_s(M_{\rm Z}^2)=0.1177^{+0.0034}_{-0.0036}$ (NNLO+NNLL).
The analysis of the low $p_T$ ($<30\,$GeV) data for the Drell-Yan produced $Z_0$ at CMS yields  \cite{Camarda:2022qdg} 
$\alpha_s(M_{\rm Z}^2)=0.1185^{+0.0014}_{-0.0015}$ (N$^3$LO+N$^3$LL). 
Several PDF global fits were tested with the result showing modest sensitivity.

The LHC ATLAS collaboration has determined $\alpha_{s}$ from transverse energy-energy correlations in multi-jet events \cite{ATLAS:2017qir}
($\alpha_{s}(M_{\rm Z}^{2})=0.1162 \pm 0.0011$(exp)$_{-0.0070}^{+0.0084}$(th), NLO) 
and their associated asymmetries 
($\alpha_{s}(M_{\rm Z}^{2})=0.1196 \pm 0.0013$(exp)$_{-0.0045}^{+0.0075}$(th), NLO). 
The collaboration also used the measurement of the multi-jet cross-section ratio 
to extract \cite{ATLAS:2018sjf}
$\alpha_{s}(M_{\rm Z}^{2})=0.1127_{-0.0027}^{+0.0063}$ (NLO).

A global analysis of inclusive jet cross-section data from CMS and STAR ($pp$ collision), CDF and D0 ($p\bar{p}$ collision), and H1 and ZEUS ($ep$ collision), is provided in Ref.\,\cite{Britzger:2017maj}. It resulted in 
$\alpha_{s}(M_{\rm Z}^{2})=0.1192 \pm 0.0012$(exp)$_{-0.0041}^{+0.0061}$(th), 
(NLO). 
The analysis studied but excluded the ATLAS $pp$ collision data as they significantly worsen the $\chi^2$ of the global fit.

An analysis of the LHC inclusive cross-sections for $Z^0$ and $W^{\pm}$ production yields \cite{dEnterria:2019aat}
$\alpha_s(M_{\rm Z}^2) = 0.1188^{+0.0019}_{-0.0013}$ (NNLO).
The impact of the corresponding Tevatron data was studied, and four different  
PDF fits were used, \textit{viz}.\ CT14 \cite{Dulat:2015mca}, HERAPDF2.0 \cite{H1:2015ubc}, MMHT14 \cite{Harland-Lang:2014zoa}, and NNPDF3.0 \cite{Ball:2018iqk}, with the conclusion that MMHT14 provides the most robust extraction. 
Namely, MMHT14-based cross-sections are most sensitive to the underlying value of $\alpha_s(M_{\rm Z}^2)$, the accord between individual and combined $\alpha_s(M_{\rm Z}^2)$ extractions is best amongst the sets, and the inferred values of $\alpha_s(M_{\rm Z}^2)$ are most stable to changes in data sets and uncertainties.  
That there is a preferred PDF fit underlines the correlation between the extraction of $\alpha_s$ and determination of PDFs and exemplifies the general conclusion, reached in Ref.\,\cite{Forte:2020pyp}, that using pre-determined PDFs biases the $\alpha_s$ extraction.

All values in this subsection are consistent within mutual uncertainties; hence, can be averaged to obtain $\bar \alpha_s(M_{\rm Z}^2 ) = 0.1173 \pm 0.0011$, which is consistent with the PDG average, Eq.\,\eqref{PDGalpha}.

\subsection{$\tau$-decay}
\label{tau-decay}
The hadronic branching ratio in $\tau$-decays:
\begin{equation}
R_{\tau,h} = \frac{\Gamma[\tau \to  \nu_\tau \mbox{\rm hadrons}]}
{\Gamma[\tau \to  \nu_\tau e^-\bar{\nu_e}]}\,,
\end{equation}
provides the most precise experimental determination of $\alpha_s$, but there are accuracy issues which are still debated \cite{Pich:2020gzz}.
This ratio provides access to $\alpha_s$ in a manner similar to that \textit{via} $W^\pm$ (and $Z^0$) decay widths because the $\tau$ decays are mediated by emission of a virtual $W^-$. 

The LO pQCD correction to $R_{\tau,h}$, which enables access to $\alpha_s$ \cite{Narison:1988ni, Braaten:1988ea}, is substantial, \textit{viz}.\ $\sim$20\%. 
It offers a good handle on $\alpha_s$; but, since the latter is extracted at the relatively low scale of Mandelstam $s=M_\tau^2=3.16$~GeV$^2$, nonperturbative effects and higher-order pQCD terms, complicate the extraction.
This caveat is partly compensated by a factor of $\sim 9$ diminution of the absolute uncertainty on 
$\alpha_s$ when evolved from $M_\tau^2$ to $M_{Z}^2$.%
\footnote{For the central value, Ref.\,\cite{Salam:2017qdl} devised a simple approximation, accurate to within 0.2\% compared to the $\beta_4$ evolution, to convert $\alpha_s(M_\tau^2)$ to $\alpha_s(M_{\rm Z}^2)$: $\alpha_s^{n_f=5}(M_{\rm Z}^2)\approx 0.1180+0.125\big(\alpha_s^{n_f=3}(M_\tau^2) -0.314\big).$} 
The reduction affects all low $Q^2$ determinations of $\alpha_s(Q^2)$ because its uncertainty, $\delta \alpha_s(Q^2)$, becomes $\delta \alpha_s(M_Z^2)= (\sfrac{\alpha^2_s(M_Z^2)}{\alpha^2_s(Q^2)}) \delta \alpha_s(Q^2)$ at the scale $M_Z^2$~\cite{Dissertori:2015tfa,Deur:2016tte,dEnterria:2022hzv}.
This enhances the impact of data at $Q^2 \ll M^2_{Z}$ that can be used to determine  $\alpha_s(M_Z^2)$.

An important difference between the extraction of $\alpha_s$ from $Z^0$ or $W^\pm$ decay widths and 
that from $R_{\tau,h}$ comes from the $\nu_\tau$ emitted during the $\tau$ decay. 
It allows the $W^-$ to span the full range of kinematically allowed momenta. 
This makes $R_{\tau,h}$ an integral from $s=0$ to $s=M_\tau^2$ over the spectral function, \textit{i.e}., the imaginary part of $\Pi(s)$, the quark current correlator polarization function:
\begin{equation}
R_{\tau, h}=12\pi S_{EW}(M_{\tau},M_{Z}) \int_0^{M_\tau^2} \frac{ds}{M_\tau^2}
\bigg(1- \frac{s}{M_\tau^2}\bigg)^2\bigg(\big[1- \sfrac{2s}{M_\tau^2}\big]{\rm Im}\Pi^T(s)+{\rm Im}\Pi^L(s) \bigg),
\end{equation}
with $S_{EW}=1.01907\pm0.0003$ accounting for electroweak radiative corrections and $L$ and $T$ being, respectively, the longitudinal and transverse angular momentum directions in the produced hadron rest frame.
In principle, the lowest $s$ values over which the integral runs forbid the pQCD treatment needed to extract $\alpha_s$; but the analyticity of $R_{\tau,h}$ allows one to rewrite it as a contour integral along $|s|=M_\tau^2$, whereupon pQCD is applicable. 
The resulting pQCD approximant is, in $\overline{\rm MS}$~\cite{Baikov:2008jh}:
\begin{multline}
R_{\tau, h}=N_C\left|V_{ud}\right|^{2}S_{EW}(M_{\tau},M_{Z})\bigl[1+A_{1}+1.63982A_{2}+6.37101A_{3} \\
+49.07570A_{4}+\mathcal{O}(A_{5})+\delta_{NP}\bigr], 
\end{multline}
where $|V_{ud}|=0.97373 \pm 0.00031$ \cite{Workman:2022ynf} is the relative probability that a down quark decays into an up quark, $\delta_{NP}\simeq-0.006$, and $A_{n=1,2,\ldots}$ contain the contour integrals:
\begin{equation}
A_{n}=\frac{1}{2\pi i}\oint_{\left|s\right|=M_{\tau}^{2}}\frac{ds}{s}\left(\frac{\alpha_{s}(s)}{\pi}\right)^{n}\left(1-2\frac{s}{M_{\tau}^{2}}+2\frac{s^{3}}{M_{\tau}^{6}}-2\frac{s^{4}}{M_{\tau}^{8}}\right)\,.
\end{equation}
At LO, they only depend on $\alpha_{s}$.

When expanding any $A_n$ into a pQCD series, it is necessary to resum the unknown higher order contributions because of the slow convergence of the series. 
Mainly, two perturbative schemes have been used.
The first is ``fixed order perturbation theory'' (FOPT), a development in $\alpha_s(M_\tau^2)$ with, 
as usual in standard pQCD series, the renormalization scale kept fixed. 
The second scheme is ``contour-improved perturbation theory'' (CIPT), where the series is estimated 
by evolving the integrand using the RGE for the behavior of $\alpha_s(s)$ along the contour $s=M_\tau^2 e^{i \phi}$.
Namely, the renormalization scale varies along the contour, resumming the running of $\alpha_s$.
This scheme yields $\alpha_s(M_\tau^2)$ results that are systematically higher by $\sim5\%$ compared to FOPT and the question of which scheme is accurate has long been debated \cite{Bethke:2011tr}. 
The difference arises from the unknown resummed higher orders, estimated differently in FOPT and CIPT. Specifically, the difference is mainly due to a sensitivity to the first IR renormalon, associated with the gluon
condensate~\cite{Hoang:2021nlz}.
Given this, other approaches are explored, such as
the Borel sum Principal Value (PV) truncation method \cite{Ayala:2021mwc} or 
the renormalization-group-summed (RGS) expansion \cite{Abbas:2012fi}, which uses a RG-improved FOPT expansion, where, in contrast to CIPT, the implemented RG-invariance is amenable to analytical analysis.
Alternatively, a new renormalon-free
scheme for the gluon condensate was recently developed~\cite{Benitez-Rathgeb:2022hfj, Benitez-Rathgeb:2022yqb}. Thus, it strongly suppresses the difference between FOPT and CIPT, with the remaining difference being attributed mainly to missing higher orders.

Although the issue of which resummation method is most accurate is not yet settled, it has been shown \cite{Hoang:2020mkw, Beneke:2008ad, Beneke:2012vb, Boito:2020hvu} that the CIPT analysis, 
possesses ambiguities that cannot be computed using standard renormalon calculus.
FOPT does not have this problem; hence, perhaps, should be preferred. 
Since a common practice is to average the extractions from FOPT and CIPT, with the difference, interpreted as the uncertainty, dominating the total error on $\alpha_s(M_\tau^2)$, then finding that FOPT is more accurate would be a crucial progress.

Aside from this question, the different extractions of $\alpha_s(M_\tau^2)$ continue to be actively debated because of the increasing importance given to isolating the nonperturbative effects in 
$\Pi(s)$ \cite{Pich:2016bdg, Boito:2016oam}. 
They are small at $s=M_\tau^2$, being suppressed as $1/M_\tau^6$ \cite{Braaten:1991qm, LeDiberder:1992zhd} and could be neglected in the past, but the progress in accuracy and order of the extraction (now reaching N$^3$LO~\cite{Baikov:2008jh}) requires careful control of nonperturbative contributions, including resonance effects (so-called ``quark-hadron duality violations'' \cite{Pich:2022tca}) that are not described by the OPE power corrections. 

Among the recent analyses extracting  $\alpha_s$ from $\tau$ decay is a comprehensive study 
using ALEPH (LEP) data  performed in Ref.\,\cite{Pich:2016bdg}, using both FOPT and CIPT. 
It yields 
$\alpha_s(M_\tau^2)=0.328 \pm 0.013$ (N$^3$LO, $n_f=3$), \textit{i.e}., $\alpha_{s}(M_{\rm Z}^{2})=0.1197 \pm 0.0015$ (N$^3$LO, $n_f=5$) 
where the uncertainty is dominated by the truncation of the approximant and just covers the difference in $\alpha_s(M_\tau^2)$ from  FOPT and CIPT. 

The analysis method of Ref.\,\cite{Pich:2016bdg} was applied to the ALEPH $\tau$-decay vector and axial-vector data using  FOPT, CIPT and Borel sum PV truncation \cite{Ayala:2021mwc}. 
The FOPT and PV results are close to each other and average to 
$\alpha_s(M_\tau^2)=0.3116 \pm 0.0073$ (N$^3$LO, $n_f=3$), or
$\alpha_{s}(M_{\rm Z}^{2})=0.1176 \pm 0.0010$ (N$^3$LO, $n_f=5$). 
The CIPT result remains higher than for FOPT and PV.
This supports the FOPT and agrees with the conclusion of the Ref.\,\cite{Hoang:2020mkw}.

Another recent analysis studied the ALEPH, OPAL (LEP) and BABAR (SLAC) $e^+ e^-$ data \cite{Boito:2020xli} and also concluded that FOPT should be favored. 
It yields 
$\alpha_s(M_\tau^2)=0.3077 \pm 0.0075$ (FOPT, N$^3$LO, $n_f=3$), or 
$\alpha_{s}(M_{\rm Z}^{2})=0.1171 \pm 0.0010$ (FOPT N$^3$LO, $n_f=5$).

These results are mutually consistent and average to 
$\bar \alpha_s(M_{\rm Z}^2 ) = 0.1181 \pm 0.0007$, in line with the PDG average, Eq.\,\eqref{PDGalpha}.

\paragraph{Hadronic $R$-ratio.}
The $R$ ratio for $e^+e^- \to \rm hadrons$ can be used as an alternate to $\tau$-decay.
Although the procedure for extracting $\alpha_s(s)$ from $R(s)$ is similar to that of $\tau$-decay, 
{\it viz}.\ using spectral functions, the extraction from $R(s)$ is less sensitive to nonperturbative 
effects than the $\tau$-decay case. 
Ref.\,\cite{Boito:2018yvl} uses $R(s)$ obtained by combining the world data on electroproduction cross-sections \cite{Keshavarzi:2018mgv}.
The  $s$  range  over which $\alpha_s$ is obtained is dominated by high statistic $M_\tau^2 \leq s < 4$ GeV  data and provides the running of $\alpha_s(s)$ over this range. 
The combined data evolved to $M_{\rm Z}^2$ then yield 
$\alpha_s(M_{\rm Z}^2) = 0.1158 \pm 0.0022 $ (FOPT) or 
$\alpha_s(M_{\rm Z}^2) = 0.1166 \pm 0.0025 $ (CIPT).
The total uncertainty is relatively large and dominated by that of experiment. 
The difference between the FOPT and CIPT results
is lower than in the $\tau$-decay case, partly owing to a higher average $s$, but perhaps also because the $R(s)$ perturbative series converges faster than in the $\tau$-decay case.

\subsection{Numerical simulations of lattice QCD}
\label{LGT determinations}
A basic introduction to LGT and how it is used to determine $\alpha_s$ in the UV and IR is given elsewhere \cite{Deur:2016tte}. 
The compilation from the flavor lattice averaging group (FLAG) \cite{FlavourLatticeAveragingGroupFLAG:2021npn} has become, together with the PDG review \cite{Workman:2022ynf}, a standard reference on our knowledge of $\alpha_s$ in the UV. 
The FLAG review is updated every two years and provides, {\it inter alia}, a comprehensive summary of LGT determinations of $\alpha_s(M_{\rm Z}^2)$ and $\Lambda_s$. 
Other recent reviews are Ref.\,\cite{DelDebbio:2021ryq}, which focuses on a pedagogical discussion of LGT methods, and Ref.\,\cite{Komijani:2020kst}, which provides an alternative to the FLAG and PDG assessments of the world's LGT calculations.

LGT currently reports the most precise determination of $\alpha_s(M_{\rm Z}^2)$, with the FLAG global LGT average quoted at better than 0.7\%: 
$\alpha_s(M_{\rm Z}^2)=0.1182 \pm 0.0008$ (2021 average, LGT only). 
It is worth recording that uncertainties were typically underestimated and not well controlled in earlier LGT determinations -- see, \textit{e.g}., the year 2015 jump in Fig.\,\ref{Flo:history}.  Today, on the other hand, such issues are said to be better managed and results from different groups generally agree. 

\paragraph{LGT calculation technique.}
LGT is based on a path (functional) integral formalism \cite{Dirac:1933xn, Feynman:1948ur}, which
determines the transition probability from an initial to a final state by summing over all possible connecting space-time trajectories.
Each path is exponentially weighted by its corresponding classical action; hence, the trajectories of largest action are strongly suppressed in comparison with that corresponding to the path of least action, \textit{i.e}., the classical trajectory, in an expression of the Fermat/Maupertuis least action
principle.
The action-weighted suppression, but not elimination, of non-classical trajectories allows for quantum processes, and thus, \textit{e.g}., the running of $\alpha_s$.

The complexity of functional integrals demands that, for realistic theories in four dimensional spacetime, the integration is performed numerically. 
Today, the approach is usually formulated in Euclidean spacetime \cite{Wilson:1974sk}, with the continuum quantum field theory reformulated on a lattice of discrete points in four dimensions and the  integrals over values of the field variables at the lattice sites (matter fields) or the interstices (gauge fields) evaluated using Monte Carlo methods -- see, \textit{e.g}., Ref.\,\cite{Gattringer:2010zz}.
A given configuration of field variable values is weighted by a factor $\exp(-S_E)$, where $S_E$ is the Euclidean action evaluated at the field configuration.  
Again, the ``classical'' configuration is favored, but all possible configurations may contribute at some (exponentially damped) level. 
A desired correlation function is computed using the collection of field configurations, each weighted according to the associated value $\exp(-S_E)$; and the statistical precision is given by the square-root of the number of configurations used to estimate the integral.


\paragraph{General considerations.}
Typically, LGT accesses $\alpha_s$ in much the same way as experiments, \textit{i.e}., a UV-dominated physics quantity (not necessarily an observable, in this case) is selected and the LGT computed result is then matched to its pQCD approximant. 
Although LGT is a nonperturbative technique, $\alpha_s$ from LGT still acquires a RS-dependence and 
other convention-dependences through the matching procedure. 
A phenomenological input, \textit{e.g}., a hadron mass or a meson decay constant, is required for LGT  to convert its result from a value expressed in units of the lattice spacing, $a$, into physical units. 
That the LGT-computed UV quantity need not be an observable can be exploited by selecting matrix elements that are most efficiently computed using the LGT framework.  
Usefully, the quantity can be computed at several $Q^2$ values, thereby enabling a check on the $Q^2$-evolution of $\alpha_s$. 
This is not always possible in experiment, where the $Q^2$ may be set, \textit{e.g}., to $M_\tau$ for $\alpha_s$ extracted from $\tau$-decay. 

Naturally, any LGT determination of $\alpha_s$ encounters the usual difficulties and limitations of the lattice method and also some additional challenges, which are specific to the $\alpha_s$ case.  We list a few of the issues here.  
\begin{itemize}
    \item For a variety of reasons, including the ensuing need to invert huge matrices, using dynamical fermions -- unquenching the computation -- complicates the LGT formulation of the problem and increases computation time~\cite{Nielsen:1980rz, Nielsen:1981xu, Nielsen:1981hk, Kogut:1974ag}.   Nevertheless, it is crucial for true QCD simulations. 
    
    \item With light dynamical quarks, LGT encounters a  {\it critical slow-down} problem, 
    whose severity grows as $1/[a^2 m_\pi^2]$, with $m_\pi$ the pion mass in the simulation.
    
    \item To sidestep critical slow-down, simulations use current-quark masses that produce larger than physical values of $m_\pi$.  Chiral effective field theories are then used to develop extrapolations to the physical pion mass~\cite{Bernard:1995dp}. 
    
    \item A reliable computation of $\alpha_s$ must use $L/a \ggg \mu/\Lambda_s$, where $L^4$ is the lattice volume.  However, available computational resources limit both the lattice spacing and volume in a given simulation; so the condition cannot be met with available resources.  This issue is typically addressed using a finite-size scaling method.
       
    \item While local operators are conveniently computed with LGT methods, it is difficult to simulate non-local operators, \textit{e.g}., those providing structure functions, because the calculation of a single path involves all lattice sites. In contrast, for local actions and operators, the only sites involved are those used to update the path and the sites' neighbors, \textit{i.e}., just 9 operations for four-dimensional lattices. 
    Another difficulty is that PDFs reside in Minkowski space, a domain for which standard Euclidean LGT is ill-suited.
    Within the scope of this perspective, these difficulties limit the type of quantities LGT can compute to extract $\alpha_s$. For instance, the Bjorken integral, $\Gamma_1^{p-n}(Q^2)$ in Eq.\,\eqref{bjorken SR}, cannot readily be computed, despite being especially well-suited for extraction of $\alpha_s$. 
    The development of procedures which may circumvent these difficulties is currently being actively pursued \cite{Lin:2017snn}.

\end{itemize} 
Notwithstanding these and kindred issues, advances in computer technology and simulation algorithms are today reported to have reduced the attendant uncertainties to a level whereat they no longer dominate the overall error in a LGT determination of $\alpha_s(M_{\rm Z}^2)$.  That burden is now said to be borne by the truncation error on the pQCD series for the quantity to which the LGT result is matched.  At this level, the inclusion of QED effects -- long neglected in LGT -- may begin to become important \cite{DiCarlo:2019thl}.


\paragraph{Diverse LGT subjects.}
Just as with experimental observables, there are many quantities upon which one might focus in order to use LGT to obtain $\alpha_s(M_Z^2)$. 
Considering a range of ``reliability'' factors -- Ref.\,\cite[Sec.\,9.2.1]{Aoki:2021kgd}, FLAG chooses to include results from the following four methods in their LGT average \cite{Aoki:2021kgd, Workman:2022ynf}: 
step-scaling methods ($\alpha_s(M_Z^2)=0.11848 \pm 0.00081$) \cite{Luscher:1991wu, Luscher:1992an, Luscher:1992zx, Luscher:1993gh, Sint:1995ch, Capitani:1998mq, ALPHA:2001tjb, DellaMorte:2004bc, PACS-CS:2009zxm, Tekin:2010mm, Fritzsch:2012wq, DallaBrida:2016uha, Ishikawa:2017xam, Bruno:2017gxd, DallaBrida:2019wur, Nada:2020jay,Francesconi:2020fgi};
small Wilson loops ($\alpha_s(M_Z^2)=0.11871 \pm 0.00128$) \cite{El-Khadra:1992ytg, Aoki:1994pc, Davies:1994ei, Wingate:1995fd, Spitz:1999tu, QCDSF-UKQCD:2001lsi, Gockeler:2005rv, Mason:2005zx, Maltman:2008bx, Davies:2008sw, McNeile:2010ji, Asakawa:2015vta, Kitazawa:2016dsl};
heavy-quark current two-point functions ($\alpha_s(M_Z^2)=0.11818 \pm 0.00156$ \cite{HPQCD:2008kxl, McNeile:2010ji, Chakraborty:2014aca, Nakayama:2016atf, Maezawa:2016vgv, Petreczky:2019ozv, Petreczky:2020tky, Boito:2020lyp};
and heavy-quark potential at short distances ($\alpha_s(M_Z^2)=0.11660 \pm 0.00160$) \cite{Michael:1992nj, Booth:1992bm, Bali:1992ru, Necco:2001xg, Necco:2001gh, Takahashi:2002bw, JLQCD:2002zto, HPQCD:2003rsu, Mason:2005zx, Davies:2008sw, Jansen:2011vv, Bazavov:2012ka, Tormo:2013tha, Karbstein:2014bsa, Bazavov:2014soa, Husung:2017qjz, Karbstein:2018mzo, Takaura:2018lpw,  Takaura:2018vcy, Bazavov:2019qoo, Ayala:2020odx, Husung:2020pxg}.  
Other possibilities, such as different QCD vertices \cite{Alles:1996ka, Boucaud:1998bq, Boucaud:1998xi, Becirevic:1999uc, Becirevic:1999hj, Boucaud:2000ey, Boucaud:2000nd, Boucaud:2001st, DeSoto:2001qx, Boucaud:2001qz, Boucaud:2005gg, Boucaud:2005xn, Sternbeck:2007br, Boucaud:2008gn, Sternbeck:2009hna, Blossier:2010ky, Ilgenfritz:2010gu, Blossier:2011tf, Blossier:2012ef, Sternbeck:2012qs, Blossier:2013ioa, Zafeiropoulos:2019flq}, Dirac operator \cite{Nakayama:2018ubk}, 
or Vacuum polarization \cite{JLQCD:2008bwj, Shintani:2010ph, Hudspith:2016yhn, Hudspith:2018bpz, Cali:2020hrj}, have been explored, but existing studies do not meet all FLAG's conditions.

\paragraph{Global LGT averages.} 
Combining the listed results, FLAG reports the following LGT average:
\begin{equation}
    \alpha_s(M_Z^2)=0.1184 \pm 0.0008\,.
\end{equation}
For comparison, another compilation reports\cite{Komijani:2020kst}:
\begin{equation}
    \alpha_s(M_Z^2)=0.11803_{-0.00068}^{+0.00047}\,.
\end{equation}
Notably, both results are reported with uncertainties less than the PDG average.  This explains why, when LGT results are excluded by the PDG, they obtain 
\begin{equation}
    \alpha_s(M_Z^2)=0.1176 \pm 0.0010\, 
\end{equation}
which may be compared with their overall average in Eq.\,\eqref{PDGalpha}.

The current FLAG results for $\alpha_s(M_Z^2)$ translate into the values of $\Lambda_s$ recorded in Table~\ref{Flo:Table of Lambda from LGT}.

\begin{table}[t]
\centering
\begin{tabular}{|c|c|c|c|c|c|}
\hline 
$n_f$ & 0  &  2 & 3 & 4 & 5 \tabularnewline
\hline 
 $\rule{0ex}{3ex}\Lambda_s^{(n_f)\,\rm LGT}$ (GeV) & 0.261(15) & $0.330(^{+21}_{-63})$ & 0.339(12) & 0.297(12) & 0.214(10)   \tabularnewline
\hline
\end{tabular}
\caption{
\label{Flo:Table of Lambda from LGT} \small Average values of $\Lambda_s$ in the $\overline{\rm MS}$
RS for different number of flavor $n_f$, as determined from the FLAG 2021 compilation \cite{Aoki:2021kgd}. }
\end{table}

\subsection{Continuum and Lattice Schwinger Function Methods}
\label{alphaCSM}
As we shall discuss in Sec.\,\ref{DSE:alpha_PI}, there is an approach to analyzing QCD's Schwinger functions, \textit{viz}.\ the combination of pinch technique (PT) \cite{Cornwall:1981zr, Cornwall:1989gv, Pilaftsis:1996fh, Binosi:2009qm, Cornwall:2010upa} and background field method (BFM) \cite{Abbott:1980hw, Abbott:1981ke}, which enables one to rigorously define a unique effective charge, $\hat \alpha(Q^2)$, from a modified gluon vacuum polarization, in direct analogy with the Gell-Mann--Low coupling in QED \cite{GellMann:1954fq}.
This effective charge is \cite{Binosi:2016nme, Cui:2019dwv}: process-independent (PI); bounded, smooth and monotonically decreasing on $Q^2>0$; and equivalent, in the UV, to that deduced from the ghost-gluon vertex \cite{Sternbeck:2007br, Boucaud:2008gn, Aguilar:2009nf} and sometimes called the ``Taylor coupling,'' $\alpha_{\rm T}$ \cite{Blossier:2011tf, Blossier:2012ef}.

The connection between $\hat \alpha(Q^2)$ and $\alpha_{\rm T}(Q^2)$ elevates the importance of information about the QCD running coupling obtained from the Taylor coupling.  This charge has long been a focus of attention and has recently been computed using a combination of continuum and LGT methods \cite{Zafeiropoulos:2019flq}, employing field configurations built with three flavors of domain wall fermions and a physical pion mass.  Expressing the results in the \mbox{$\overline{\rm MS}$} RS, one finds $\Lambda_s^{n_f=2+1} = (0.320 \pm 0.014)\,$GeV 
and 
\begin{equation}
    \alpha_s(M_Z^2)=0.1172 \pm 0.0011\,,
\end{equation}
in good agreement with other cited estimates and averages.  Having drawn this connection, it becomes possible to deliver parameter-free predictions for hadron properties using the PI charge, $\hat \alpha(Q^2)$ -- see, \textit{e.g}., Refs.\,\cite{Cui:2020tdf, Zhang:2021mtn, Raya:2021zrz, Cui:2021mom, Yao:2021pdy, Lu:2022cjx}.

\subsection{Gauge-gravity duality}
\label{AdS/CFT determinations}
The scale $\Lambda_s$ has been estimated \cite{Deur:2014qfa, Deur:2016opc} using the Holographic Light-Front QCD (HLFQCD) approach \cite{Brodsky:2014yha}.  
HLFQCD, a nonperturbative method based on a presumed gauge-gravity duality \cite{Maldacena:1997re},
employs a semiclassical potential to account for the gluon interaction. 
This makes HLFQCD applicable only in the IR and at the interface between IR and UV. 
Therefore, we postpone the detailed description of HLFQCD to Section\,\ref{HLFQCD}. 
Here, we will only outline how $\Lambda_s$ is determined within HLFQCD and how it relates to hadron masses. 

One considers the strong force coupling, $\alpha_{g_1}(Q^2)$, defined following the effective charge prescription of Grunberg \cite{Grunberg:1980ja, Grunberg:1982fw, Grunberg:1989xf} with the $g_1$ scheme. 
It can be calculated in the UV with the usual techniques to obtain the running coupling or by using the known relation between $\alpha^{\mathrm{pQCD}}_s(Q^2)$ and $\alpha_{g_1}(Q^2)$, determined via commensurate scale relations (CSRs) \cite{Brodsky:1994eh}. 
In the IR, on the other hand, it can be computed using HLFQCD -- see Sec.\,\ref{alpha_g1 from HLFQCD}. 
We denote the resulting coupling $\alpha^{\rm HLF}_{g_1}(Q^2)$.
The $Q^2$ evolution of $\alpha^{\rm HLF}_{g_1}(Q^2)$ is controlled by a hadronic scale, \textit{e.g}., the $\rho$ meson mass, $M_\rho$, or the proton mass, $M_{\rm p}$, calculated in the model and referred to experiment.

Above $Q^2 \simeq 1$ GeV$^2$, HLFQCD ceases to be applicable because its semiclassical potential 
does not, by definition, include the short distance quantum effects that are responsible for the running of the perturbative $\alpha_s(Q^2)$.
Hence, HLFQCD cannot predict the coupling in the UV.%
\footnote{
This statement applies to HLFQCD, a bottom-up approach. 
Ref.\,\cite{Dubovsky:2018vde} provides a top-down discussion, sketching how the QCD $\beta$-function might emerge from string theory's worldsheet.} 
Nevertheless, the applicability domains of HLFQCD and pQCD seem to overlap on
$1 \lesssim Q^2 \lesssim 2$ GeV$^2$ \cite{Deur:2016cxb}.
Supposing they do, then one can match $\alpha^{\rm HLF}_{g_1}$ to the pQCD coupling $\alpha^{\rm pQCD}_{g_1}$. 
Approximating the matching to occur at a single point, $Q_0$, one may require:
\begin{eqnarray} 
\alpha^{\rm HLF}_{g_1}(Q_0) &=& \alpha^{\rm pQCD}_{g_1}(Q_0) \,,  \nonumber \\
\frac{d\alpha^{\rm HLF}_{g_1}(Q)}{dQ}|_{Q=Q_0} &= & \frac{d\alpha^{\rm pQCD}_{g_1}(Q)}{dQ} |_{Q=Q_0}.
 \label{eq:matching eqs.}
\end{eqnarray}  
Defined thus, the scale $Q_0$ represents the interface between the nonperturbative and perturbative 
domains, with, \textit{e.g}., DGLAP and ERBL \cite{Efremov:1979qk, Lepage:1979zb, Lepage:1980fj} evolution becoming valid upon the latter.
Clearly, validity conditions for the method are
A) $\Lambda_s \ll Q_0$ and
B) Small HT effect in the approximant of the observable that defines $\alpha_{g_1}$, {\it viz} $\Gamma_1^{p-n}$, Eq.~(\ref{bjorken SR}). The smaller the HT, the less stringent condition A becomes.
The smallness of HT for $\Gamma_1^{p-n}$ is expected from various calculations
and was verified experimentally\cite{Deur:2004ti, Deur:2014vea, Deur:2008ej}.

The supposed existence of an overlap domain, whereupon both HLFQCD and pQCD descriptions of the strong force are valid, is supported {\it a posteriori} by the availability of a solution to Eqs.\,\eqref{eq:matching eqs.} and successful HLFQCD predictions of other nonperturbative observables using the same IR-UV matching as described here.
The solution to Eqs.\,\eqref{eq:matching eqs.} provides the value of $Q_0$ and an algebraic relation between $\Lambda_s$ and $M_\rho$ \cite{Deur:2014qfa}, \textit{e.g}., at LO:
\begin{equation}  \label{eq: Lambda LO analytical relation}
\Lambda_s=\frac{M_\rho {\rm e}^{-a}}{\sqrt{a}},
\end{equation}
with $a=4\big[(\ln^2(2)+\beta_0/4+1)^{\sfrac{1}{2}}-\ln(2)\big]/\beta_0\simeq 0.55$
for $n_f = 3$, yielding  $\Lambda^{(3)}_s \approx 0.603$~GeV at LO. 
At 5-loop, $\Lambda_s$ is RS-dependent and the evaluation of $\alpha^{\rm pQCD}_s$ is most readily completed numerically. 
In the  $\overline{\rm MS}$ RS, this yields 
$\Lambda^{(3)\rm HLF}_{\overline{\rm MS}}=0.339\pm0.019$~GeV \cite{Deur:2016opc} or 
$\alpha^{\rm HLF}_s(M_z^2)=0.1190\pm0.0006$ (N$^3$LO, $\overline{\rm MS}$). 
These values are consistent with world averages \cite{Workman:2022ynf, Aoki:2021kgd, Komijani:2020kst}.

Conversely, one can use a given determination of $\Lambda_{\overline{\rm MS}}$ and the relations between hadron masses obtained in HLFQCD \cite{deTeramond:2014asa} to analytically determine the hadron mass spectrum, with $\Lambda_s$ as the sole input \cite{Deur:2014qfa}.

Similar to the above method providing $\alpha^{\rm HLF}_s(M_{\rm Z}^2)$, HLFQCD has been used to compute PDFs and GPDs characterizing the nonperturbative structure of hadrons in 
the UV \cite{deTeramond:2018ecg, deTeramond:2021lxc, Liu:2019vsn, Chang:2020kjj}. 
In this approach, the PDFs and GPDs are calculated at the IR-UV interface and then DGLAP-evolved to a perturbative scale, typically $Q^2=5$ GeV$^2$, whereat results from global fits to data are available.
The HLFQCD predictions agree well with the phenomenologically extracted PDFs and GPDs. 
It is unclear whether this is a feature or flaw of HLFQCD, however, because those phenomenological fits are not systematically consistent with QCD endpoint ($x \simeq 0, 1$) constraints, e.g., in the case of the pion~\cite{deTeramond:2018ecg, Chang:2020kjj}.

\subsection{Future prospects in precision determination of $\alpha_s(M_{\rm Z}^2)$ }
Our knowledge of $\alpha_s(M_{\rm Z}^2)$ will continue to improve as new data becomes available, from existing and anticipated facilities, pQCD theory is developed further, and progress is made with nonperturbative methods for strong QCD.

Collection of data on observables directly relevant to $\alpha_s$ is continuing at LHC and RHIC (both $pp$ and heavy-ion colliders) and has started at the upgraded JLab (11 GeV electron beam on fixed targets). 
New facilities relevant to $\alpha_s(M_{\rm Z}^2)$ extractions, and either approved for construction or anticipated, include:
\begin{description}
    \item [EIC] Variable $e+p$ center-of-mass energies from 20-100\,GeV and high electron-nucleon luminosity ($\mathcal{L}=10^{33}$-$10^{34}\,$cm$^{-2}$s$^{-1}$) \cite{Arrington:2021biu, AbdulKhalek:2021gbh}.  For example, one expects that the Bjorken sum $\Gamma_1^{p-n}(Q^2)$ measured at the EIC will enable extraction of $\alpha_s$ with a precision better than 2\%~\cite{EIC-alpha_s}.
    \item [EicC] Variable $e+p$ center-of-mass energies from 15-20\,GeV and $\mathcal{L}= 2-3 \times 10^{33}$cm$^{-2}$s$^{-1}$ \cite{Chen:2020ijn, Anderle:2021wcy}
\end{description}

Improvements from experimental data will also come with upgrades at existing facilities, including the various proposals for LHC luminosity \cite{EuropeanStrategyforParticlePhysicsPreparatoryGroup:2019qin} and fixed target programs \cite{Brodsky:2012vg, Lansberg:2016urh, Kikola:2017hnp, Hadjidakis:2018ifr, Aidala:2019pit}, 
and a possible 22\,GeV upgrade of the accelerator at JLab~\cite{Accardi:2023chb}. 
A study similar to that carried out for the EIC \cite{EIC-alpha_s} shows that $\Gamma_1^{p-n}$ data from the upgraded JLab and EIC can reach a precision $\Delta \alpha_s/\alpha_s \simeq0.6\%$~\cite{Accardi:2023chb}.

On the theory front, observables whose approximants are presently only known up to NLO can be computed to higher orders. 
That will not only provide additional constraints but also further test the universality of $\alpha_s$, thereby constraining the search for physics beyond the standard model. 
Methods are also becoming available or being refined to minimize systematics, such as that
associated with setting the renormalization scale (\textit{e.g}., the PMC -- see Sec.\,\ref{BLM/PMC}), solving the FOPT/CIPT issue in $\tau$-decay observables -- Sec.\,\ref{tau-decay}, or understanding/computing/measuring power corrections and other nonperturbative contributions.  
All this will allow for phenomenological determinations of $\alpha_s$ below the $\%$ level. 

However, it may nevertheless be impossible to reach an accuracy on par with that of the other fundamental couplings, such as $\Delta G_{\rm N}/G_{\rm N}\simeq 10^{-5}$, $\Delta G_{\rm F}/G_{\rm F} \simeq 10^{-8}$ and $\Delta \alpha/\alpha \simeq$$10^{-10}$. 
That would demand knowledge of clean observables at least to N$^6$LO (and leading log equivalent, as necessary), including interference from electroweak effects; and construction and operation of $e^+e^-$ colliders -- the reaction offering the cleanest observables -- with luminosities many 
orders-of-magnitude greater than achieved at LEP. 
On the other hand, reaching the $\permil$ level is a challenging but achievable goal. 
In order to achieve such success, one may expect that, as with the current sub-\% goal, continuum and LGT Schwinger function methods and, possibly, other nonperturbative 
approaches to QCD, such as AdS/QCD -- Sec.\,\ref{AdS/CFT determinations}, may need to play a leading role.  

\chapter{Long distance behaviour of \mbox{$\mathbf \alpha_s$}}
\label{low Q regime}
\section{Importance of $\alpha_s$ in the nonperturbative domain, difficulties and progress}
\label{low Q regime:intro}
Perturbative methods have been central to establishing QCD as the primary candidate for a (possibly  effective) theory of strong nuclear interactions; and pQCD continues to be actively studied and developed as increasingly accurate pQCD predictions are needed for high-energy physics -- see, \textit{e.g}., Refs.\,\cite{Anastasiou:2016cez, Heinrich:2020ybq}. 
Yet, Nature is largely dominated by low-energy/IR aspects of QCD. 
For instance, the Higgs mechanism of mass generation is directly responsible for only $\simeq 1$\% of the visible mass in the Universe, with the remaining 99\% emerging as a consequence of IR dynamics in QCD \cite{Roberts:2021nhw, Binosi:2022djx, Papavassiliou:2022wrb, Ding:2022ows, Roberts:2022rxm, Ferreira:2023fva}.
QCD's emergent hadron mass (EHM) is tightly linked with the infrared behavior of $\alpha_s$, as may be seen in a variety of ways \cite{Deur:2014qfa, Binosi:2014aea, Binosi:2016xxu, Binosi:2016wcx, Binosi:2016nme}. 
Another crucial phenomenon, critically connected to the magnitude of $\alpha_s$ in the IR, is dynamical chiral symmetry breaking (DCSB) \cite{Binosi:2016wcx}. 
This was realized even before the advent of QCD, following studies based on a nonrenormalizable four-fermion interaction \cite{Nambu:1961tp}, and translated into the QCD context following Refs.\,\cite{Lane:1974he, Politzer:1976tv}.
Viewed from the continuum perspective, precise knowledge of $\alpha_s$ at IR momenta is a prerequisite for any reliable calculation of hadron properties.
Illustrative examples abound \cite{Deur:2016tte, Eichmann:2016yit, Fischer:2018sdj, Qin:2020rad, Roberts:2021nhw, Binosi:2022djx, Papavassiliou:2022wrb, Ding:2022ows, Roberts:2022rxm, Ferreira:2023fva}.

Evaluation of $\alpha_s$ in the IR, which we will call $\alpha_s^{\rm IR}$, requires a sound nonperturbative treatment of QCD. 
This is evident, for instance, in Eq.\,\eqref{eq:alpha_s}, \textit{i.e}., the pQCD prediction that $\alpha^{\mathrm{pQCD}}_s(Q^2) \to \infty$ as $Q^2\to\Lambda_s^2$ -- see Sec.\,\ref{Landau pole}.
This is internally inconsistent because $\alpha^{\mathrm{pQCD}}_s(Q^2) \gg1 $ conflicts with the use of a perturbative expansion in $\alpha^{\mathrm{pQCD}}_s$ to compute the $\beta$-series, Eq.\,\eqref{eq:alpha_s beta series}. 
It is also unphysical because observables, like elastic and transition form factors and structure functions, have been measured on $Q^2$ domains that cover the pQCD Landau pole region: no divergences, discontinuities, or other unusual behaviors have been found.
Thus, the $\alpha^{\mathrm{pQCD}}_s(Q^2)$ divergence is merely an artifact of perturbation theory whose appearance signals that all perturbative approximants to any given quantity become invalid at some $Q^2 > \Lambda_s^2$.

Naturally, the internal inconsistency of perturbation theory and, \textit{a fortiori}, the unphysical divergence in $\alpha^{\mathrm{pQCD}}_s(Q^2)$ cannot be cured by computing the coupling to higher orders.  In fact, higher orders can only make the problems worse because pQCD approximants are all examples of asymptotic series (Poincar\'e expansions) \cite{Dyson:1952tj}.
Writing such an approximant for a measurable quantity in the form $O_N(\alpha_s) \sim \sum_{n=0}^N a_n \alpha_s^n$, then the sequence of associated partial sums begins to depart from the true result for $N>N_\alpha$, where the value of $N_\alpha$ may be estimated from the condition $\alpha_s \, a_{N_\alpha+1}= {\rm O}(a_{N_\alpha})$.  Adding terms with $N>N_\alpha$ leads to partial sums that rapidly proceed to diverge.  (Additional discussion can be found elsewhere \cite{HARRIS201443}.)
To illustrate, consider $Q^2 \simeq Q_0^2 = 1\,$GeV$^2$, \textit{i.e}., in a neighbourhood of the UV-IR transition; then, $\alpha^{\mathrm{pQCD},\overline{\rm MS}}_s(Q^2)/\pi\simeq 0.2$ and the asymptotic character of pQCD series begins to become apparent at order N$^{\sim4}$LO.  
Different RSs can delay or hasten the onset of this problem, but it cannot be avoided because one may always express a running coupling computed in one RS to that obtained using another via a power series in the latter coupling, as discussed in connection with Eq.\,\eqref{eq:alpha_rel_2RS}.
On the other hand, if QCD is truly a theory with a connection to strong interactions in Nature, then including nonperturbative effects must eliminate the Landau pole \cite{Deur:2016tte, Binosi:2016nme, Burkert:2017djo} because Nature does not exhibit such a defect.

As will be detailed below, studying $\alpha_s$ is much more challenging in the IR than in the UV. 
This is not because nonperturbative calculations are necessarily more difficult than those using perturbation theory; the impediments are rather of a different character.
Namely, at this stage in the development of QCD theory, there is no single agreed prescription for defining and calculating an IR completion of QCD's running coupling.
Further, \textit{prima facie}, it is possible that more than one coupling may be needed to fully characterize strong QCD.
Notwithstanding these issues, it is crucial to deliver a nonperturbative solution of QCD, and calculating $\alpha_s^{\rm IR}$ is a key piece in that puzzle.  
Indeed, with a single $\alpha_s$ known at all spacelike momenta, one would have in hand that quantity which describes the strength of QCD's interaction at all momentum scales \cite{Dokshitzer:1998nz}, thereby completing a big step toward an economical and fundamental description of low-energy hadronic phenomena.

There are many additional, complementary reasons for determining the IR behavior of $\alpha_s$.
For instance, given that some definitions of $\alpha_s$ involve ratios of field and vertex renormalization constants -- see, \textit{e.g}., Eqs.\,\eqref{Eq:alternate_Zs}, \eqref{eq:zg}, then knowing $\alpha_s^{\rm IR}$ can deliver insights into the long-wavelength properties of (dressed) gluons and quarks, and even ghosts, if the formulation involves them. 
Of course, their UV properties are well known, but emergent, nonperturbative phenomena are expected to vastly modify the nature of QCD's parton fields in the IR.
Hence, any sound understanding of a nonperturbative extension of $\alpha_s$ into the IR will deliver information on the quasiparticle degrees-of-freedom that replace partons in descriptions of low-energy phenomena.

In contrast to studies at UV momenta, no one approach to the calculation of $\alpha_s^{\rm IR}$ has yet acquired the status of ``preferred''.  
This is largely because a nonperturbative solution of QCD is still being sought; and in this hunt, many complementary approaches currently seem viable.
Indeed, using an amalgam of methods might be the only way to solve QCD.
Hereafter, we will describe a number of widely used approaches to the study of $\alpha_s^{\rm IR}$.

As noted previously \cite{Deur:2016tte}, there is currently no consensus on what might be called the ``correct'' scheme for the calculation of $\alpha_s^{\rm IR}$.  
Furthermore, some may not even consider that $\alpha_s^{\rm IR}$ is universal because there are four individual UV-divergent interaction vertices in the perturbative treatment of QCD.
So, there could be up-to four distinct couplings at infrared (IR) momenta; and when considering $n\geq 3$ point functions, one has an uncountable infinity of choices for the momentum with which the coupling runs.  
In our view, such considerations are spurious because if QCD is a theory, then BRST symmetry will be preserved nonperturbatively; hence, a unique IR completion of $\alpha_s$ exists.
As we subsequently elucidate, this position is supported by developments in the past quinquennium.

Today, definitions and determinations of $\alpha_s^{\rm IR}$ are available that ensure its universality and deliver consistent results.  
A universal PI charge, $\hat\alpha(Q^2)$, has been defined \cite{Binosi:2016nme} using continuum Schwinger function methods (CSMs) \cite{Eichmann:2016yit, Fischer:2018sdj, Qin:2020rad} and recently computed using a combination of continuum and LGT inputs \cite{Cui:2019dwv}.
For reasons that are understood both mathematically and phenomenologically, the prediction for $\hat\alpha(Q^2)$ is pointwise almost identical to $\alpha_{g_1}(Q^2)$, the experimental coupling \cite{Deur:2005cf, Deur:2008rf, Deur:2022msf} derived from the Bjorken sum rule \cite{Bjorken:1966jh, Bjorken:1969mm} using the method of effective charges \cite{Grunberg:1980ja, Grunberg:1982fw, Grunberg:1989xf}, and also practically indistinguishable from the effective charge, $\alpha^{\rm HLF}_{g_1}(Q^2)$, obtained using HLFQCD \cite{Brodsky:2010ur}. 

A characterizing feature of all three charges is that they ``freeze'' in the IR, saturating to a value $\alpha_s(Q^2=0)\approx \pi$, \textit{i.e}., the couplings become $Q^2$-independent at IR momenta and express a $Q^2=0$ fixed point, something which was long ago conjectured -- see, \textit{e.g}., Refs.\,\cite{Caswell:1974gg, Sanda:1979xp, Banks:1981nn}.
The $Q^2=0$ value of the coupling now becomes ``critical'' in the sense that it must be large enough to ensure DCSB.
Allowing for self-reinforcement \textit{via} the gluon-quark vertex, $\alpha_s(Q^2=0)\approx \pi$ is sufficient \cite{Binosi:2016wcx}.

The universal (PI) feature of $\hat\alpha$, the consistency between effective charge measurements and CSM and HLFQCD predictions, and the fact that $\alpha_{g_1}$, $\alpha^{\rm HLF}_{g_1}$, $\hat \alpha$ have all been used to make sensible predictions for key low-energy hadron observables%
\footnote{For instance: 
the QCD scale $\Lambda_s$ -- Secs.\,\ref{alphaCSM}, \ref{AdS/CFT determinations};
hadron wave functions and spectra \cite{Brodsky:2006uqa, deTeramond:2008ht, Chang:2011ei, Chang:2013pq, Shi:2015esa, Ding:2015rkn, Ding:2019qlr, Ding:2019lwe,Ding:2018xwy};
elastic and transition form factors \cite{Sufian:2016hwn, Ding:2018xwy},
polarized and unpolarized quark and gluon PDFs and GPDs for different hadrons \cite{deTeramond:2018ecg, deTeramond:2021lxc, Chang:2020kjj, Raya:2021zrz};
leptonic decay constants of Nambu-Goldstone bosons \cite{Chang:2013pq , Shi:2015esa, Ding:2015rkn, Ding:2019qlr, Ding:2019lwe, Gao:2017mmp};
dressed-quark mass functions \cite[Fig.\,2.5]{Roberts:2021nhw}, \cite{Binosi:2016wcx};
and a unified picture of the soft and hard pomeron \cite{Dosch:2022mop}.
}
form three pillars in a compelling argument that supports the existence of a canonical QCD coupling.
This coupling not only describes the strength of QCD's interaction at all momentum scales but also serves as the basis for predictions of a prodigious array of hadronic phenomena.  
The past few years have also seen important progress toward an understanding of the physical mechanism which underlies the IR behavior (running) of QCD's effective charge, \textit{viz}.\ a deep understanding of how a Schwinger mechanism \cite{Schwinger:1962tn, Schwinger:1962tp} in QCD leads to the emergence of a gluon mass-scale \cite{Binosi:2022djx, Papavassiliou:2022wrb, Aguilar:2021uwa, Aguilar:2022thg}.

\section{\mbox{$\mathbf{\alpha_s}$} and emergent phenomena}
\label{alpha_s and confinement}
As we saw at the beginning of Ch.\,\ref{high Q regime}, in the UV context, the dressing of classical tree-diagrams by short-distance quantum loops leads to the running of $\alpha_s$.
This is only one part of the picture, however.
Infrared completions of $\alpha_s$ may include additional phenomena, including those which cannot be captured by any summation of a finite number of diagrams, such as confinement, DCSB, EHM, \textit{etc}. 
These inherently nonperturbative phenomena are likely connected; even, perhaps, essentially indistinguishable in origin.  
It follows that any discussion of $\alpha_s^{\rm IR}$ will probe into the very core of the most fundamental questions in the Standard Model.

We have explained why the Landau pole is artificial and unphysical.
As a consequence, any suggestion that confinement -- a physical phenomenon -- owes to an IR slavery stemming from the Landau pole is erroneous.
Notwithstanding this, it is widely held that $\alpha_s^{\rm IR}(Q^2\simeq 0)$ should be large (exceeding unity) in order for confinement to be realized.
Yet confinement scenarios exist which do not require a large coupling, \textit{e.g}., the supercritical binding model discussed in Ref.\,\cite{Gribov:1999ui}, which realizes a form of confinement with $\alpha_s^{\rm IR}(Q^2\simeq 0) \approx 0.4$.
Some other small-coupling confinement models are described elsewhere \cite{Deur:2016tte}.

Given the almost certain connection between $\alpha_s^{\rm IR}$ and Standard Model emergent phenomena, the following discussion will range over a variety of approaches to low-energy strong-interaction observables. 
Much work has focused on pure gauge QCD, largely because this eliminates Nambu-Goldstone bosons from the analysis.  
However, the attendant simplifications, while useful for some forms of analysis, probably come at too high a price because light-quark physics is at the heart of Nature: the proton, absolutely stable owing to real-world confinement, is defined by its valence light-quark content; hence, no discussion of proton stability can be conclusive in pure gauge QCD.  
In our view, despite the complications they bring, it is better to include light quarks from the outset, especially if the method pretends to an explanation of hadron observables.

One may readily highlight the manifold character of the approaches being used in the pursuit of $\alpha_s^{\rm IR}$ by listing just some of the scales employed to characterize nonperturbative effects.  
For instance, as we have already often noted, $\Lambda_s$ is a popular choice; it is also reasonable, so long as proponents admit the caveat that, being RS-dependent, it is not an objective quantity.
Other examples include 
the mass of a given hadron, with the proton being a good choice because, since its stable, the proton mass-squared is a real number;
QCD string tension, $\sigma$;
AdS/QCD scale, $\kappa$;
chiral condensate;
gluon mass-scale; 
and Gribov horizon parameter.
If wielding a unifying tool, then some or all of these scales can be related.  
For instance, Eq.\,\eqref{eq: Lambda LO analytical relation} provides a link between $\Lambda_{\overline{\rm MS}}$ and $\kappa$; Eq.\,\eqref{eq:string tension from HLFQCD} connects $\sigma$ and $\kappa$; and Ref.\,\cite{Gao:2017uox} argues that the gluon mass scale and Gribov horizon parameter are identical.
Additional discussions can be found elsewhere \cite{Deur:2016tte, Roberts:2021nhw, Binosi:2022djx, Papavassiliou:2022wrb, Ding:2022ows, Roberts:2022rxm, Ferreira:2023fva}.
  
\section{Effective charge method}
\label{EffectiveCharge}
\subsection{Generalities}
The effective charge scheme described by Grunberg \cite{Grunberg:1980ja, Grunberg:1982fw, Grunberg:1989xf} associates a QCD coupling with the expansion of any given observable restricted to 
first order of a perturbative expansion in $\alpha_s^{\rm pQCD}$.
Such an effective charge implicitly incorporates terms of arbitrarily high order, $n>1$, in the perturbative coupling, $[\alpha_s^{\rm pQCD}]^n$, expressed in gluon bremsstrahlung and gluon vertex corrections at the probe--quark vertex and also pair creations, {\it viz}. all the contributions that lead to DGLAP evolution at $n>1$. 
A related prescription -- in spirit -- is the 't Hooft scheme \cite{tHooft:1977xjm, Dokshitzer:1998nz}, which defines the coupling from the QCD $\beta$-function truncated at 2 loops, Eq.\,\eqref{eq:beta4 list}.  Like the 1-loop form, the result is RS-independent.  However, this scheme does not eliminate the Landau pole, and \cite{Dokshitzer:1998nz}: ``the fake infrared problem -- obviously limits the use of $\alpha_s$ so defined, to sufficiently large momentum scales.''
Herein, therefore, we focus on the Grunberg prescription, which, being defined by an observable, has no such problem.

We illustrate the Grunberg scheme with an example based on the Bjorken sum rule \cite{Bjorken:1966jh, Bjorken:1969mm}, which connects the nucleon axial charge, $g_A$, to the integral of the isovector part of the nucleon spin structure function $g_1(x ,Q^2)$:  
\begin{eqnarray} 
 \Gamma_1^{p-n}(Q^2) := \int_0^{1} dx\,\big[g^p_1(x ,Q^2)-g^n_1(x ,Q^2)\big] \stackrel{\Lambda_s^2/Q^2 \simeq 0}{=} \frac{g_A}{6}.
\label{Eq: original bjorken SR}
\end{eqnarray} 
where we have highlighted that this simple form is valid on $\Lambda_s^2/Q^2 \simeq 0$.
At any value of $Q^2$ outside this neighborhood, there are corrections, which lead to the pQCD series \cite{Gribov:1972ri, Kataev:1994gd, Kataev:2005hv, Baikov:2008jh}:
\begin{align} 
\Gamma_1^{\rm p-n}& (Q^2 )  
= \frac{g_{\rm A}}{6}\bigg[1-\frac{\alpha^{\rm pQCD}_{{\rm {s}}}(Q^2)}{\pi}
-3.58\left(\frac{\alpha^{\rm pQCD}_{{\rm {s}}}(Q^2)}{\pi}\right)^2  
-20.21\left(\frac{\alpha^{\rm pQCD}_{{\rm {s}}}(Q^2)}{\pi}\right)^{3} \nonumber \\
&
\qquad - 175.7\left(\frac{\alpha^{\rm pQCD}_{{\rm {s}}}(Q^2)}{\pi}\right)^{4}+
\sim -893.38\left(\frac{\alpha^{\rm pQCD}_{{\rm {s}}}(Q^2)}{\pi}\right)^{5} \nonumber \\
& \qquad +  
\mathcal O\left(\big(\alpha^{\rm pQCD}_{\rm {s}} \big)^6\right) \bigg]+\sum_{n > 1} \frac{\mu_{2n}(Q^2)}{Q^{2n-2}}.
 \label{bjorken SR}
\end{align} 
Here, the series coefficients are calculated in the $\overline {\rm MS}$ RS. 
The expression in the V-scheme is also available up to $\big(\alpha_V^{pQCD}\big)^4$~\cite{Kataev:2023sru}.
The generalization of the Bjorken sum is derived in the DGLAP framework and is thus a solid prediction of pQCD.
HT corrections are provided by the coefficients $\{\mu_{2n}(Q^2)\}$. 
The $Q^2$-dependence of the $\{\mu_{2n}\}$ also arises from pQCD radiative corrections.
While the $\mu_{2n}$ can be phenomenologically determined -- see Refs.\,\cite{Deur:2004ti, Deur:2014vea, Deur:2008ej}, pQCD radiative corrections also make them dependent on $\alpha_s^{\rm pQCD}$.
Consequently, the extraction of  $\alpha_s^{\rm pQCD}$ from Eq.\,\eqref{bjorken SR} is difficult on domains where HT contributions are not negligible.
Finally, since HT contributions are nonperturbative, their computations is challenging \cite{Martinelli:1996pk}; and for the Bjorken sum, they are currently only estimated with models \cite{Stein:1995si, Balitsky:1989jb, Stein:1994zk, Lee:2001ug, Sidorov:2006vu}.
All such issues are circumvented \textit{via} the effective charge definition.

{\allowdisplaybreaks}
Implementing the Grunberg scheme, Eq.~(\ref{bjorken SR}) is written
\begin{subequations}
\label{eqn:alphadefG}
\begin{align}
\Gamma_1^{p-n}(Q^2)  & =: \frac{g_A}{6}\bigg(1-\frac{\alpha_{g_1}(Q^2)}{\pi}\bigg)\,, 
\label{alpha_g1}\\
\mbox{or, equivalently,}\quad \alpha_{g_1}(Q^2) & = \pi \left(1-\frac{6}{g_A}\Gamma_1^{p-n}(Q^2) \right),
\label{eqn:alphadef}
\end{align}
\end{subequations}
where the $g_1$ subscript indicates the observable chosen for the effective charge. 

With this definition of the coupling, both the short distance pQCD effects -- those within the bracket in Eq.\,\eqref{bjorken SR} and the $Q^2$-dependence of the $\{\mu_{2n}\}$ terms, 
and the long distance effects -- the $\{{\mu_{2n}}/{Q^{2n}}\}$ terms -- are incorporated into $\alpha_{g_1}$. 
This is analogous to the procedure that transforms the coupling {\it constant} in a classical Lagrangian into a {\it running} effective coupling during the renormalization process -- see Sec.\,\ref{introduction}, with the additional feature that long distance effects are also folded into $\alpha_s$.
It is their inclusion, as well as other non-perturbative mechanisms that may be omitted in Eq.\,\eqref{bjorken SR}, which suppresses the Landau pole -- see Sec.\,\ref{Landau pole}). 
One may therefore interpret the effective charge approach as a renormalization prescription, wherewith $\alpha_s$ becomes an observable, with all attendant features.

Although Grunberg's prescription was initially intended only for the pQCD domain, effective charges, when evaluated from measurements, naturally extend to cover the nonperturbative domain. 
This feature will be our primary focus herein.
(For a recent use of effective charges in the UV, see the discussion of LGT calculations of  $\alpha_s(M_{\rm Z}^2)$ \cite[Eq.\,(335)]{Aoki:2021kgd}.)
In addition to being defined in the IR, effective charges also have other merits, \textit{e.g}.:
they acquire the RS-independence of every first order pQCD approximant -- see the discussion of Eq.\,\eqref{eq:alpha_rel_2RS};
on the perturbative domain, they nonperturbatively complete the pQCD series;
and in being defined by an observable, they do not exhibit a Landau pole. 

A seeming disadvantage of effective charges is a loss of predictability; namely, since such a charge is defined \textit{via} one particular observable, then there are as many distinct effective charges as there are physical processes. 
However, owing to the strengths of CSRs \cite{Brodsky:1994eh}, predictive power is preserved on the perturbative domain, at least, because a choice of process is not qualitatively different from a choice of RS. 
Extension of CSRs to the IR is nontrivial.  
One attempt is described in Ref.\,\cite{Brodsky:1994eh}; and Refs.\,\cite{Deur:2014qfa, Deur:2016cxb, Deur:2016opc} describe a nonperturbative method that relates, one to another, different effective charges or RS-dependent couplings.

\subsection{The effective charge \mbox{$\mathbf \alpha_{g_1}(Q^2)$}}
\label{BjorkenPDCharge}
The generalized Bjorken sum rule, Eq.\,\eqref{eqn:alphadefG}, is especially well suited to defining 
an effective charge for several reasons:\label{ag1 assets}
\begin{enumerate}[(i)]
    \item The pQCD series for $\Gamma_1^{p-n}$ has no $x$-dependence and, compared to many other observables, is easier to compute. 
    It is currently known up N$^{3}$LO with a N$^4$LO estimate \cite{Bjorken_a5} -- see also Ref.\,\cite{Goriachuk:2021ayq}. 
    \item \label{Note2} $\Gamma_1^{p-n}$ is an isovector quantity; so some potential contributions, the $\Delta$-resonances and, at amplitude level, disconnected diagrams, are eliminated. 
    This facilitates calculations of $\Gamma_1^{p-n}$, \textit{e.g}., those of chiral effective field Theory \cite{Burkert:2000qm} and LGT.
    Moreover, $\Gamma_1^{p-n}$ is an $x^0$-moment (local operator); hence, directly computable using LGT, as opposed to the structure functions themselves (nonlocal quantities). 
    Contemporary LGT analyses of $g_A$ report results with sub-percent precision \cite[Sec.\,10.3.1]{Aoki:2021kgd}.
    \item $g_A$ is precisely measured from neutron $\beta$ decay: $ g_A=1.2762(5)$ \cite{Workman:2022ynf}. 
    \item \label{Note4} $\Gamma_1^{p-n}$ is measured over a wide $Q^2$ domain: $Q^2/{\rm GeV}^2 \in [0.02,10]$ \cite{Deur:2014vea, Deur:2021klh,
    Ackerstaff:1997ws, Ackerstaff:1998ja, Airapetian:1998wi, Airapetian:2002rw, Airapetian:2006vy,
    Kim:1998kia,
    Adeva:1993km, Alexakhin:2006oza, Alekseev:2010hc, Adolph:2015saz,
    Anthony:1993uf, Abe:1994cp, Abe:1995mt, Abe:1995dc, Abe:1995rn, Anthony:1996mw, Abe:1997cx, Abe:1997qk, Abe:1997dp, Abe:1998wq, Anthony:1999py, Anthony:1999rm, Anthony:2000fn, Anthony:2002hy}. 
    \item \label{Note5} Robust sum rules allow extractions of $\alpha_{g_1}(Q^2)$ on domains not covered by measurements: on $Q^2 \simeq 0$ and the pQCD domain, $\alpha_{g_1}(Q^2)$ can be computed from Gerasimov-Drell-Hearn (GDH) \cite{Gerasimov:1965et, Drell:1966jv}  and Bjorken \cite{Bjorken:1966jh, Bjorken:1969mm} sum rules, respectively, see Eqs.\,\eqref{alphag10}, \eqref{eq:GDH limit of alpha_g_1}, \eqref{eq:msbar to g_1}.
    \item \label{Note6} $\alpha_{g_1}$ appears to be interpretable as a standard coupling, {\it even in the nonperturbative domain}~\cite{Deur:2005rp}. Moreover, crucially, $\alpha_{g_1}(Q^2)$ connects to several calculations  of $\alpha_s$ in the IR, including those from CSMs and AdS/QCD, see below.
\end{enumerate}

Given both that cross-sections are finite quantities and $Q^2\to 0 \Rightarrow x=Q^2/(2 M_N\nu)\to 0$, where $M_N$ is the nucleon mass and $\nu$ is the energy transferred to this target, then the support of the integrand in Eq.\,\eqref{Eq: original bjorken SR} must also vanish in this limit; namely, 
\begin{equation}
\lim_{Q^2\to 0}\Gamma_1(Q^2) = 0 \quad  
\mbox{and} \quad 
\alpha_{g_1}(0) = \pi\,.
\label{alphag10}
\end{equation}
Furthermore, the GDH sum rule can be written as follows \cite{Drechsel:2000ct, Ji:1999mr, Deur:2018roz}: 
\begin{equation}
\frac{8}{Q^2}\int_{0}^{1^{-}} g_1\left(x,Q^2\right)dx~ \xrightarrow[Q^2 \to 0] ~ \frac{-\kappa_N^2}{M_N^2},
\label{eq:GDH limit of alpha_g_1}
\end{equation}
where $\kappa_N$ is the nucleon anomalous magnetic moment and $1^-$ indicates that the elastic contribution ($x =1$) to the integral is excluded.
Hence, Eqs.\,\eqref{eqn:alphadef}, \eqref{eq:GDH limit of alpha_g_1} entail:
\begin{equation}
\alpha_{g_1}\left(Q^2 \right) \stackrel{Q^2/M_N^2 \simeq 0}{\approx} \pi + \frac{3\pi}{4g_{A}}\left(\frac{\kappa_{p}^2}{M_{p}^2}-\frac{\kappa_{n}^2}{M_{n}^2}\right)Q^2\,.
\label{eq:GDH on alpha_g_1}
\end{equation}
Finally, the empirical facts $M_p \approx M_n$ and $\kappa_p + \kappa_n \lesssim 0 $ entail that, with increasing $Q^2$, $\alpha_{g_1}\left(Q^2 \right)$ falls slowly away from its maximum value at $Q^2=0$:
\begin{equation}
\beta(Q^2) \propto \frac{d\alpha_{g_1}}{dQ^2} \stackrel{Q^2/M_N^2 \simeq 0}{\approx} \frac{3\pi}{4g_{A}}\left(\frac{\kappa_{p}^2}{M_{p}^2}-\frac{\kappa_{n}^2}{M_{n}^2}\right) \lesssim 0\,.
\end{equation}
Evidently, therefore, the Bjorken sum rule effective charge has an IR stable fixed point.  

It is also worth noting that, on any domain for which pQCD is valid, the generalized Bjorken sum rule, Eq.\,\eqref{bjorken SR}, together with the pQCD determination of $\alpha^{\rm pQCD}_s(Q^2)$ yield, in $\overline{\rm MS}$:
{\allowdisplaybreaks
\begin{eqnarray}
\alpha_{g_1}(Q^2)=\alpha^{\rm pQCD}_s(Q^2)+3.58\frac{({\scriptstyle\alpha^{\rm pQCD}_s(Q^2)})^2}{\pi}+
20.21\frac{({\scriptstyle\alpha^{\rm pQCD}_s(Q^2)})^{3}}{\pi^2}+ \nonumber \\
175.7\frac{({\scriptstyle\alpha^{\rm pQCD}_s(Q^2)})^{4}}{\pi^{3}}
\sim -893.38 \frac{({\scriptstyle\alpha^{\rm pQCD}_s(Q^2)})^{5}}{\pi^{4}} + 
\mathcal{O}\left((\alpha^{\rm pQCD}_s)^{6}\right).
\label{eq:msbar to g_1}
\end{eqnarray}
}

One may alternatively use the PMC (Sec.\,\ref{BLM/PMC}) to express $\alpha_{g_1}(Q^2)$ as a conformal series, where the argument at order $n$ evolves with a calculable PMC scale $Q^2_n$.
The series is nearly independent of the choice of renormalization scale.
The PMC, a rigorous procedure which generalizes the BLM method \cite{Brodsky:1982gc}, eliminates 
the occurrences of the $\beta_n$-terms in the series coefficients, evident in Eq.\,\eqref{eq:alpha_s}. 
By definition, these terms are non-conformal and the origin of the renormalon divergence problem.
The PMC is the non-Abelian generalization of the standard Gell-Mann-Low method used for fixing the renormalization scale in QED. It is thus consistent with the grand unification of QCD and electroweak theory. 
It also satisfies the analytic condition that pQCD and QED must match in the limit $N_C \to 0$ \cite{Brodsky:1997jk, Brodsky:2015eba} -- see also the discussion on page~\pageref{NC-->limit}.

Using the PMC, perturbative approximants are expressed with different scales, $Q^2_n$, at each order, 
in contrast to the standard scale-setting with a single scale, $Q^2$, for any order.
When using  conventional scale setting, the renormalization scale  is heuristically estimated as the momentum transfer characterizing the process, with an arbitrary uncertainty $0.5Q^2 - 2Q^2$ assigned.
In contrast, the PMC scales, $Q^2_n$, correspond to the  effective virtuality in the processes contributing at order $n$.
As in QED, the PMC renormalization scales only depend on the choice of renormalization scheme by a single universal factor.

For $\alpha_{g_1}$ \cite{Deur:2017cvd},
\begin{eqnarray}
\alpha_{g_1}(Q^2)=\alpha^{\rm pQCD}_s(Q^2_1)+1.146\frac{({\scriptstyle\alpha^{\rm pQCD}_s(Q^2_2)})^2}{\pi}+
0.144\frac{({\scriptstyle\alpha^{\rm pQCD}_s(Q^2_3)})^{3}}{\pi^2}+ \nonumber \\
0.763\frac{({\scriptstyle\alpha^{\rm pQCD}_s(Q^2_4)})^{4}}{\pi^{3}}+
\mathcal{O}\left((\alpha^{\rm pQCD}_s)^{5}\right),
\label{eq:g_1 PMC}
\end{eqnarray}
where the series coefficients and PMC scales are computed at NNLO, for $n_f=3$ using $\overline{\rm MS}$ as the intermediate RS for $\alpha^{\rm pQCD}_s$.
The scale values are $Q^2_1= 0.338Q^2$, $Q^2_2= 0.047Q^2$ and $Q^2_3= 1600Q^2$. 
The last scale is undetermined: it can be set to $Q^2_4=Q^2$ or $Q^2_4=Q^2_3$ with negligible differences.
Residual $\beta$-dependences of the series in Eq.~(\ref{eq:g_1 PMC}) are discussed in Refs.~\cite{Deur:2017cvd, Shen:2016dnq, Kataev:2022iqf}. Nevertheless, comparing Eqs.\,\eqref{eq:msbar to g_1} and \eqref{eq:g_1 PMC} reveals that the PMC series has significantly better convergence properties and exhibits no obvious sign of a renormalon divergence.

On the other hand, the low PMC scales, notably $Q^2_2=0.047\, Q^2$, restrict applicability of the PMC series to higher values of $Q^2$ than the $\overline{\rm MS}$ series: in practice, one must take   $Q^2>2.2\,$GeV$^2$, at least \cite {Deur:2017cvd}.
To address this issue, $\alpha_{g_1}$ has been expressed using the PMC with a single scale $Q^2_\ast$ and the $V$-scheme as intermediate renormalization prescription for $\alpha^{\rm pQCD}_s$ \cite{Yu:2021yvw}:
{\allowdisplaybreaks
\begin{eqnarray}
\alpha_{g_1}(Q^2)=\alpha^{\rm pQCD, V}_s(Q^2_*)+3.15\frac{({\scriptstyle\alpha^{\rm pQCD, V}_s(Q^2_*)})^2}{\pi}+
20.46\frac{({\scriptstyle\alpha^{\rm pQCD, V}_s(Q^2_*)})^{3}}{\pi^2}+ \nonumber \\
51.36\frac{({\scriptstyle\alpha^{\rm pQCD, V}_s(Q^2_*)})^{4}}{\pi^{3}}+
\mathcal{O}\left((\alpha^{\rm pQCD, V}_s)^{5}\right),
\label{eq:g_1 PMC, V}
\end{eqnarray}
with $Q^2_*=Q^2 e^{0.58 + 2.06{\scriptscriptstyle\alpha^{\rm pQCD, V}_s(Q^2)} - 7.41({\scriptscriptstyle\alpha^{\rm pQCD, V}_s(Q^2)})^2} \simeq 1.8Q^2$.
The expression for $\alpha^{\rm pQCD, V}_s$ can be found elsewhere \cite[Eq.\,(3.42)]{Deur:2016tte}.
Plainly, the convergence of the series in Eq.~(\ref{eq:g_1 PMC, V}) is not as good as that in Eq.\,\eqref{eq:g_1 PMC}, but it is improved compared to Eq.\,\eqref{eq:msbar to g_1} and can be applied over a similar $Q^2$ range. 
}

\begin{figure}[!t]
\begin{center}
\centerline{\includegraphics[width=0.6\textwidth, angle=0]{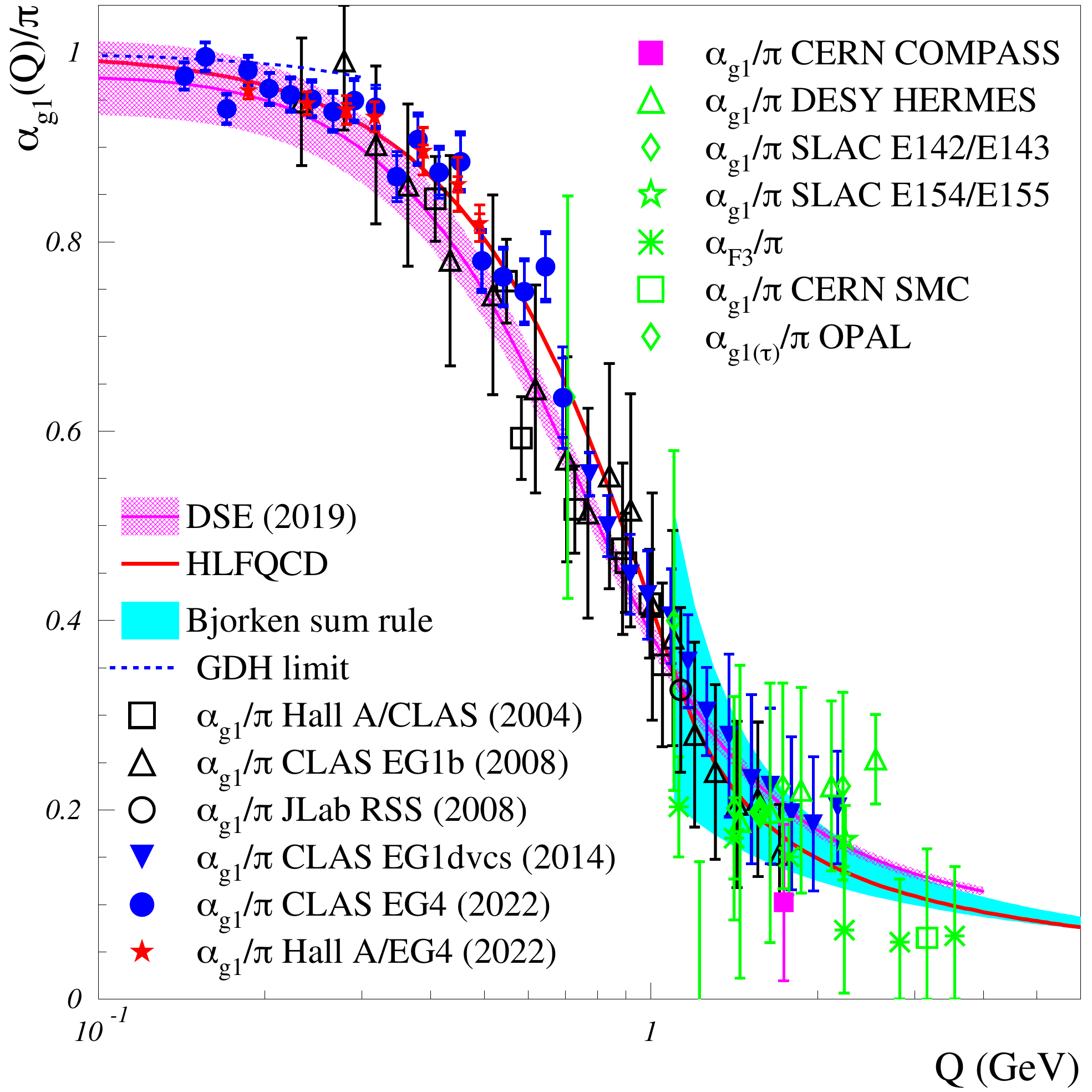}}
\end{center}
\vspace{-1.3 cm}
\caption{\small 
Effective charge $\alpha_{g_1}(Q)/\pi$. 
The most recent extractions from data \cite{Deur:2022msf} are shown by the filled blue circles and red stars.   
Their inner error bars give the statistical uncertainties and the outer ones represent the quadratic sum of the systematic and statistical uncertainties. 
The open symbols show earlier extractions \cite{Deur:2005cf, Deur:2008rf}, with the error bars being the quadratic sum of the 
systematic and statistical uncertainties.  Recent theoretical predictions are also shown:
CSM \cite{Binosi:2016nme, Cui:2019dwv} (magenta curve and shaded band); 
and HLFQCD~\cite{Brodsky:2010ur} (red line, using $\kappa=0.534$~GeV).
The cyan band and associated dashed curve are computed using the  Bjorken and GDH sum rules, respectively.
\label{fig:alpha_g1}}
\end{figure}

The advantages of notes (\ref{Note4}), (\ref{Note5}) on page~\pageref{Note4} make $\alpha_{g_1}$  known at any $Q^2$, and note (\ref{Note6}) indicates that it may be interpreted in terms of calculations and comparable therewith. 
The Bjorken sum rule effective charge, extracted \cite{Deur:2005cf, Deur:2008rf, Deur:2022msf} from measurements of $\Gamma_1^{\rm p-n}(Q^2)$ 
at the world's accelerator facilities \cite{Deur:2014vea, Deur:2021klh, Ackerstaff:1997ws, Ackerstaff:1998ja, Airapetian:1998wi, Airapetian:2002rw, Airapetian:2006vy, Kim:1998kia, Adeva:1993km, Alexakhin:2006oza, Alekseev:2010hc, Adolph:2015saz, Anthony:1993uf, Abe:1994cp, Abe:1995mt, Abe:1995dc, Abe:1995rn, Anthony:1996mw, Abe:1997cx, Abe:1997qk, Abe:1997dp, Abe:1998wq, Anthony:1999py, Anthony:1999rm, Anthony:2000fn, Anthony:2002hy}, is shown in Fig.~\ref{fig:alpha_g1}. 

To understand the meaning of $\alpha_{g_1}$, one needs to recall the origin of each term in Eq.\,\eqref{bjorken SR}.
On $\Lambda_s^2/Q^2 \simeq 0$, $g_1(x ,Q^2)$ represents the net polarization%
\footnote{
That is the net number density in light-front longitudinal momentum-space of quarks with spin pointing along the nucleon spin minus the net density of quarks with spin pointing oppositely.} 
of quarks with light-front momentum fraction $x$, making clear the reason for the presence of $g_A$ in Eq.\eqref{Eq: original bjorken SR}.
The spin-dependent part of the scattering amplitude stems from the matrix elements of the axial current  $\bar \psi \vec{\tau}\gamma^\mu \gamma^5 \psi$, where $\psi$ is the nucleon state and $\vec{\tau}$ are the isospin Pauli matrices.
By Noether's theorem, the conserved axial current is associated with the chiral symmetry $\psi \to e^{i\vec{\phi}\cdot \vec{\tau}\gamma^5} \psi$. 
In an elastic reaction, the axial current generates $g_A(Q^2)$, the nucleon axial form factor, 
just like the electromagnetic current $\bar \psi \gamma^\mu \psi$ generates the nucleon electromagnetic form factors $G_E(Q^2)$ and $G_M(Q^2)$.  
A Fourier transform of $G_E(Q^2)$ gives the electric charge spatial distribution and likewise 
a Fourier transform of $g_A(Q^2)$ represents the nucleon spin spatial distributions,  {\it viz}.\ 
the evolution from the nucleon center to its boundary of the net valence-quark polarization \cite{Chen:2022odn}. 
Thus, $g_A$ is the net valence-quark polarization, without spatial resolution,
 \textit{i.e}., the spatial average, which directly connects to the average valence-quark polarization in momentum-space, $\int dx g_1$.

Using $\overline {\rm MS}$, the leading-twist $\alpha_s^{\rm pQCD}$-dependence originates from gluon bremsstrahlung and gluon vertex correction at the probe-photon+struck-quark interaction vertex.
HT terms comprise kinematical and dynamical corrections. 
The former arise from the nonzero nucleon mass, which provides a scale violating the UV conformality of QCD, \textit{i.e}., Bjorken scaling. 
The latter represent interactions, via one or several hard gluons, of the struck quark with the rest of the nucleon made of two valence-quarks, sea quarks and gluons (nearly-spectator partons). 
This part of the nucleon is nonperturbative;
hence, the  $\{\mu_{2n}\}$ express nonperturbative distributions. 
In this picture, some HT contributions may implicitly be related to confinement forces \cite{Burkardt:2008ps, Abdallah:2016xfk}. 
However, HT terms do not include 3-nucleon coherent reactions (resonances). 

With the origin of the terms in the generalized Bjorken sum rule thus explicated, $\alpha_{g_1}(Q)$ can be interpreted.
Short-distance quantum effects, such as vertex correction and vacuum polarization, cause the running of $\alpha^{\rm pQCD}_s$.
In addition, the $\alpha_{g_1}$ effective charge includes 
gluon bremsstrahlung;
gluon exchange between the incoming and outgoing active quark;
and, at low-$Q^2$, HT terms and resonance effects, not formalized by the OPE and therefore not written in Eq.\,\eqref{bjorken SR}.

On the nonperturbative domain, where pQCD radiative corrections and HT have merged, possibly for $Q^2 \lesssim 0.6\,$GeV$^2$ \cite{Deur:2014qfa}, the meaning of $\alpha_{g_1}(Q)$ is more speculative~\cite{Deur:2005rp}. 
If the contribution to $\Gamma_{1}^{\rm p-n}$ from nucleon resonances dominates, the connection between $\alpha_{g_1}(Q)$ and the QCD coupling becomes unclear at low $Q^2$ since in the former case at least three color sources are involved, whilst a coupling is defined from a two-body interaction; 
but if the contribution to $\Gamma_{1}^{\rm p-n}$ of incoherent reactions dominates, like at large $Q^2$, 
$\alpha_{g_1}(Q)$ may, at least approximately, still be interpreted as a coupling. 
Note~(\ref{Note2}) on page~\pageref{Note2} makes this plausible because the 
contribution from $\Delta$ resonances, which dominates for $\Gamma_{1}^{\rm p}$ and $\Gamma_{1}^{\rm n}$ separately, 
vanishes in $\Gamma_{1}^{\rm p-n}$ \cite{Burkert:2000qm}.
The issue is then whether this ensures that incoherent reactions dominate over the remaining resonances in $\Gamma_{1}^{\rm p-n}$.

\begin{figure}[!t]
\hspace*{-1ex}\begin{tabular}{ll}
{\sf A} & {\sf B} \\[-0ex]
\includegraphics[clip, width=0.5\textwidth]{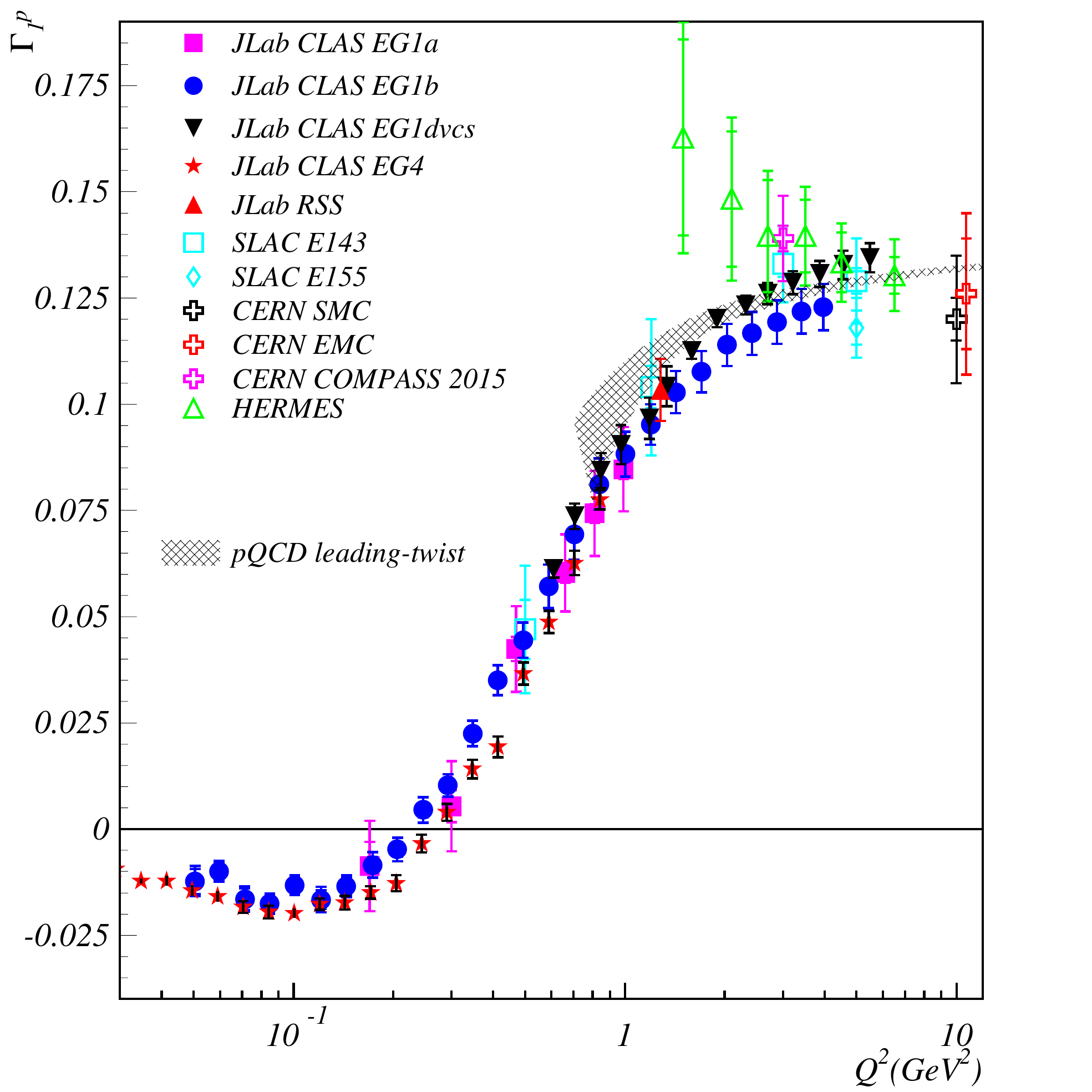} &
\includegraphics[clip, width=0.5\textwidth]{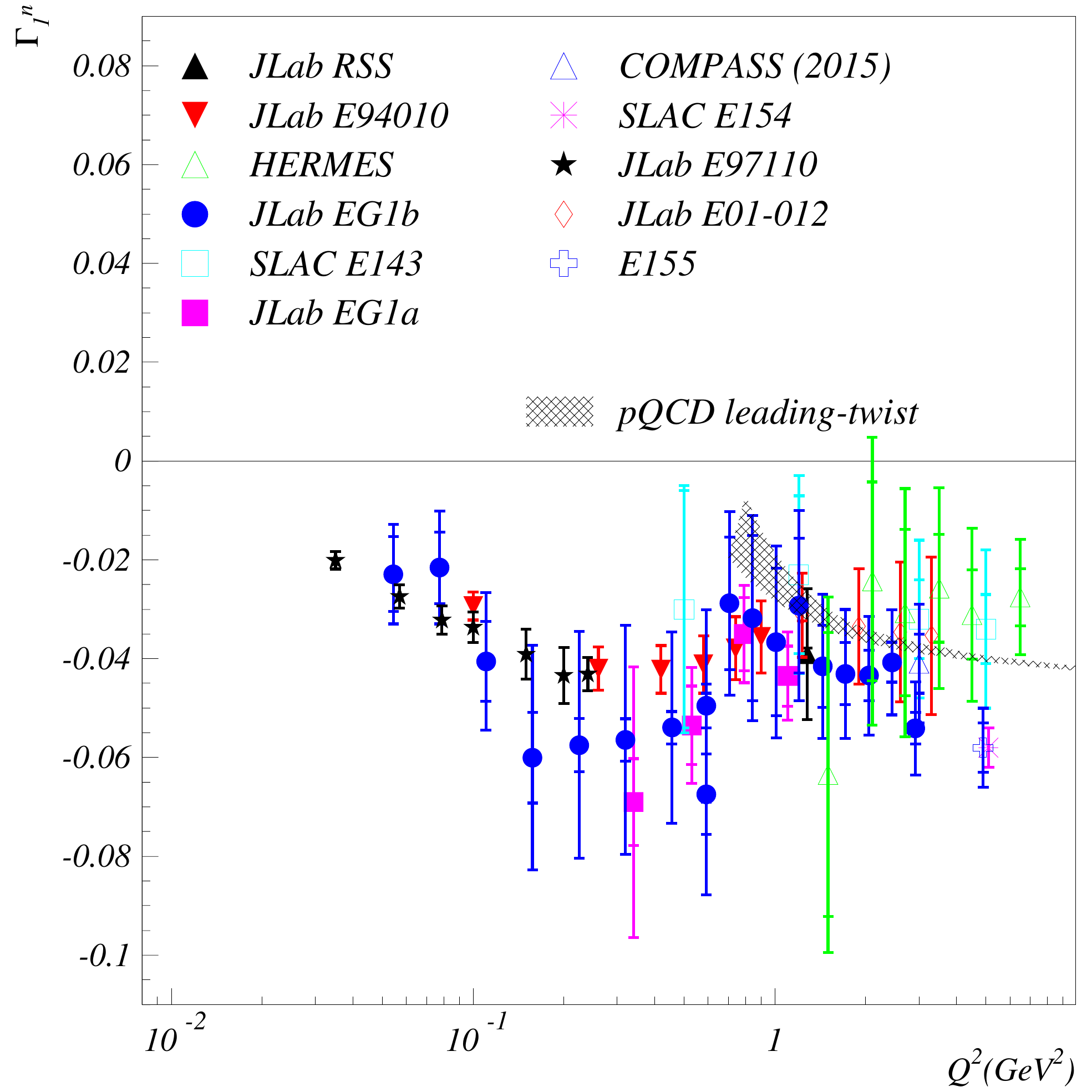}
\end{tabular}

\leftline{\rule{8ex}{0ex}{\sf C}}
$\,$\\[-2ex]
\centering{\includegraphics[width=0.70\textwidth, height=0.60\textwidth]{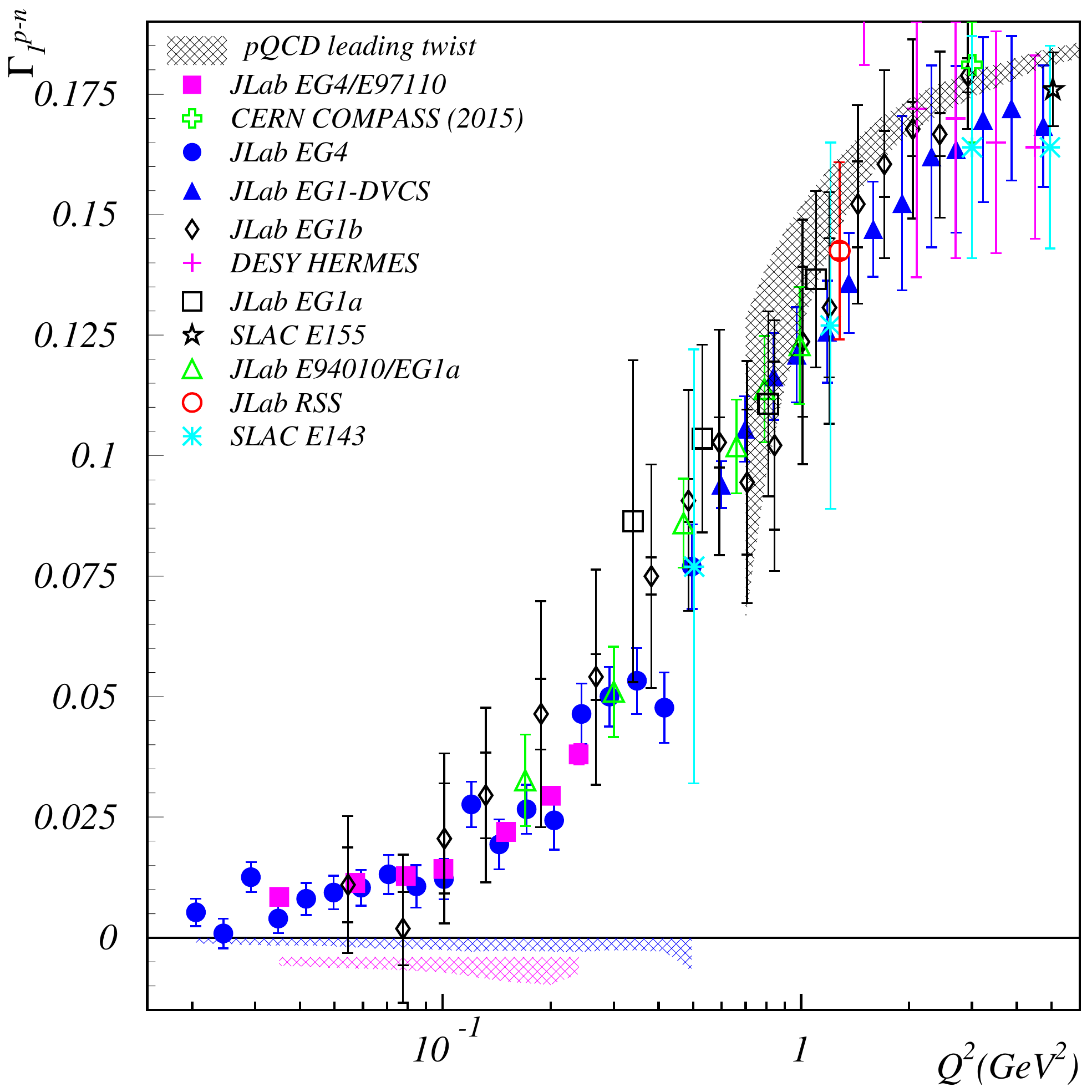}}

\caption{\footnotesize Experimental data (symbols) and pQCD prediction at leading twist (grey bands) for the first moment 
$\Gamma_{1}(Q^2)$. Top left: proton $\Gamma_{1}^{\rm p}$, Top right: neutron 
$\Gamma_{1}^{\rm n}$. Bottom: isovector part $\Gamma_{1}^{\rm p-n}$.}
\label{fig:gamma1}
\end{figure}

That it may be the case can be seen by comparing $\Gamma_{1}^{\rm p}$ and $\Gamma_{1}^{\rm n}$ to $\Gamma_{1}^{\rm p-n}$ -- see Fig.~\ref{fig:gamma1}.

With decreasing $Q^2$, the incoherent reaction contributions (DIS and, at low $Q^2$, nonresonant background) drop, as predicted by pQCD  (grey bands). 
On the pQCD domain, $\Gamma_{1}^{\rm p-n}$ is large because the incoherent reactions contribute to the  
proton and neutron with opposite sign, owing to isospin symmetry and the fact that up and down quarks contribute to the nucleon spin with opposite sign (the up quarks tending to (anti)align with the proton(neutron) spin, whilst down quarks have the opposite tendency -- see Ref.\,\cite[Table~I]{Chen:2022odn}). 
In fact, $\Gamma_{1}^{\rm p}$ and $\Gamma_{1}^{\rm n}$ display opposite signs.

On the nonperturbative domain, down to $\sim0.6$ GeV$^2$, the nonresonant contributions to $\Gamma_{1}^{\rm p}$, $\Gamma_{1}^{\rm n}$ and $\Gamma_{1}^{\rm p-n}$ are still relatively large, with no sign of structure that would signal a change of the mechanism driving their $Q^2$-evolution and the importance of the resonance contribution.
Such structure appears for $\Gamma_{1}^{\rm p}$ and $\Gamma_{1}^{\rm n}$ for $0\leq Q^2 \leq 0.5\,$GeV$^2$, with a negative peak for $\Gamma_{1}^{\rm n}$ centered around $Q^2 \simeq 0.2\,$GeV$^2$ and a $\Gamma_{1}^{\rm p}$ that becomes negative and then turns-over presumably owing to the same negative resonance contribution as for the neutron. 
Yet, $\Gamma_{1}^{\rm p-n}$ continues to decrease smoothly, with no sign of this structure. 
These observations suggest that resonances have similar effects on the proton and the neutron, but cancel in their difference, as expected if the $\Delta$ resonance is the main mechanism driving the low $Q^2$-evolution of $\Gamma_{1}^{\rm p}$ and $\Gamma_{1}^{\rm n}$ \cite{Burkert:2000qm}. 

Below $Q^2=0.2\,$GeV$^2$, the resonance contribution dominates in the individual $\Gamma_{1}^{\rm p}$ and $\Gamma_{1}^{\rm n}$. 
They are both negative and therefore partly cancel in $\Gamma_{1}^{\rm p-n}$. 
Consequently, the nonresonant reactions may be the dominant contribution to $\Gamma_{1}^{\rm p-n}$ at any $Q^2$, with the residual resonance effects further suppressed by global quark-hadron duality \cite{Bloom:1970xb}.

Neglecting any residual coherent (resonance) effects, 
$\alpha_{g_1}$ can be interpreted as in the large $Q^2$ domain, \textit{i.e}., the usual  coupling of an interaction between two force sources \cite{Deur:2009zy}. 
The first source is the struck quark and the second source is either undefined (quasi-real gluon bremsstrahlung) or the residual hadronic structure for high and low $Q^2$, respectively. 
These observations highlight that the interaction need not be carried by a single gluon, \textit{e.g}., $H$-graphs (Ref.\,\cite[Fig.\,4]{Ding:2022ows}) are permitted; and that this does not forbid the definition of a coupling because there is no requirement that it be associated with single boson exchange.
An explicit example is provided in Refs.\,\cite{Binosi:2014aea, Binosi:2016xxu, Binosi:2016wcx, Binosi:2016nme, Cui:2019dwv}, in which the PI running coupling, owing to its derivation using the PT-BFM, cannot be associated with single gluon exchange, if that exchange is viewed in the textbook sense.

In contrast, effective charges drawn from a coupling to more than two resolved sources, thereby allowing for several vertices, or that involve coherence effects, may not easily be interpreted 
as a force coupling. 

All in all, $\alpha_{g_1}$ is defined to follow, on the perturbative domain, the Gell-Mann Low coupling prescription \cite{GellMann:1954fq}. 
In the IR, it may still be argued to represent a force coupling.
The latter is a crucial and nontrivial property, which may be peculiar to $\alpha_{g_1}$ as it largely filters out coherent scattering effects. 
At  high $Q^2$, $\alpha_{g_1}$ acquires its $Q^2$-dependence from small distance 
quantum loops (vacuum polarization; quark self-energy; vertex corrections), just like $\alpha^{\rm pQCD}_s$, to which are added multi-gluon quasi-real emissions and photon-quark vertex 
corrections. 
At lower $Q^2$, but still upon a domain for which pQCD is applicable, HT terms contribute.  
At the lowest $Q^2$ values, incoherent reactions still appear to dominate, perhaps involving the full panoply of nonperturbative gauge sector interactions.
In the past quinquennium, the viability of this picture of $\alpha_{g_1}$ has received material support from its agreement with theoretical calculations using definitions consistent with $\alpha_{g_1}$ -- see Fig.~\ref{fig:alpha_g1} and Secs.\,\ref{HLFQCD}, \ref{DSE:alpha_PI}.

\section{$\alpha_s$ and a confining linear potential}
\label{EffectivePotential}
The idea of strings to describe hadronic structure emerged in the 1960s \cite{Susskind:1969ha, Susskind:1970xm, Frye:1970bu, Fairlie:1970tc, Nambu:1974zg}, following a suggestion that hadron mass spectra can be organized as sets of Regge trajectories \cite{Regge:1959mz},
{\it viz}.\ that the masses-squared depend linearly on both the principal quantum number $n$ (which gives the number of nodes possessed by a hadron wavefunction) and the internal orbital angular momentum, $L$. 
Plainly, given that neither of these quantities is Poincar\'e invariant, some caution ought to be used in going forward from here \cite{Brodsky:2022fqy}.  
Nevertheless, it was subsequently suggested that, for mesons, such behavior might indicate that these states are constituted from two ``constituent-like'' quarks (QCD connection unknown) interacting via a potential that grows linearly with their separation, $r$. 
Further phenomenology yielded a nonrelativistic, static potential of the form:
\begin{equation}
V(r)=-\frac{C_F\alpha_V\left(r\right)}{r}+\sigma r, \label{eq:Q-Q stat pot.}
\end{equation}
known as the Cornell potential \cite{Eichten:1974af, Eichten:1978tg, Eichten:2002qv}. 
Here, $C_F=\sfrac{4}{3}$ is a QCD color factor, $\sigma \approx (0.42\,{\rm GeV})^2$  is the string tension parameter and $\alpha_V(r)$ is the three-dimensional Fourier transform to coordinate space of $\alpha_s(Q^2)$ in the $V$ RS. 

The first term in Eq.\,\eqref{eq:Q-Q stat pot.} dominates at short distances and is a Coulomb-like static potential attributed to perturbative one-gluon exchange.
It affects the hadronic wavefunctions and the fine structure of the hadron mass spectrum. Refs.\,\cite{Takaura:2018vcy, Karbstein:2018mzo} provide the expression of the perturbative static source+sink potential up to O$(\alpha_s^3)$, including logarithmic corrections and $\alpha_V$ to 4-loop. 
The second term in Eq.\,\eqref{eq:Q-Q stat pot.} acts like a string potential between the 
static source+sink pair. 
The string was historically interpreted as stemming from gluons being condensed into a flux tube, although other pictures have emerged that also generate a linear static potential at long distance, \textit{e.g}., 
the harmonic oscillator LF potential discussed in Sec.\,\ref{HLFQCD}, and 
the center vortices and instantons seen in LGT studies \cite{Langfeld:2003ev, Bowman:2010zr, OCais:2008kqh, Leinweber:2022ukj, Trewartha:2015ida, Biddle:2022zgw, Virgili:2022ybm}.

This second term serves to produce confinement for the nonrelativistic constituent-like quarks: in practice, it provides the basis for a useful phenomenology of heavy quarkonium systems, $Q\bar Q$, with $m_{Q}\gg\Lambda_s$. 
It also determines the slopes and intercepts of the Regge trajectories. 
These can then be straightforwardly interpreted within the model: $L$ expresses how fast the $Q\bar Q$ rotates around the hadron center of mass. 
The faster they orbit, the larger is $L$ and the tension in the string that balances the centrifugal force. 
This produces a larger hadron binding energy, \textit{i.e}., a mass increase. Whether the trajectory is linear or not depends on model details \cite{Inopin:1999nf, Tang:2000tb, Masjuan:2012gc}.
Nevertheless, something like the widely-used Cornell potential is a feature of strong interactions between static source+sink pairs and should be reproduced in any nonperturbative approach that aims to describe such systems.
It is seen in LGT studies of static source+sink pairs \cite{Bali:1994de, Bali:2000gf} -- as, too, is string breaking when dynamical fermions are introduced \cite{Bali:2005fu, Prkacin:2005dc};
and Regge trajectories are found in anti-de Sitter/conformal field theory (AdS/CFT) models of high-energy fixed-angle scattering of glueballs \cite{Polchinski:2001tt}.
In HLFQCD -- see Sec.\,\ref{HLFQCD}, one obtains a prediction for $\sigma$ \cite{Trawinski:2014msa}:
\begin{equation}
\sigma_{AdS} = 2\kappa^2/\pi =  (0.42\,\mbox{GeV})^2,
\label{eq:string tension from HLFQCD}
\end{equation}
matching phenomenology-based expectations.

Working within a framework defined by Eq.\,\eqref{eq:Q-Q stat pot.}, one can reach the position that the hadron spectrum is sensitive to the long distance behavior of $\alpha_s$ when that spectrum is interpreted in terms of an effective charge.
It has been long recognized that this indicates that $\alpha_V$, taken from a potential defined without the confinement term, $\sigma r$, freezes in the IR; albeit, there was no agreement on the value of the IR-fixed point -- see Ref.\,\cite{Deur:2016tte} for an overview of different, early works on the subject.

Following Refs.\,\cite{Booth:1992bm, Michael:1992nj,  Bali:1992ru}, the static quark-quark potential has been employed extensively by LGT practitioners to determine $\alpha_V$ in the UV\,\cite{Necco:2001xg, Necco:2001gh, Takahashi:2002bw, JLQCD:2002zto, HPQCD:2003rsu, Mason:2005zx, Davies:2008sw, Jansen:2011vv, Bazavov:2012ka, Tormo:2013tha, Karbstein:2014bsa, Bazavov:2014soa, Husung:2017qjz, Karbstein:2018mzo, Takaura:2018lpw,  Takaura:2018vcy, Bazavov:2019qoo, Ayala:2020odx, Husung:2020pxg} -- see Sec.\,\ref{LGT determinations}. 
Recent LGT determinations of $V(r)$ up to $\sim 1$~fm are available in Refs.\,\cite{Takaura:2018vcy, Karbstein:2018mzo, Brambilla:2022het}.

Alternatively, one may  include $\sigma r$ in the coupling, so that the potential becomes:
\begin{equation}
V(r)=-\frac{4}{3}\frac{\alpha_{V}\left(r\right)}{r}+\sigma r =: -\frac{4}{3}\frac{\alpha_{\rm Rich}\left(r\right)}{r}
\end{equation}
or, in momentum space:
\begin{equation}
V(|\vec{Q}|)=\frac{1}{\left(2\pi\right)^{3}}\int d^{3}re^{-ir.Q}V(r)=-\frac{2}{3\pi^{3}}\frac{\alpha_{V}\left(|\vec{Q}|^2\right)}{Q^2}-\frac{\sigma}{\pi^{3}|\vec{Q}|^{4}}
=: -\frac{2}{3\pi^{3}}\frac{\alpha_{Rich}(|\vec{Q}|^2)}{|\vec{Q}|^2},
\label{eq:Q-Q stat pot. Mom space.}
\end{equation}
where the subscript refers to Richardson, who first proposed this type of definition \cite{Richardson:1978bt}.
In the IR, $V \propto 1/|\vec{Q}|^4$; so, $\alpha_{\rm Rich}$ diverges as $1/|\vec{Q}|^2$.
One can understand why $\alpha_{\rm Rich}$ does not freeze, despite the 
general argument that confinement suppresses, at hadronic scales, the quantum effects causing $\alpha_s$ to run \cite{Brodsky:2008be, Brodsky:2012ku, Gao:2017uox}: in connection with static potentials, such as $\sigma r$, there are no well-defined finite, nonzero wavelengths. 
The wavelengths of the force carriers, whatever they are, tend to infinity and static sources have zero wavelength. 
Therefore, the long wavelength suppression mechanism \cite{Brodsky:2008be, Brodsky:2012ku, Gao:2017uox}, whereby $\alpha_s$ freezes, is obscured by the loss of an explanation of the potential in terms of (even dressed) quanta.
Regular IR behaviour can be obtained by introducing a gluon mass scale as, \textit{e.g}., in Ref.\,\cite{Cucchieri:2017icl}, but this does not typically lead to improved phenomenological outcomes. 
Little progress has been made with an effective charge based on Richardson-like ideas in the past quinquennium; so, we refer to Ref.\,\cite{Deur:2016tte} for a fair discussion of the status of this scheme. 

\section{Holographic Light-Front  QCD }
\label{HLFQCD}
\subsection{Background sketch}
Holographic Light-Front QCD (HLFQCD) \cite{Brodsky:2014yha} is a nonperturbative approach to QCD phenomena based on AdS/CFT and light-front (LF) quantization.
The latter means that canonical quantization is performed using light-front time, $x^+ \equiv  x^0+x^3$, rather than ordinary time $x^0$ \cite{Dirac:1949cp}. 
A general introduction to LF Hamiltonian QCD is given in Ref.\,\cite{Brodsky:1997de} and the LF computational rules are given in Ref.\cite{Lepage:1979zb}.
LF quantization offers many advantages, some of which we now list. 
\begin{enumerate}[(a)]
    \item It is effectively Poincar\'e invariant and therefore free of pseudo-dynamical effects, akin to inertial forces, in contrast to the standard canonical quantization based on $x^0$ \cite{Brodsky:2022fqy}.
    \item It is generally considered to lead to a trivial vacuum structure \cite{Brodsky:2022fqy}.
    \item LF wavefunctions have a clear and straightforward interpretation, which provides for a rigorous derivation and interpretation of the parton model \cite{Brodsky:1997de}. 
    \item \label{Noted} It offers a systematic method for solving nonperturbative bound-state problems \cite{Brodsky:1997de}.
\end{enumerate}
    
Regarding note~\eqref{Noted}, we remark that LF quantization provides a rigorous formulation of QCD with hadron structure described by a relativistic Schr\"{o}dinger equation. 
In principle, the components of the equation can be derived directly from the QCD Lagrangian.
However, in practice, the confining potential term, $U(\zeta^2)$, where $\zeta$ is the transverse parton
separation in LF coordinates, has only been analytically computed in $(1+1)$ dimensions \cite{Hornbostel:1988fb}.
In (3+1) dimensions, a first-principles computation of $U(\zeta^2)$ has thus far proved too difficult. 
One method, currently being explored, aims to capitalize on the strengths of LGT 
~\cite{Burkardt:2001jg}.
Another approach follows the path of gauge-gravity duality \cite{Maldacena:1997re, Brodsky:2006uqa, deTeramond:2008ht}, 
which has proven fruitful for delivering nonperturbative information about the QCD coupling.

Gauge-gravity, or AdS/CFT, duality posits that a classical
gravitational theory in a $(d+1)$ dimensional space-time, slightly curved (or, in interaction language, a weakly coupled gravitational theory) projects, on the boundary of the $(d+1)$ space, into a strongly coupled CFT in $d$ dimensional flat Minkowski spacetime. 
One crucial application of AdS/CFT duality is that it provides a method for the nonperturbative solution of some QFTs. 
Such programs are classified via two categories.
The {\it top-down} approach starts with a superstring theory, whose background is chosen so that the desired properties of the CFT are reproduced, in our context, \textit{e.g}., some form of confinement for QCD. 
The {\it bottom-up} approach starts with the CFT and determines the gravitational theory. 
Once the dual theories are matched, weakly coupled gravitational calculations can be performed to obtain strongly coupled gauge CFT results. 

Herein, we will be concerned with the  bottom-up method.
Applied to QCD, it yields an AdS/QCD correspondence, which stems from the duality between the group of isometries of a 5-dimensional AdS spacetime and the $SO(4,2)$ conformal group that approximately describes QCD. 
In fact, the Lagrangian in Eq.\,\eqref{eq:QCD Lagrangian}, indicates that classical chromodynamics is a CFT in the chiral ($m_q=0$) limit.
Observation, however, reveals that a $\sim 1\,$GeV scale emerges in QCD, which breaks the classical conformal invariance and chiral symmetry. 
This scale can be expressed as, {\it inter alia}, the confinement scale, $\Lambda_s$, a hadron (proton or $\rho$) mass, the chiral symmetry breaking scale, $\chi_B$, the HLFQCD scale $\kappa$, or the string tension $\sigma$.

In contrast, away from this emergent scale, the CFT character of QCD holds approximately. 
On $M_N^2/Q^2 \simeq 0$, where short-distance quantum fluctuations and nonzero hadron mass effects are immaterial, it manifests as Bjorken scaling \cite{Bjorken:1969ja, Feynman:1969wa}. 
Within the complementary domain, $Q^2/M_N^2 \lesssim 1$, the experimental or calculated behavior of $\alpha_s$ suggests that it freezes, asymptotically approaching an IR-fixed point at $Q^2=0$.  
This, together with AdS/CFT duality, offers a nonperturbative method for QCD calculations. 

Classical gravitational calculations in 5-dimensional AdS-space, projected onto the 4-dimensional AdS-space boundary, which is identified with physical Minkowski spacetime, may be linked with solutions of QCD equations in which quantum (and quark mass) effects are absent. 
From this perspective, AdS/QCD provides a semiclassical approximation to QCD that incorporates many key aspects, among them the conformal invariance of the chromodynamics Lagrangian. 
If QCD is quantized on the LF, then AdS/QCD yields the HLFQCD approach \cite{deTeramond:2008ht, Brodsky:2006uqa}. 

Numerous methods for determining $U(\zeta^2)$ in the LF QCD Schr\"{o}dinger-like equation 
all yield the same potential, \textit{e.g}., the following schemes have been considered. 
\begin{enumerate}[(A)]
    \item Require the approximate conformal symmetry of QCD \cite{Brodsky:2014yha}.
    \item Impose the approximate chiral symmetry of QCD, in practice requiring that $U(\zeta^2)$ yields massless pions \cite{Dosch:2015nwa}.
    \item \label{NoteCC}Require that $U(\zeta^2)$, after transformation to a framework based on instant-time $x^0$ quantization, yields the Cornell potential for static quark-quark systems~\cite{Trawinski:2014msa}.
    \item Apply the de~Alfaro-Fubini-Furlan procedure (dAFF) \cite{deAlfaro:1976vlx, Fubini:1984hf, Akulov:1983hjq}, which allows for a scale in a Lagrangian while its corresponding Action remains conformally invariant.
    \item Use the RGE flow equation \cite{Gao:2022ojh}.
\end{enumerate}
Note~\eqref{NoteCC} can intuitively be understood from the fact that the $\zeta^2$-dependence of $U(\zeta^2)$ is independent of the space dimension. In the classical picture, applicable to HLFQCD, the force from a static source is given by the field flux crossing a small boundary element, \textit{e.g}., a surface in 3-dimensional space. 
That the $\zeta^2$-dependence of $U(\zeta^2)$ is independent of the space dimension
implies that the flux must be that of a field propagating freely in the minimal number of dimensions, \textit{i.e}., one dimension. 
Otherwise, if the flux were that of a field propagating freely in $n$-dimensions, then one could have the same flux arrangement, \textit{i.e}., the same $\zeta^2$-dependence in $d>n$ dimensions, but not for $d<n$. 
Since, a force propagating in one dimension is constant -- the flux has nowhere to spread, the instant-time potential is linear.

All methods indicated above lead to the result that $U(\zeta^2)$ has the form of a harmonic oscillator \cite{Brodsky:2013ar}:
\begin{equation} 
U(\zeta^2)= \kappa^4 \zeta^2 +b, 
\label{eq: HLFQCD potential}
\end{equation} 
where $\kappa$ depends only on AdS space dimension and $b$ is determined 
by the spin $J$ representations in AdS space. For a 2-body system in AdS$_{5}$, $b=2\kappa(J-1)$. 
With $m_q=0$ and Eq.~(\ref{eq: HLFQCD potential}), 
a wide range of phenomena are reproduced and predicted, including the following.
\begin{enumerate}[(i)]
    \item Nonperturbative running of $\alpha_s(Q^2)$ with the prediction of freezing in the IR.\label{HLFfreeze}
    \item Regge trajectory descriptions of hadron spectra \cite{deTeramond:2014asa, Brodsky:2016rvj}. 
    (Within HLFQCD, this property might be linked with the IR-freezing of $\alpha_s(Q^2)$ \cite{Gao:2022ojh}.)
    \item A connection between the soft and hard pomerons \cite{Dosch:2022mop}.
    \item Prediction of a symmetry between the masses of baryons, mesons, and tetraquarks with universal Regge trajectories \cite{deTeramond:2014asa, Brodsky:2016rvj}.
    \item Predictions for hadron form factors \cite{Brodsky:2014yha, Sufian:2016hwn},  PDFs \cite{deTeramond:2018ecg, deTeramond:2021lxc, Chang:2020kjj}, and GPDs \cite{Liu:2019vsn}.
\end{enumerate}
In regard to prediction~\eqref{HLFfreeze}, one may ask how freezing happens since a classical instantaneous potential is employed, as for $\alpha_{\rm Rich}$ in Sec.\,\ref{EffectivePotential}. 
The answer may simply be that, like $\alpha_{\rm Rich}$, the maximum wavelength argument \cite{Brodsky:2008be, Brodsky:2012ku, Gao:2017uox} does not apply; but this does not preclude a freezing of $\alpha_s$. 
In fact, the reason for the coupling's freezing in HLFQCD is plain: 
it is a conformal theory except in the vicinity of its scale $\kappa$ and, as a semi-classical theory,  other scales cannot be generated by quantum anomalies. 
Conformality manifests itself for the $Q^2\ll \kappa^2$ (and $Q^2\gg \kappa^2$) as the freezing of coupling.
So, one may argue that although maximum wavelength argument does not apply explicitly, HLFQCD expresses enough of QCD to ensure that $\alpha^{HLF}_s$ freezes, effectively realizing the wavelength cutoff. 
The manner by which this is physically realized is a complex question that involves multi-gluon dynamics. 
There is {\it a priori} no contradiction with the fact that, in the non-relativistic case, HLFQCD reproduces the linear potential included in $\alpha_{\rm Rich}$, because this necessitates the introduction of massive quarks that break the conformality of QCD and therefore directly void the cause of $\alpha^{HLF}_s$ freezing. 
If so, such violation of the freezing would be a quark mass effect, which can be excluded in the definition of the running coupling in order to preserve a universal behavior independent of a quark specific mass. 

Some of the other predictions  above require several parameters in addition to $\kappa$,%
\footnote{
In the natural unit system $\hbar=c=1$, one free parameter is the minimum required for a model or theory to describe the strong force since absolute measurements are expressed in the conventional, human-chosen, unit, \textit{e.g}., GeV. 
For pQCD, that parameter is $\Lambda_s$, while it is $\kappa$ for HLFQCD. 
The relation between $\Lambda_s$ and $\kappa$ \cite{Deur:2014qfa} is given in Sec.\,\ref{AdS/CFT determinations}.} 
introduced to:
describe higher Fock states in the LF wavefunction solutions of the Schr\"{o}dinger-like equation;
parameterize SU(3)-flavor symmetry breaking;
and characterize a universal function \cite{deTeramond:2018ecg, deTeramond:2021lxc, Chang:2020kjj} from which the PDFs and GPDs are derived. 
Yet, HLFQCD describes a great deal of strong force phenomenology with remarkably few 
parameters. 
In fact, only $\kappa$ is needed to predict $\alpha_s(Q^2)$, as will be shown next.

\subsection{HLFQCD computation of \mbox{$\boldmath \alpha_{g_1}(Q^2)$}}
\label{alpha_g1 from HLFQCD}
The measured $\alpha_{g_1}$ -- Fig.\,\ref{fig:alpha_g1} -- is nearly constant in the deep IR, \textit{i.e}., QCD is approximately conformal on $Q^2/M_N^2 \simeq 0$. 
This can also be deduced by using the relation between the Bjorken sum $\Gamma_1^{p-n}(Q^2)$ and the generalized GDH sum rule, as detailed below Eq.\,\eqref{eq:GDH on alpha_g_1}.
%
%
Such near conformality encourages the use of HLFQCD to derive $\alpha_{g_1}(Q^2)$ \cite{Brodsky:2010ur}. 

The five dimensional AdS$_5$ Action has the same form as that of general relativity: 
\begin{equation} 
S_{AdS_5} = -\frac{1}{4} \int \sqrt{g}\frac{1}{\hat{a}^2_5}F^2~d^5x. 
\end{equation} 
Here, $F$ is the gauge field, $\hat{a}_5$ its self-coupling, and $g=\det(g_{\mu \nu})$, with $g_{\mu \nu}$, the AdS space metric, yielding the invariant interval:
\begin{equation} 
ds^2 = \frac{R^2}{z^2} \big(\eta_{\mu \nu} dx^\mu dx^\nu -dz^2 \big)\,,
\label{Eqds2}
\end{equation} 
where $R$ is the AdS radius and $\eta_{\mu \nu}$ is the Minkowski metric. 
The fifth (holographic) dimension, $z$, is associated with the scale at which the hadron is probed, \textit{i.e}., it corresponds to $1/Q$. 

$S_{AdS_5}$ generates a CFT, devoid of any scale.  
To introduce 
a characterizing scale, a factor $e^{+\kappa^2 z^2}$ can be used to distort the geometry of AdS space, thereby breaking the conformal invariance of the theory so that Eq.\,\eqref{Eqds2} becomes
\begin{equation} 
ds^2 = \frac{R^2}{z^2} e^{\kappa^2 z^2}\big(\eta_{\mu \nu} dx^\mu dx^\nu -dz^2 \big).
\label{eq:HLFQCD interval}
\end{equation} 
Now $\kappa$ provides the theory's scale, like $\Lambda_s$ does for QCD. 
It is worth remarking here that only an exponential term $e^{\pm \kappa^2 z^2}$ can yield the harmonic oscillator LF potential in Eq.\,\eqref{eq: HLFQCD potential} \cite{deTeramond:2008ht, Brodsky:2006uqa}.
The positive exponent is selected by requiring a consistent AdS-LFQCD mapping \cite{Brodsky:2014yha}.
Indeed, the derivation of the spin term in the LF potential depends directly upon the positive sign of the profile.
Moreover, as will become apparent in connection with Eq.\,\eqref{alpha_s from HLFQCD}, the positive sign is also necessary to ensure the expected decrease of $\alpha_s$ with $Q^2$: whilst one should not be prejudiced about the behavior of $\alpha_s(Q^2)$ at low $Q^2$, consistency between pQCD and HLFQCD on the intermediate $Q^2$ domain requires that $\alpha_s^{HLF}(Q^2)$ decrease with $Q^2$.
%

In short, $e^{+\kappa^2 z^2}$ is the potential dual in Minkowski space, retains the conformal symmetry of the action, and yields the uniquely suited harmonic oscillator LF potential. 
The action in the distorted AdS geometry becomes:
\begin{equation} 
S_{HLF}=  -\frac{1}{4} \int \sqrt{g}\frac{1}{{\hat{a}_5}^2}F^2 ~e^{\kappa^2 z^2}d^5x.
\label{eq:AdS distorted action}
\end{equation} 
In the chiral limit theory discussed here, $\kappa$ is universal and can thus be extracted from hadron masses, \textit{e.g}.\,\cite{Brodsky:2014yha}, $\kappa=M_{\rm p}/2$ or $\kappa=M_\rho / \sqrt{2}$, where $M_\rho$ is the $\rho$-meson mass;
or connected with $\Lambda_s$ \cite{Deur:2016opc};
or from some dimension associated with hadron form factors \cite{Brodsky:2014yha, Sufian:2016hwn}, 
etc.
The $\kappa$ values determined by these means are consistent and combining them yields 
$\kappa= 0.523\pm0.024$ GeV \cite{Brodsky:2016yod}.

Recall now that effective charges generalize the concept of running coupling by including, in addition to the short distance quantum loops, short distance gluon radiative effects (higher order DGLAP corrections), long distance parton distribution correlations (HT corrections), and EHM contributions.
Analogously, in the AdS coupling of Eq.\,\eqref{eq:AdS distorted action}, one can incorporate 
the $e^{+\kappa^2 z^2}$ term that is dual to the HLFQCD confinement potential: ${a_5}(z^2) \equiv {\hat{a}_5} e^{-\kappa^2 z^2/2}$. 
Transforming ${a_5}(z^2)$ to Minkowski 4-momentum space yields $\alpha^{\rm HLF}_s(Q^2) = \alpha^{\rm HLF}_s(0) e^{-\frac{Q^2}{4\kappa^2}}$, with $\alpha^{\rm HLF}_s(0)$ an undetermined parameter.  
It may be fixed by a scheme choice, \textit{e.g}., requiring agreement with Eq.\eqref{alphag10} imposes the $g_1$ scheme~\cite{Brodsky:1994eh}, in which case
\begin{equation} 
\alpha^{\rm HLF}_{g_1}(Q^2) = \pi {\rm e}^{-Q^2/[4\kappa^2]}.
\label{alpha_s from HLFQCD}
\end{equation} 
The value of $\alpha^{\rm HLF}_s(0)$ encompasses the observable-dependence of an effective charge -- see Sec.\,\ref{EffectiveCharge},  analogous to the RS-dependence of $\alpha^{\rm pQCD}_s$.
The pointwise behavior of the prediction in Eq.\,\eqref{alpha_s from HLFQCD} agrees well with low $Q^2$ data, see Fig.~\ref{fig:alpha_g1}, despite having no adjustable parameters after the scheme choice is imposed.

\section{Continuum Schwinger Function Methods}
\label{SecCSMs}
Implicit in many aspects of the preceding discussion of effective charges and running couplings is an appreciation of the role that can be played by calculating and understanding QCD's Schwinger functions -- loosely, Euclidean-space Green functions \cite{GJ81, Dedushenko:2022zwd}. 
 These are the primary quantities computed in LGT; and, of course, results and insights have also long been forthcoming from an array of continuum methods.  
 The key point is that in any rigorously well-defined Poincar\'e-invariant quantum theory, knowledge of all its Schwinger functions is necessary and sufficient to complete a solution of the theory.
 Naturally, this does not mean it is the only way to solve the problem, but it is one rigorous approach. 
 As discussed above, LF QCD offers another.

One of the merits of CSMs is that some of the simpler Schwinger functions calculated are also accessible to LGT, \textit{e.g}., two-point functions (propagators) are generally accessible and so are certain projections of three-point functions.  
This opens the door to useful synergies, with combined results from both approaches leading to greater insights than are achievable with either alone. 

\subsection{Defining \mbox{$\boldmath \alpha_s$} via three-point functions}
\label{CSMThreePoint}
One of the most direct routes to $\alpha_s$ using CSMs is paved by identities such as that in Eq.\,\eqref{eq:zg}.  An example of their use is provided by the ghost-gluon vertex, which has long been a focus because, when evaluated in Landau gauge at zero incoming ghost momentum, the renormalization constant of the gluon-ghost vertex is unity \cite{Taylor:1971ff}: $\tilde Z_1 \equiv 1$.  Capitalizing on this, then one can define a strong running coupling from the ghost-gluon vertex as follows:
\begin{equation}
\alpha_s^{\rm gh}\left(Q^2\right)= \alpha_s^{\rm gh} \left(\mu\right)G^2
\left(Q^2,\mu\right)Z\left(Q^2,\mu\right),
\label{eq:alpha_s DSE ghost--gluon}
\end{equation}
where $\mu$ is the renormalization scale.  Here, $G(Q^2,\mu)$ and $Z(Q^2,\mu)$ are the ghost and (transverse) gluon propagator dressing functions, respectively, which are obtained from calculated ghost and gluon two-point functions:
\begin{equation} 
\delta^{bc} G(Q^2,\mu) = Q^2 D^{\rm G}_{bc}(Q^2,\mu),
\quad 
\delta^{bc}Z(Q^2,\mu) :=
- \tfrac{1}{3} Q^2 [\delta_{\mu\nu}-Q_{\mu}Q_{\nu}/Q^2] D^{bc}_{\mu\nu}(Q^2,\mu),
\label{eq:ghost, gluon props}
\end{equation}
where $b$ and $c$ are color indices.  The effective charge in Eq.\,\eqref{eq:alpha_s DSE ghost--gluon} is known as the \textit{Taylor coupling} \cite{Boucaud:2011ug, Huber:2018ned}; and, as discussed in Sec.\,\ref{alphaCSM}, it has been used to obtain a precise value of $\alpha_s(M_Z^2)$.

Other vertices, combined with a choice of kinematics, may also be used to define a strong running coupling, \textit{viz}.\  
the 3-gluon vertex -- $\alpha_s^{\rm 3g}$, 4-gluon vertex -- $\alpha_s^{\rm 4g}$, and gluon-quark vertex -- $\alpha_s^{\rm gq}$.
There also exist 2-ghost--2-gluon and 4-ghost vertices; but they cannot be used to define a strong coupling because there are no graphs involving solely a bare vertex -- see, \textit{e.g}., Ref.\,\cite{Huber:2018ned}). 
This exhausts the set of vertex-based definitions of $\alpha_s$. 

A natural extension of the Gell-Mann--Low effective charge to QCD would work with the gluon vacuum polarization.  However, it is not possible to do this directly owing to the non-Abelian character of QCD, expressed in the STIs.  Notwithstanding that, a simple attempt is discussed in connection with Eq.\,\eqref{eq:alpha_s cornwall} below.  Furthermore, as explained in Sec.\,\ref{DSE:alpha_PI}, one can improve upon that by combining the pinch technique \cite{Cornwall:1981zr, Cornwall:1989gv, Pilaftsis:1996fh, Binosi:2009qm, Cornwall:2010upa} and background field method \cite{Abbott:1980hw, Abbott:1981ke} to arrive at a unique QCD analogue of the QED running coupling, which is drawn in Fig.\,\ref{fig:alpha_g1}.

Regarding definition via the 3-gluon vertex, a symmetric kinematic configuration is often chosen \cite{Pascual:1980yu}; namely, with each momentum, $k_{1,2,3}$, understood to be entering the vertex, then $k_{1,2,3}^2=-Q^2$, $2 k_i\cdot k_j=Q^2$, $i,j=1,2,3$.  In this case,  
\begin{equation}
\alpha_s^{\rm 3g}\left(Q^2\right)=
\alpha_s^{\rm 3g}\left(\mu\right)
\frac{\Gamma^{\rm 3g}(Q^2,\mu)^2 Z^{3}(Q^2,\mu)}
{\Gamma^{\rm 3g}(\mu^2,\mu)^2 Z^{3}(\mu^2,\mu)},
\label{eq:alpha_s 3g vertex}
\end{equation}
where 
$\Gamma^{\rm 3g}$ is the function that dresses the leading term in the $3$-gluon vertex, \textit{i.e}., the term which exists even at tree-level.  In many renormalization schemes, the denominator in Eq.\,\eqref{eq:alpha_s 3g vertex} is unity.

If the 4-gluon vertex is chosen, then there are many possible choices of kinematics. With choice ``$C$'' in Ref.\,\cite{Cyrol:2014kca}, 
\begin{equation}
\alpha_s^{\rm 4g}\left(Q^2\right) = 
\alpha_s^{\rm 4g}\left(\mu\right)
\frac{\Gamma^{\rm 4g}(Q^2,\mu) Z^2(Q^2,\mu)}
{\Gamma^{\rm 4g}(\mu^2,\mu) Z^2(\mu^2,\mu)},
\label{eq:alpha_s 4g vertex}
\end{equation}
where $\Gamma^{\rm 4g}$ is the analogue of $\Gamma^{\rm 3g}$ in Eq.\,\eqref{eq:alpha_s 3g vertex}.

Using the gluon-quark vertex, working with the so-called ``1-2-3'' or ``totally asymmetric'' kinematic configuration $Q^2=k_1^2=k_2^2/2=k_3^2/3$ \cite{Alkofer:2008tt}, one has
%
\begin{equation}
\alpha_s^{\rm gq}\left(Q^2\right) = \alpha_s^{\rm gq}\left( \mu \right) 
\frac{Z(Q^2,\mu) [\Gamma^{\rm gq}\left( Q^2,\mu \right)Z_f(Q^2,\mu)]^2} 
{Z(\mu^2,\mu) [\Gamma^{\rm gq}\left( \mu^2,\mu \right) Z_f(\mu^2,\mu)]^2}
\label{eq:alpha_s q-g vertex}
\end{equation}
%
where $\Gamma^{\rm gq}$, $Z_f$ are, respectively, the analogues of $\Gamma^{\rm 3g}$, $Z$ for the gluon-quark vertex and quark two-point function.



At UV momenta and using any RS that is independent of current-quark mass, all vertex choices yield the same running coupling owing to STIs like those in Eq.\,\eqref{Eq:alternate_Zs}.  
Specifically, the STIs ensure that the couplings calculated using different vertex choices are equal at the renormalization point.  Then, since pQCD evolution is characterized by a single scale, $\Lambda_s$, the running of those couplings also coincides.   
On the other hand, progressing into the IR, distinct vertices are solutions of different integral equations with vertex-specific integrands, and each offers an infinite variety of possible momentum-flow assignments.  So, in concrete CSM calculations, which typically employ MOM-like RSs, practitioner-dependent choices obscure the effect of STIs and yield inequivalent vertex couplings.
%
%
Furthermore, the value of $\alpha_s^{\rm gq}$ can explicitly depend on the current-quark mass, \textit{e.g}., Ref.\,\cite{Williams:2014iea} produced results for $\alpha_s^{\rm gq}$ that are strongly dependent on quark current-mass -- see Fig.\,11 therein, and Ref.\,\cite[Fig.\,17]{Fu:2019hdw} displays values of $\alpha_s^{\rm qg}$ that are smaller for the $s$ quark than for the lighter quarks.  The dependence on quark current-mass diminishes with increasing $Q^2$.

Comparing Eq.\,\eqref{eq:alpha_s DSE ghost--gluon} with Eqs.\,\eqref{eq:alpha_s 3g vertex} -- \eqref{eq:alpha_s q-g vertex}, it becomes clear that the Taylor coupling presents the simplest choice if one elects to define a running coupling from a vertex.  The simplicity is, in fact, even greater because the number of diagrams that must be computed is also smallest in this case.


Color-carrying correlation (Schwinger) functions are not directly observable.  
Further, they need not individually be renormalization group invariant or independent of the gauge parameter.  
In most CSM analyses, the (Poincar\'e-covariant) Landau gauge is used because it is itself a fixed point of the renormalization group and its transverse-projector character assists in reducing the number of independent tensor structures that contribute to a given result. 

For instance, in general, the 3-gluon vertex involves 14 independent tensors, but 10 of them are purely longitudinal; so, in Landau gauge, it is sufficient to retain only the 4 remaining (transverse) structures \cite{Williams:2014iea}. 
Moreover, the longitudinal component of the gluon 2-point function is not dynamical and decouples from the transverse piece.  
Finally, $\tilde Z_1 \equiv 1$ in Landau gauge \cite{Taylor:1971ff}, in consequence of which $\alpha_s^{\rm gh}$ in Eq.\,\eqref{eq:alpha_s DSE ghost--gluon} only involves two-point functions. 
Others choices have also been explored, including axial, Coulomb, Feynman, light-cone,
maximal Abelian, and general linear covariant gauges. 
Some analyses of the gauge-dependence of $\alpha_s^{\rm gh}$ claim a weak sensitivity for gauges interpolating between the Landau and Coulomb gauges, with a variation of less than 30\% of the IR-fixed point around the Landau gauge result \cite{Deur:2016tte, Siringo:2018uho}.
However, the quantitative reliability of such estimates is difficult to judge because the results can depend on practitioner-dependent choices, made in formulating the Schwinger function computations, whose magnitude cannot readily be determined. 

\subsection{Complementary continuum formulations}
There are two commonly used approaches to the continuum calculation of Schwinger functions: Dyson-Schwinger equations (DSEs) and the functional renormalization group (FRG).  Hereafter we provide a brief sketch of both schemes.

\subsubsection{Dyson-Schwinger equations}
The DSEs can be interpreted as the quantum equations-of-motion for the theory under consideration \cite{Dyson:1949ha, Schwinger:1951ex, Schwinger:1951hq, Roberts:1994dr, Alkofer:2000wg, Maris:2003vk, Fischer:2006ub}.  
Typically formulated in Euclidean metric -- for reasons explained elsewhere \cite[Sec.\,1]{Ding:2022ows}, their solutions are the theory's Schwinger functions: with all such Schwinger functions known, then the theory itself is solved.
A Poincar\'e-covariant formulation is always possible in a Poincar\'e-invariant quantum gauge field theory. 

The DSEs form an infinite tower of coupled nonlinear integral equations, with the equation for a given $n$-point Schwinger function coupled to kindred equations for higher-$n$-point functions: in QCD, this coupling can reach to $n+2$. 
The DSE tower includes, \textit{inter alia}, the gap equations for gluons and quarks (and ghosts), the Bethe-Salpeter equations for meson bound states, and Faddeev-like equations for baryon bound states.

The basic strength of this continuum approach is that it allows for an essentially nonperturbative treatment of the theory.  
No diagrammatic expansion of the Schwinger functions need be assumed to exist because the DSEs can be derived directly from the theory's generating functional. 
In principle, the DSEs and their solutions deliver an exact representation of the complete information content of the QFT under consideration.  

In practice, of course, the complexity of quantum gauge field theory requires that ``sacrifices''  be made in order to define a tractable problem: the tower is truncated.
A carelessly implemented truncation will typically lead to violations of the theory's principal symmetries.
This was a problem in the last millennium.
However, during the past vicennium, material progress has been made with the development of systematic, symmetry-preserving truncation schemes \cite{Munczek:1994zz, Bender:1996bb, Chang:2009zb, Chang:2011ei, Binosi:2014aea, Binosi:2016xxu, Binosi:2016wcx, Williams:2015cvx, Binosi:2016rxz, Qin:2020jig, Xu:2022kng}.
Thus, today, in an array of cases, it is even possible to identify and estimate truncation-dependent uncertainties; then, the choice of truncation specifies a level of approximation. 
The resulting integral equations are typically solved numerically, something which is straightforward with contemporaneous computer resources.

A typical, viable truncation scheme will ensure, amongst other things,  
unitarity; 
multiplicative renormalizability; 
the Euclidean space equivalent of Poincar\'e-covariance; 
preservation of Ward-Green-Takahashi identities; and 
that perturbative results are preserved -- especially, so far as we are concerned herein, the UV behavior of the running coupling.
Exploiting these features, one ensures that the basic qualities of the underlying theory are preserved, \textit{e.g}., the number and properties of dressing functions (form factors) for each $n$-point function.
In QCD, this latter means that 
for quarks, there are two dynamical form factors; 
gluons, one, \textit{viz}.\ the vacuum polarization; 
ghost fields, one; 
and a limited, specified number of dynamical form factors for each vertex.  
These dressing functions are the outputs of the DSE solution procedure; and, as stressed above, the inputs to any DSE calculation of a running coupling. 

Another merit of the DSEs is that since they deliver QCD Schwinger functions, which are also the subject of LGT computations, then synergies between these two nonperturbative approaches can be identified and exploited. Insofar as $\alpha_s$ is concerned, relevant examples are provided by the following Schwinger functions \cite{Boucaud:2018xup, Aguilar:2021lke, Aguilar:2021okw, Aguilar:2022thg}: 
ghost and gluon propagators; and ghost-gluon and 3-gluon vertices.
DSE solutions for these Schwinger functions can be used to isolate and understand the body of artefacts deriving from lattice regularization in LGT analyses and, vice versa, LGT results can be used to expose and reduce the effects of DSE truncations.

\subsubsection{Functional Renormalisation Group}
\label{FRG}
Complementing the DSE approach, the FRG provides another in-principle exact scheme for computing a theory's Schwinger functions. 
Its origins may be traced to the momentum shell renormalization group method developed in Refs.\,\cite{Wilson:1971bg, Wilson:1971dh, Wilson:1973jj}. 
By progressively integrating the fast (large momentum) components of a field, the method sums the field's quantum fluctuations into high-momentum Wilson shells -- see Ref.\,\cite{Fradkin:2021} for an elucidation of the momentum shell renormalization group and Refs.\,\cite{Pawlowski:2005xe, Gies:2006wv, Rosten:2010vm, Braun:2011pp, Dupuis:2020fhh} for the FRG.

The FRG begins by setting an IR cutoff, $k$, at a boundary above which the theory is truly perturbative.  
The value of $k$ is then progressively lowered toward and into the nonperturbative domain.  
Thus, $k$ serves as a regulator term, added to the classical action, which effectively generates a momentum-dependent mass for the associated field.
That mass vanishes in the UV.%
\footnote{
That the quantum field (the gluon field in the context of this discussion) becomes effectively massive is not surprising. 
Indeed, the purpose of some of the counter terms that are added to the classical bare action during the renormalization procedure is to ensure that the symmetries characterizing the classical action are not broken, \textit{e.g}., by a momentum cut-off. 
Consequently, a mass term may be necessary to counter the artifacts associated with such a renormalization procedure.}
The effect of the running mass is to decouple the slow/IR components of the field -- those of momenta lower than $k$, thereby suppressing quantum fluctuations below the scale $k$.

Working with the thus regularized generating functional for one-particle irreducible Schwinger functions, $\Gamma_k[\phi]$, where $\phi$ is here some generic field, and differentiating with-respect-to $k$, one arrives at the basic flow equation \cite{Pawlowski:2005xe, Gies:2006wv, Rosten:2010vm, Braun:2011pp, Dupuis:2020fhh}:
\begin{equation} \label{FRGE}
\frac{\partial}{\partial k} \Gamma_k[\phi] =   \tfrac{1}{2} {\rm Tr} \,  \frac{\frac{\partial }{\partial k}R_k}{\Gamma_k^{(2)}[\phi] + R_k},
\end{equation}
where $R_k$ is the regulator function used to suppress the (IR) modes below $k$, and $\Gamma_k^{(2)}$ is 
the functional second derivative of the action, \textit{i.e}., the inverse of the $\phi$-propagator after its modification by $R_k$. 
The trace, $\rm Tr$, sums over internal indices, momenta and fields. 

Eq.\,\eqref{FRGE} describes the $k$-dependence of the theory, {\it viz},\ of all the theory's Schwinger functions.
It provides an {\it exact} flow equation for $\Gamma_k[\phi]$ with a one-loop structure, 
in contrast to the multi-loop structure of generic perturbation theory.
A system of flow equations for $n$-point Schwinger functions is then constructed by successively differentiating Eq.\,\eqref{FRGE} with respect to the field or source terms in the action.

Taking the $k\to 0$ limit makes the cut-off disappear into the far IR, integrating out the slow modes, {\it viz}.\ the QFT's IR quantum fluctuations.
It results in an analytic, nonperturbative method that allows one to study QFTs in a strong-coupling regime.
In particular, running couplings can be studied because they may be defined via correlation functions -- see Sec.\,\ref{CSMThreePoint}. 
By construction, the FRG ensures the RG evolution expected from the QFT.
This was verified for $\alpha_{s}(Q^{2})$ in the pure Yang--Mills case in the UV regime \cite{Gies:2002af}.
In practice, the infinite system of flow equations must be truncated.
As with the DSEs, this introduces uncertainties whose magnitude can be difficult to assess.
Additionally, the choice of regulator is not unique and may be an extra source of arbitrariness. 

Similarities between the DSE and FRG schemes are now plain.  
Their principal foci are a theory's Schwinger functions: 
the DSEs obtain them from a theory's quantum equations-of-motion, 
whereas the FRG works with the flow equation for the effective action.
Each scheme is identified with an infinite tower of coupled equations; and supposing a sound approach is employed to make the problem tractable, they deliver the complete solution of the QFT.
In principle, the DSE and FRG approaches must yield the same results for each Schwinger function.
In practice, different truncation schemes can lead to mismatches that are artifacts of the choices made.
Nevertheless, adding the FRG to the DSE+LGT mix, then one has three distinct nonperturbative tools whose results can be used to mutually check and improve the results of any one. 

\subsection{CSM results for vertex-based couplings.}
\label{vertexcouplingsnumerical}
Once results for the relevant Schwinger functions are in hand, then the couplings described in Sec.\,\ref{CSMThreePoint} can be computed.  This is an active field and numerous studies exist \cite{Huber:2018ned, Roberts:2021nhw, Binosi:2022djx, Papavassiliou:2022wrb, Ding:2022ows, Roberts:2022rxm, Ferreira:2023fva}.  Herein, we exploit those related to vertex-based running couplings from the past quinquennium.  Earlier contributions are discussed elsewhere \cite{Deur:2016tte}.

\subsubsection{IR behavior of Schwinger functions}
%

Following Refs.\,\cite{vonSmekal:1997ohs, vonSmekal:1997ern, Lerche:2002ep, Fischer:2002eq, Fischer:2002hna, Alkofer:2000wg, Zwanziger:2002ia, Pawlowski:2003hq}, it was considered plausible by some groups that the ghost dressing function diverged as an irrational power of $Q^2$ at IR momenta, \textit{viz}.\ $G(Q^2,\mu)\propto 1/Q^{2\kappa}$, $0<\kappa \in \mathbb P$; the gluon dressing function vanished as $Q^{4\kappa}$; and the leading dressing function of the 3-gluon vertex diverged as $1/Q^{6\kappa}$.  This position came to be called the ``scaling scenario'' and is discussed at length in Ref.\,\cite{Deur:2016tte}.


Considering existing contrary evidence from both continuum and LGT analyses \cite{Cornwall:1981zr, Mandula:1987rh}, subsequent studies, involving distinct collaborations from diverse locations, carefully reexamined the scaling possibility -- see, \textit{e.g}., Refs.\,\cite{Boucaud:2006if, Boucaud:2008ji, Boucaud:2008ky, Aguilar:2008fh, Aguilar:2009pp, Cucchieri:2008mv, Bogolubsky:2009dc, Boucaud:2011ug}.  Those continuum and LGT analyses arrived at a different set of conclusions, \textit{viz}.\ 
the ghost dressing function is IR finite and, 
qualitatively matching the original work \cite{Cornwall:1981zr, Mandula:1987rh}, the transverse gluon propagator function is nonzero at $Q^2=0$, increases a little to a maximum in the far infrared, then decreases monotonically toward zero with increasing $Q^2$.  (See, \textit{e.g}., Ref.\,\cite[Secs.\,7, 9]{Ferreira:2023fva}.)
This is the ``decoupling scenario'', so named because IR massive gluons decouple from interactions on $Q^2\lesssim \Lambda_s^2$.
These conclusions have been confirmed in more recent analyses \cite{Williams:2015cvx, Cyrol:2016tym, Binosi:2019ecz, Fischer:2020xnb, Falcao:2020vyr, Boito:2022rad}.  
Furthermore, the leading 3-gluon vertex dressing function has been found to exhibit a zero crossing to negative values at $Q^2 \approx 1\,$GeV$^2$ and to diverge logarithmically on $Q^2\simeq 0$ \cite{Eichmann:2014xya, Athenodorou:2016oyh, Papavassiliou:2022umz, Pinto-Gomez:2022brg}.

Evidently, there is tension between these two representations of gauge sector dynamics.  It is therefore worth sketching recent developments in understanding QCD's gauge sector made using the the pinch-technique + background field method (PTBFM) \cite{Cornwall:1981zr, Cornwall:1989gv, Pilaftsis:1996fh, Binosi:2009qm, Cornwall:2010upa, Abbott:1980hw, Abbott:1981ke}. 
The PTBFM provides a continuum approach to the analysis of QCD that manifestly ensures the preservation of key symmetries.  For instance, it provides for analyses of QCD's gauge sector that are entirely free of quadratic divergences, which spoil renormalizability, by virtue of the ``seagull cancellation''.%
\footnote{In this context, seagull diagrams are all those contributions to a given Schwinger function that are either explicitly or effectively quadratically divergent and the seagull cancellation is that sequence of operations which guarantees that such divergences have no impact on the final result.}    
This crucial cancellation hinges on the use of a symmetry-preserving regularization scheme in which, \textit{inter alia}, translational invariance is preserved.  The precise combination of integrals, which conspires to secure the seagull cancellation, is achieved by ensuring that the theory's fully-dressed vertices satisfy the STIs.  Importantly, the seagull identity guarantees that, in the absence of a Schwinger mechanism \cite{Schwinger:1962tn, Schwinger:1962tp}, the gluon propagator remains strictly massless \cite{Aguilar:2016vin}, \textit{i.e}., no gluon mass can be generated.

It is worth highlighting that although the PTBFM makes the action of the seagull cancellation manifest, the identity also operates in other approaches to analyzing QCD, albeit in a less transparent manner.  To capitalize on the seagull cancellation in such other approaches, one must simply ensure that the chosen scheme preserves key symmetries. 

The Schwinger mechanism is unambiguously identified with the appearance of a simple pole at $Q^2=0$ in the vacuum polarization scalar of a theory's gauge boson(s).  (For this discussion, consider the gluon propagator to be written in terms of a denominator with the form $Q^2 [ 1 + \Pi(Q^2)]$, where $\Pi(Q^2)$ is the polarization.) The connection between the emergence of a (Schwinger mechanism) simple pole in $\Pi(Q^2)$ and the seagull cancellation is direct.  In fact, as established in Ref.\,\cite{Aguilar:2021uwa}, the emergence of a Schwinger mechanism leads to a displacement (modification) of the relevant STI, in which case the seagull cancellation can be evaded and the generation of a running gluon mass becomes possible.  

Pursuing these features further, Ref.\,\cite{Aguilar:2022thg} used results for elementary Schwinger functions obtained via the numerical simulation of lattice-regularized QCD to demonstrate that a dynamically-generated, massless, colour-carrying, scalar gluon + gluon correlation emerges as a feature of the dressed three-gluon vertex.  This is a necessary and sufficient condition for the existence of a Schwinger mechanism in QCD; and the consequent emergent gluon mass scale is finite because the function that characterizes the STI displacement falls sufficiently rapidly with increasing momentum, \textit{i.e}., in the UV.  This outcome expresses a unique decoupling solution to the problem of QCD gauge sector dynamics. 

Notably, the displacement function is an identifying feature of QCD.  It is known quantitatively -- see Fig.\,8 in Ref.\,\cite{Aguilar:2022thg}.  However, it is not recovered in approaches associated with the scaling solution. 

Notwithstanding these developments, hereafter we will typically include both decoupling and scaling scenario results for the purpose of illustrating the sensitivity of vertex couplings to the character of Schwinger functions.

\subsubsection{Discussion of vertex-based couplings}
It has been known for fifty years that the Landau gauge ghost-gluon vertex function is finite \cite{Taylor:1971ff}. 
Hence, whether an IR extension of the QCD running coupling can meaningfully be defined via Eq.\,\eqref{eq:alpha_s DSE ghost--gluon} depends on the IR behavior of the gluon 2-point function.  
This question is settled in the affirmative by the modern studies which have demonstrated that $Z(Q^2,\mu)/Q^2$ is nonzero at $Q^2=0$ and decreases monotonically with increasing $Q^2$ from a maximum value at far IR momenta.%
\footnote{The effect of light quarks is not completely settled, but existing analyses indicate that they do not alter this result \cite{Binosi:2016xxu}.}

As first argued in Ref.\,\cite{Cornwall:1981zr}, the underlying cause of the decoupling behavior of $Z(Q^2,\mu)$ and, therefore, the nonzero IR-finite value of $\alpha_s^{\rm gh}$, is dynamical generation of a running gluon mass, $m_g(Q^2)$, which regularizes the gluon propagator at IR momenta whilst preserving the STIs.  
This mass function eliminates the Landau pole and forces gluons to decouple from all interaction processes at length-scales greater than $1/m_g(0)$.
As explained in a more general context elsewhere \cite{Brodsky:2008be}, the coupling therefore loses its 
$Q^2$-dependence, \textit{i.e}., it ``freezes''. 
A running gluon mass is also expressed in other analysis methods, as discussed below; 
albeit, the interpretation of the gluon mass-scale is sometimes different. 

%
The source of the gluon running-mass has long been debated.  
Naturally, it may depend on the framework used to analyze strong-field QCD.  
It was originally supposed \cite{Cornwall:1981zr} to connect with formation of a gluon condensate, and that possibility is still explored \cite{Horak:2022aqx}. 
However, given the elusiveness of the gluon condensate \cite[Sec.\,4.5]{Cloet:2013jya}, 
it is more likely to be the result of another, better established phenomenon, \textit{viz}.\ the Schwinger mechanism of gauge boson mass generation \cite{Schwinger:1962tn, Schwinger:1962tp}.
As remarked above, recent efforts \cite{Aguilar:2021uwa, Aguilar:2022thg}, combining continuum and LGT results, have provided strong support for this position by demonstrating with confidence 
that a dynamically-generated, massless, colour-carrying, scalar gluon+gluon correlation emerges as a feature of the dressed 3-gluon vertex. 
This massless pole triggers the Schwinger mechanism in QCD. 
Alternative explanations typically rely on a realization of the scaling scenario. 
In any event, should a source for the gluon mass scale be proposed that is unconnected with the Schwinger mechanism, then, without artifice, that scheme must also simultaneously explain and reproduce the appearance of the massless pole in the 3-gluon vertex, which, as shown in Ref.\,\cite{Aguilar:2022thg}, is a feature of QCD. 

In Ref.\,\cite{Cornwall:1981zr}, Feynman diagrams are resummed using the pinch technique, which was actually developed for this calculation in order to ensure that a gauge-independent result was obtained for the gluon dressing function.  
It was then found that the numerical solution of the gap equation is well described by the following simple analytic form, which preserves QCD's one-loop renormalization group behavior:
\begin{equation}
\alpha_s(\mu^2) \frac{Z(Q^2,\mu)}{Q^2} = 
\frac{4\pi}{\beta_0 \ln\bigg[ \big(Q^2 + 4 m_g^2(Q^2) \big)/\Lambda_s^2  \bigg]}\, \frac{1}{Q^2+m_g(Q^2)}\,,
\label{eq:alpha_s cornwallA}
\end{equation}
where the gauge-independent running gluon mass is 
\begin{equation}
m_{g}^2\left(Q^2\right)=\frac{m_0^2}{\left[\mbox{ln}\left(\frac{Q^2+4 m_0^2}{\Lambda_s^2}\right)/\mbox{ln}\left(\frac{4 m_0^2}{\Lambda_s^2}\right)\right]^{12/11}},
\label{eq:gluon mass Cornwall}
\end{equation}
with $m_0=0.5\pm0.2\,$GeV.  It is worth recalling that whilst adding a mass term directly to the QCD Lagrangian would violate gauge invariance, this mass contributes only to the transverse part of the gluon 2-point function; hence, is consistent with the STIs and therefore, gauge invariance. 

The running mass in Eq.\,\eqref{eq:gluon mass Cornwall} vanishes on $\Lambda_s^2/Q^2\simeq 0$, ensuring compatibility with pQCD.  Moreover, Eq.\,\eqref{eq:alpha_s cornwallA} suggests a straightforward continuation of the one-loop running coupling into the IR:
\begin{equation}
\alpha_s^{\Pi}(Q^2) = 
\frac{4\pi}{\beta_0 \ln \bigg[ \big(Q^2 + 4 m_g^2(Q^2)/\Lambda_s^2 \big) \bigg] }\,,
\label{eq:alpha_s cornwall}
\end{equation}
where the superscript $\Pi$ indicates that the coupling behavior stems from the gluon vacuum polarisation.
This coupling freezes at a maximum value in the IR and decreases monotonically with increasing $Q^2$.

The result in Eq.\,\eqref{eq:alpha_s cornwall} is gauge-independent, having been obtained using the PT; but details depend on the 1-loop analysis used in Ref.\,\cite{Cornwall:1981zr} and the form chosen in Eq.\,\eqref{eq:alpha_s cornwall}. 
Different values of $m_0$ may be obtained if the pinch technique is not used and a different gauge is chosen and/or another RS, going beyond 1-loop, were to be employed.  
Nevertheless, using the quoted central value of $m_0$ and $\Lambda_s=0.63$~GeV, so that $\alpha_s^{\Pi}$ matches $\alpha_s^{\rm pQCD}$ in the UV, one finds $\alpha_s^\Pi(Q^2=0)/\pi\approx 0.4$.
%
%
This is compatible with the value of the IR fixed point from HLFQCD (Sec.\,\ref{HLFQCD}) expressed in $\overline{\rm MS}$: 
$\alpha^{\mathrm{HLF}}_{\overline{\rm MS}}(0)/\pi = 0.388 
\pm 0.047$ \cite{Deur:2016cxb}.

Eqs.\,\eqref{eq:alpha_s cornwallA}-\eqref{eq:gluon mass Cornwall} have been much discussed in the literature -- see, e.g., Ref.\,\cite{Deur:2016tte} for a review of the many studies; developments or alterations of their forms; and also the generic constraints on $m_g(Q^2)$, its proposed values, and $Q^2$-dependences. 
Typically, determinations agree with the picture drawn in Ref.\,\cite{Cornwall:1981zr}.

Although Refs.\,\cite{Cornwall:1975aq, Cornwall:1975ty, Cornwall:1976ii, Cornwall:1979hz, Cornwall:1981zr} laid a foundation for discussions of $m_g(Q^2)$ and $\alpha_s^{\rm IR}$, delivering predictions that have subsequently been refined but not truly replaced \cite{Roberts:2021nhw, Binosi:2022djx, Papavassiliou:2022wrb, Ding:2022ows, Roberts:2022rxm, Ferreira:2023fva}, they are peculiar in relying entirely on 1-loop expressions for the gluon 2-point function and running coupling, and using a ghost-free gauge.  As noted above, many studies have instead focused on vertex-based definitions of $\alpha_s^{\rm IR}$; and, generally, the couplings thus defined are inequivalent: on $ Q^2 \lesssim m_N^2$, $\alpha_s^{\Pi} \neq \alpha_s^{\rm gh} \neq \alpha_s^{\rm qg} \neq \alpha_s^{\rm 3g} \neq \alpha_s^{\rm 4g}$.

\begin{figure}[!t]
\includegraphics[width=0.46\textwidth, height=0.5\textwidth]{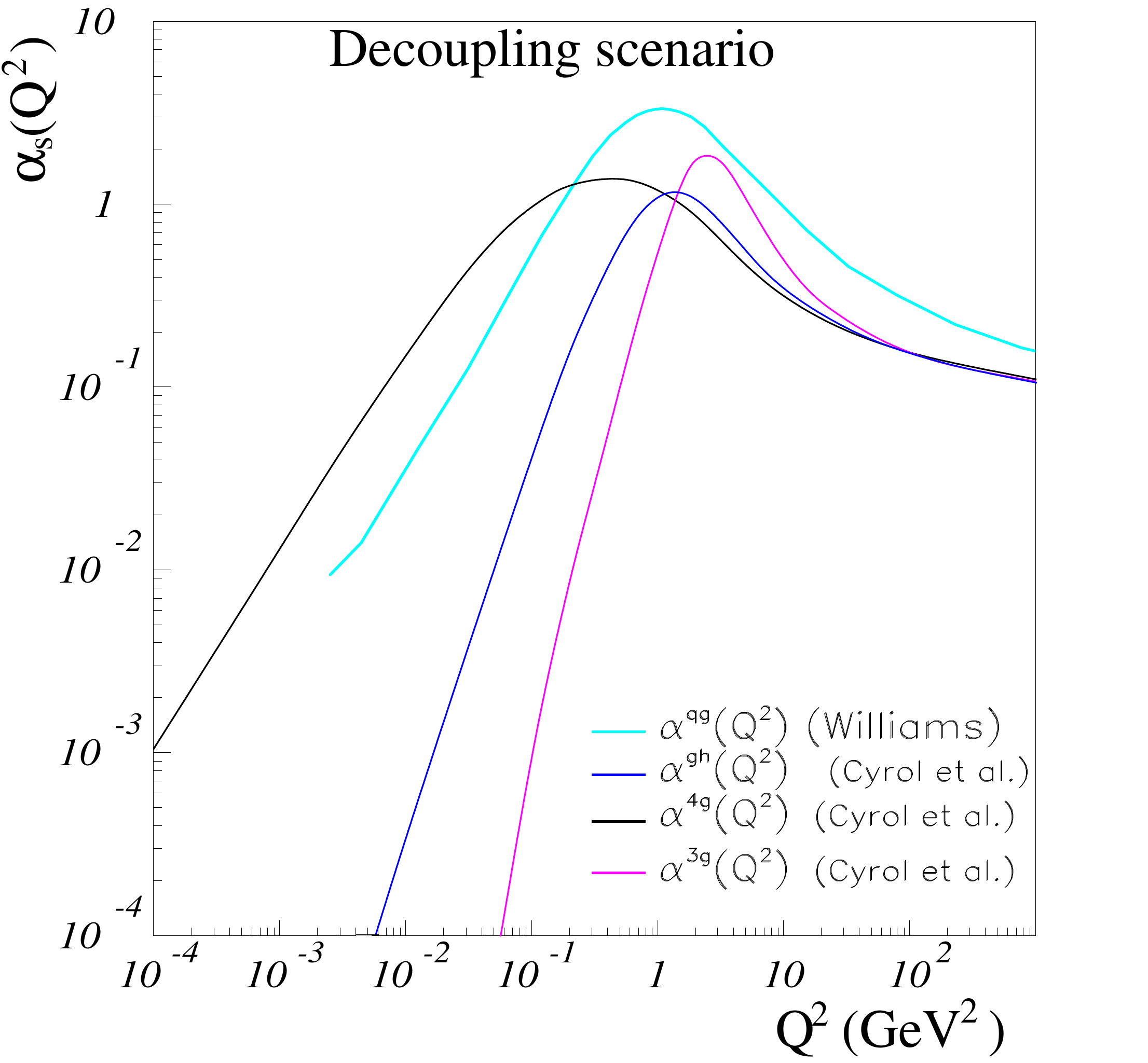}
\includegraphics[width=0.53\textwidth, height=0.5\textwidth]{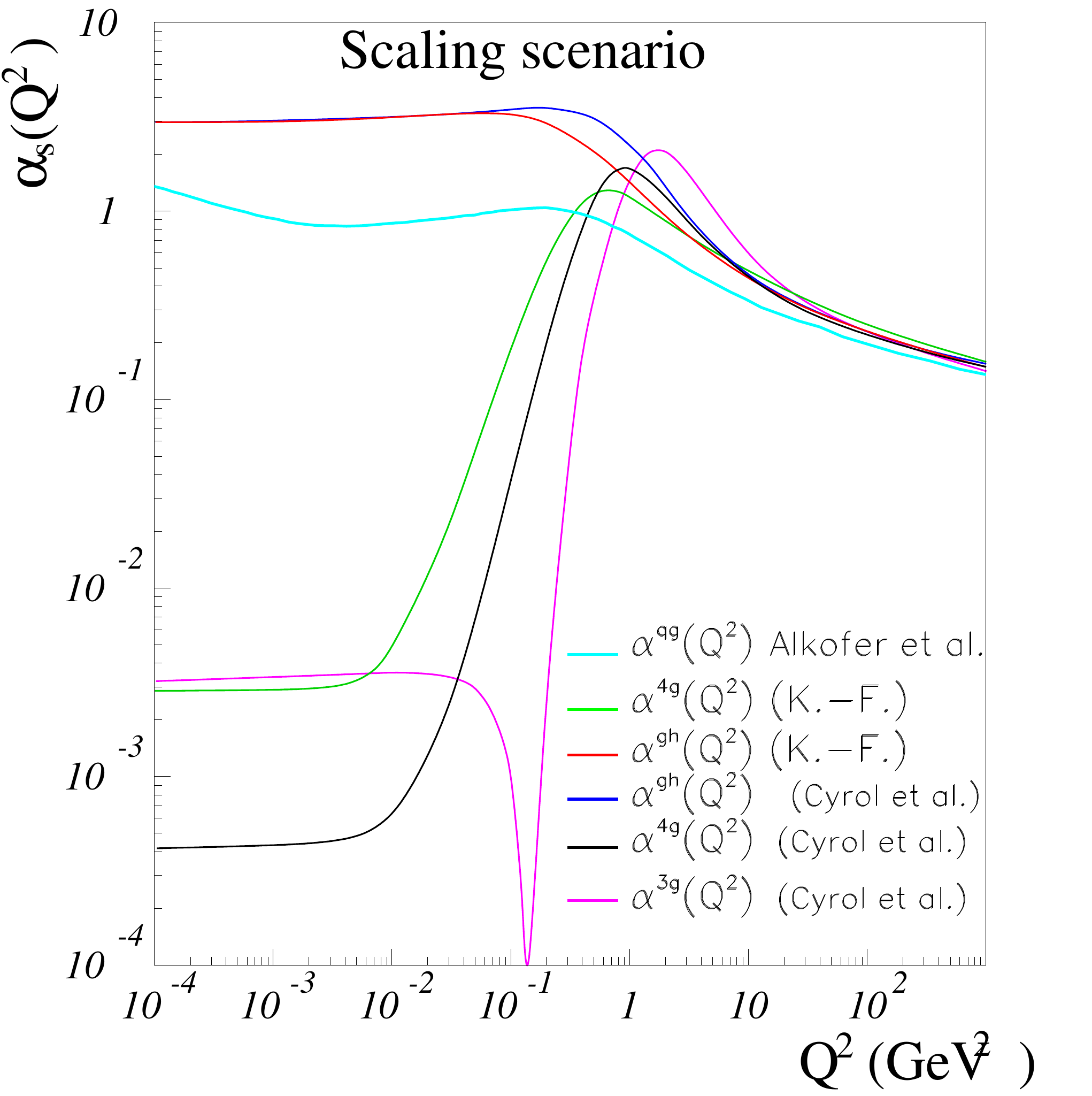}
\caption{  \small \label{Flo:lowq alpha. DSE} 
Representative DSE results for vertex-based couplings.
$\alpha^{\rm gh}_{s}$: blue curve from Ref.\,\cite{Cyrol:2014kca} and red curve from Ref.\,\cite{Kellermann:2008iw}).
$\alpha^{\rm qg}_{s}$: cyan curves -- Ref.\,\cite{Williams:2014iea}, decoupling solution; Ref.\,\cite{Alkofer:2008tt}, scaling.
(Quark degrees-of-freedom are only included in this coupling.  Ref.\,\cite{Williams:2014iea} -- $m_q=30\,$MeV, with results also provided up to $m_q=3.6\,$GeV; and Ref.\,\cite{Alkofer:2008tt} -- chiral limit.)
$\alpha^{\rm 3g}_{s}$: magenta curves \cite{Cyrol:2014kca}.
$\alpha^{\rm 4g}_{s}$: black curves \cite{Cyrol:2014kca}; and green curve (right panel) \cite{Kellermann:2008iw}).
Calculations are performed in Landau gauge and using the MOM RS.
The decoupling scenario -- left panel -- is consistent with the emergence of a gluon mass via the Schwinger mechanism \cite{Boucaud:2006if, Boucaud:2008ji, Boucaud:2008ky, Aguilar:2008fh, Aguilar:2009pp, Cucchieri:2008mv, Bogolubsky:2009dc, Boucaud:2011ug, Williams:2015cvx, Cyrol:2016tym, Binosi:2019ecz, Fischer:2020xnb, Falcao:2020vyr, Boito:2022rad}.
The $Q^2$-location of the maxima of the various running couplings depends on the particular value $G(0,\mu^2)$ at which the ghost dressing function saturates. 
The vanishing of $\alpha_s^{\rm 3g}$ at a nonzero $Q^2$ value owes to the IR zero-crossing of $\Gamma^{\rm 3g}(Q^2)$, the 3-gluon vertex dressing function \cite{Eichmann:2014xya, Athenodorou:2016oyh, Duarte:2016ieu, Boucaud:2017obn, Papavassiliou:2022umz, Pinto-Gomez:2022brg}.
This is an unavoidable consequence of the emergence of a gluon mass, which entails that IR gluon modes are screened and ghost fields dominate gauge sector dynamics at IR momenta: finiteness of the gluon DSE requires a zero-crossing in some of the dressing functions.
IR fixed points in vertex-based couplings are characteristic of the scaling scenario.
}
\end{figure}

As displayed in Fig.\,\ref{Flo:lowq alpha. DSE}, DSE results for the ghost--gluon coupling express $\alpha_s^{\rm gh}(0)=0$ in the decoupling scenario.  On the other hand, $\alpha_s^{\rm gh}(0)\neq 0$ in the scaling case.
Several calculations of the quark--gluon coupling $\alpha_s^{\rm qg}$ are also depicted: qualitatively, the results are similar to those for $\alpha_s^{\rm gh}$.
Notably, $\alpha_s^{\rm qg} > \alpha_s^{\rm 3g}$ on the entire IR domain.  
A more recent calculation, building upon LGT results for the gluon propagator ($n_f=2+1$, one-loop, Landau gauge, MOM RS), yields a similar coupling \cite{Gao:2021wun}.
Owing to DCSB, the impact of dynamical quark loops is small, so long as the number of light quarks does not exceed some critical value \cite{Fischer:2003rp, Fischer:2005en, Fischer:2007ze, Binosi:2016xxu}.  

Calculations of $\alpha_s^{\rm 3g}(Q^2)$ and $\alpha_s^{\rm 4g}(Q^2)$ are also drawn in Fig.\,\ref{Flo:lowq alpha. DSE}.  
At far IR momenta, these couplings are markedly suppressed when compared with $\alpha_s^{{\rm gh},{\rm qg}}$.
On the other hand, with increasing $Q^2$, they reach a maximum value $\alpha_s^{\rm 3g,4g} \sim 1$ \cite{Deur:2016tte, Campagnari:2010wc, Aguilar:2013vaa, Pelaez:2013cpa, Blum:2014gna, Binosi:2014kka, Cyrol:2014kca, Eichmann:2014xya, Mitter:2014wpa, Cyrol:2016tym, Williams:2015cvx, Athenodorou:2016oyh, Duarte:2016ieu, Boucaud:2017obn, Aguilar:2019uob}
before falling away to reproduce pQCD behavior.

Actually, it is evident in Fig.\,\ref{Flo:lowq alpha. DSE} that all vertex-based couplings exhibit a (local) maximum at some nonzero value of $Q^2$.
In order for CSM and HLFQCD to be consistent, we judge that, in all cases, the locations of these maxima must lie below the perturbative-nonperturbative transition scale determined using HLFQCD, \textit{viz}.\ $Q_0^2=1.32$~GeV$^2$ in the MOM RS and using Landau gauge \cite{Deur:2016cxb}.
This condition is satisfied (albeit, only just for $\alpha_s^{\rm 3g}$), irrespective of the particular CSM calculation considered \cite{Campagnari:2010wc, Pelaez:2013cpa, Blum:2014gna, Eichmann:2014xya, Williams:2015cvx, Cyrol:2014kca, Mitter:2014wpa, Cyrol:2016tym, Duarte:2016ieu, Athenodorou:2016oyh, Boucaud:2017obn, Aguilar:2021okw}.

The IR suppression of $\alpha_s^{\rm 3g}(Q^2)$ and $\alpha_s^{\rm 4g}(Q^2)$ in MOM-like RSs occurs because, unlike gluons, no dynamical mass emerges for ghost fields.
Consequently, in the calculation of these couplings, ghost loop contributions dominate in the IR, delivering logarithmic IR divergences.  
These ghost contributions to the leading three- and four-gluon Schwinger functions are negative, whereas the UV dominant gluon contributions are positive.  
Thus, a clear signal for entry into a ghost dominance domain is a sign-change of the gluon vertex functions.
This sign-change has been verified explicitly in Landau gauge studies of QCD's pure-gauge sector 
\cite{Campagnari:2010wc, Aguilar:2013vaa, Binosi:2013rba, Pelaez:2013cpa, Blum:2014gna, Eichmann:2014xya, Binosi:2014kka, Williams:2015cvx, Cyrol:2014kca, Mitter:2014wpa, Cyrol:2016tym, Duarte:2016ieu, Athenodorou:2016oyh, Boucaud:2017obn, Huber:2020keu}: $\Gamma^{\rm 3g}(Q^2)$ crosses zero around $Q^2\sim 0.05$~GeV$^2$. 
It leads to $\alpha_s^{\rm 3g}=0$ at that point, as can be seen from Fig.\,\ref{Flo:lowq alpha. DSE}. 

As also found for $\alpha_s^{\rm qg}$, light quark loops do not seem to qualitatively alter this picture, 
only modifying the location of the zero-crossing. 
Those analyses that have provided information on this issue indicate that the zero survives, but is shifted somewhat lower in $Q^2$ \cite{Williams:2015cvx, Aguilar:2019uob, Figueroa:2021sjm}.
Efforts to elucidate the full impact of dynamical quark loops are continuing \cite{Binosi:2016xxu, Cyrol:2017ewj, Fu:2019hdw, Gao:2021wun, Biddle:2022acd}.


In Landau gauge, with two flavors ($n_f=2$) of physical-mass light quarks, the FRG has also been used \cite{Braun:2014ata} to compute the Schwinger functions needed to arrive at results for 
$\alpha_s^{\rm gh}$, $\alpha_s^{\rm qg}$, $\alpha_s^{\rm 3g} $ and $\alpha_s^{\rm 4g}$, Eqs.\,\eqref{eq:alpha_s DSE ghost--gluon}--\eqref{eq:alpha_s 4g vertex}.
The couplings were subsequently calculated, albeit, assuming $\alpha_s^{\rm 3g} =\alpha_s^{\rm 4g}$.
In this analysis, the input ghost and gluon propagators were taken from earlier pure-gauge DSE/FRG calculations and the transition from parton to hadron degrees-of-freedom was modeled using a dynamical hadronization technique that involves the addition of a scale-dependent effective potential to the effective action \cite{Gies:2001nw}.
The study delivers IR-vanishing couplings, with $\alpha_s^{\rm gh} > \alpha_s^{\rm qg} \gg \alpha_s^{\rm 3g}$ in the deep IR, but $\alpha_s^{\rm qg} > \alpha_s^{\rm gh} > \alpha_s^{\rm 3g}$ on the higher momentum part of the IR domain, in agreement with DSE, LGT and Curci-Ferrari results. 
The couplings vanish in the IR owing to the generation of a gluon mass scale, which leads to increasing IR suppression as the number of gluon legs grows.  
However, the result $\alpha_s^{\rm qg} > \alpha_s^{\rm gh}$ at moderate IR momenta was identified as a truncation artefact \cite{Braun:2014ata}, in the absence of which one might expect $\alpha_s^{\rm gh} > \alpha_s^{\rm qg}$ everywhere.

An expanded body of FRG analyses may be found in Refs.\,\cite{Mitter:2014wpa, Cyrol:2016tym, Cyrol:2017ewj}: the former does not involve dynamical quarks and the latter works with $n_f=2$.
Once again, all vertex-based couplings are computed: $\alpha_s^{\rm gh}$, $\alpha_s^{\rm 3g}$, $\alpha_s^{\rm 4g}$  and $\alpha_s^{\rm qg}$.
The results generally agree qualitatively and (semi-)quantitatively with those obtained using the DSEs, drawn in Fig.\,\ref{Flo:lowq alpha. DSE}, except that these FRG studies do not find a zero in $\Gamma^{\rm 3g}(Q^2)$. 
The results for the ghost and gluon propagators from Ref.\,\cite{Mitter:2014wpa} are used in the FRG study of Ref.\,\cite{Horak:2022aqx} in an attempt to develop an interpretation of the gluon mass scale
as the result of gluon condensation {\it via} a Higgs-like mechanism. 
The calculation of the gluon effective mass, performed using the MOM RS, agrees with LGT results. 

With all vertex-based couplings vanishing in the far IR, one arrives at a picture of long-wavelength QCD as a noninteracting theory.  
Were this behavior really a statement about a running coupling that describes QCD's effective charge at all accessible momentum scales \cite{Dokshitzer:1998nz}, then one would be faced with many conundrums. 
For instance, DCSB -- the origin of constituent-like quark masses -- may even be absent, because that phenomenon requires a sufficiently large integrated interaction strength on $Q^2\lesssim M_N^2$ \cite{Nambu:1961tp}.
A resolution of these puzzles is discussed in Sec.\,\ref{DSE:alpha_PI}, where a connection between $\alpha_s^{\rm gh}$ and QCD's Gell-Mann--Low effective charge is explained.
The result is a freezing coupling whose behavior is (practically) indistinguishable from that obtained using HLFQCD. 

\subsection{Process-independent, effective charge}
\label{DSE:alpha_PI}
It will readily be recalled that the archetypal running coupling is that computed in quantum electrodynamics (QED) \cite{GellMann:1954fq}.
It is now known with great accuracy \cite{Workman:2022ynf} and the running has directly been observed \cite{VENUS:1998drt, Mele:2006ji}.  
Significantly, this Gell-Mann--Low effective charge is a renormalization group invariant and process-independent (PI) running coupling.
It is obtained simply by computing the photon vacuum polarisation.  
Such a definition is straightforward because ghost fields decouple in Abelian theories; hence, one has the Ward identity \cite{Ward:1950xp, Takahashi:1957xn}, in consequence of which the electric-charge renormalization constant is equivalent to that of the photon field.  
Physically, this means that the impact of dressing the interaction vertices is entirely absorbed into the vacuum polarisation.

In QCD, that is not usually true because ghost fields do not generally decouple from interactions, so the renormalization constants associated with the running coupling and the gluon vacuum polarisation are different.%
\footnote{
It is worth recalling that QCD can be formulated in one of the few ghost-free gauges \cite{KONETSCHNY1980263}.  In that case, however, the number of independent tensors required to fully express any given $n$-point function is considerably multiplied; so, in practice, one arrives at a problem of at least equal complexity.}
This fact underlies the introduction and analysis of the vertex-based definitions of $\alpha_s$ described above.  
The problems with such definitions are now plain, however. 
Whilst they are equivalent in the UV, there is a huge array of distinct vertex-based couplings whose IR momentum-dependence depends on the choices of vertex and kinematic configuration.
This presents the fundamental question: 
With $\alpha_s^{\rm gh} \neq \alpha_s^{\rm qg} \neq \alpha_s^{\rm 3g} \neq \alpha_s^{\rm 4g}$ and the behavior of any such coupling potentially depending on the arrangement of momenta flowing into the vertex, can any coupling of this type be identified as the single $\alpha_s$ that describes QCD's interaction strength at all momentum scales \cite{Dokshitzer:1998nz}?
Furthermore, there are also issues with 
the IR suppression (zero crossing) exhibited by these couplings, which makes DCSB problematic,
and the probable dependence on gauge choice.
These problems are resolved by a notable advance that begins with ideas developed using QCD's DSEs and then exploits results from LGT.

The key point is that there is an approach to analyzing QCD's Schwinger functions that preserves some of QED's simplicity; namely, the combination of BFM \cite{Abbott:1980hw, Abbott:1981ke} and PT \cite{Cornwall:1981zr, Cornwall:1989gv, Pilaftsis:1996fh, Binosi:2009qm}.  
This approach acts to make QCD ``look'' somewhat Abelian by enabling a systematic rearrangement of classes of diagrams and their sums in order to arrive at modified Schwinger functions that satisfy linear STIs \cite{Taylor:1971ff, Slavnov:1972fg}.

In the gauge sector, working with Landau gauge, this BFM+PT approach yields a modified gluon dressing function, from which one can compute a unique QCD running coupling \cite{Binosi:2016nme, Cui:2019dwv}, denoted $\hat\alpha_s$ herein, {\it i.e.}, the associated polarization captures all required features of the renormalization group -- just as is the case in QED \cite{GellMann:1954fq}.
One may express the result concretely as follows: 
\begin{equation}
\label{alphahatequation}
\hat\alpha_s(Q^2) {\mathpzc D}(Q^2) = \alpha_0 
\left[\frac{G(Q^2;\mu^2)/G(0;\mu^2)}{1-L(Q^2;\mu^2)G(Q^2;\mu^2)}\right]^2
{\mathsf D}(Q^2)
\,,
\end{equation}
where the hitherto undefined quantities are
the RGI product ${\mathsf D}(Q^2)= [Z(Q^2,\mu)/Q^2] \times$ $m_g^2(Q^2,\mu) / m_0^2$, with $m_g(Q^2,\mu)$  the running gluon mass, $m_0$ a RGI gluon mass scale, and ${\mathsf D}(0)=1/m_0^2$; 
${\mathpzc D}(Q^2)$ is a RGI function, which behaves as the free propagator for a boson with mass $m_0$ in both the far-IR and -UV and satisfies ${\mathsf D}(Q^2)={\mathpzc D}(Q^2)$ on $Q^2\lesssim \mu^2$;
and $L(Q^2;\mu^2)$ is that longitudinal part of the gluon-ghost vacuum polarisation which vanishes at $Q^2=0$ \cite{Aguilar:2009nf}.
Evidently, $\alpha_0=\hat\alpha_s(0)$ is RGI.
Thus defined, $\hat\alpha_s$ is process independent; namely, the same result is obtained independent of the scattering process considered, whether gluon+gluon$\,\to\,$gluon+gluon, quark+quark$\,\to\,$quark+quark, {\it etc}.
Furthermore, and importantly, it depends only on the momentum flowing through the vacuum polarization; so, is free of the kinematic momentum-flow ambiguities attached to vertex-based definitions of a running coupling -- see the discussion of $\alpha_s^{\rm gh}$, $\alpha_s^{\rm 3g}$, $\alpha_s^{\rm 4g}$, or $\alpha_s^{\rm gq}$ in Sec.\,\ref{CSMThreePoint}.

The clean connection between coupling and gluon vacuum polarization is facilitated by another particular feature of QCD, already mentioned often above, \textit{viz}.\ in Landau gauge the renormalization constant of the gluon-ghost vertex is unity \cite{Taylor:1971ff}, 
Consequently, the effective charge obtained from the modified gluon vacuum polarisation is directly connected with $\alpha_s^{gh}$, the coupling deduced from the gluon-ghost vertex \cite{Sternbeck:2007br, Boucaud:2008gn, Aguilar:2009nf}, and usually called the ``Taylor coupling,'' $\alpha_{\rm T}$ \cite{Blossier:2011tf, Blossier:2012ef}.

Studying Eq.\,\eqref{alphahatequation} and reviewing the character of $L$, $G$, one recognizes that it is the inclusion of an infinite series of ghost-gluon scattering contributions which transmogrifies the standard gluon vacuum polarization into a form suitable for defining a Gell-Mann--Low-like effective charge for QCD. 
Importantly, each of the Schwinger functions in Eq.\,\eqref{alphahatequation} can be computed using both
continuum and LGT methods.

It is worth remarking here that $\hat{\alpha}_s$ is PI and RGI in any gauge; but it is sufficient to know the result in Landau gauge, which is a convenient calculational choice for both continuum and LGT studies.  This is because $\hat{\alpha}_s$ is form invariant under gauge transformations \cite{Binosi:2016nme, Cui:2019dwv} and gauge covariance ensures that any such transformations can be absorbed into the Schwinger functions of the quasiparticles whose interactions are described by $\hat{\alpha}_s$ \cite{Aslam:2015nia}.

To date, the most refined calculation of $\hat\alpha_s$ is described in Ref.\,\cite{Cui:2019dwv}, which combined contemporary results from continuum analyses of QCD's gauge sector and LGT configurations generated with three domain-wall fermions at the physical pion mass \cite{Blum:2014tka, Boyle:2015exm, Boyle:2017jwu} to obtain a parameter-free prediction of $\hat\alpha_s(k^2)$, characterized by 
\begin{equation}
\alpha_0/\pi=0.97(4)\,, \quad m_0 = 0.43(1)\,{\rm GeV}.  
\end{equation}
It is now plain that this coupling saturates to a large, nonzero, finite value at $Q^2=0$, with the dynamical generation of a running gluon mass having eliminated the Landau pole.  
Notably, the gluon mass scale falls within the window estimated in Ref.\,\cite{Cornwall:1981zr} -- see the discussion of Eq.\,\eqref{eq:alpha_s cornwallA}.
The resulting charge is drawn as the magenta curve and associated uncertainty band in Fig.\,\ref{fig:alpha_g1}.
It has been shown to deliver a unified description of many observables \cite{Roberts:2021nhw, Ding:2022ows}.

The results drawn in Fig.\,\ref{fig:alpha_g1} reveal a remarkable fact: 
\begin{equation}
\hat\alpha_s(Q^2) \approx \alpha_{g_1}^{\rm HLF}(Q^2) \approx \alpha_{g_1}(Q^2)\,.  
\end{equation}
There are small quantitative differences between these charges; but to all practical intents and purposes, they are equivalent.  
Crucially, within mutual uncertainties, they all freeze to the same IR value.
These effectively identical couplings therefore provide the best available candidate for that running coupling/effective charge which describes QCD's interaction strength at all momentum scales \cite{Dokshitzer:1998nz}.

One recognizes in Eq.\,\eqref{alphahatequation} the expression of a force, 
{$F$=(coupling$\times$propagator)}.
This strengthens the correspondence between $\hat\alpha(Q^2)$, $\alpha_{g_1}(Q^2)$ and $\alpha^{\rm HLF}_{g_1}(Q^2)$ because the last two are defined in just this way.  
Moreover, it highlights that EHM effects, such as confinement, are included in all three couplings. 

%


\section{Lattice Gauge Theory}
 \label{LGT in IR}
As a nonperturbative method, in which the computation of Schwinger functions is conceptually straightforward, LGT is another approach to studying the behavior of $\alpha_{s}$ in the IR. 
Unsurprisingly, therefore, many LGT calculations of $\alpha_{s}$ use the same formulae and definitions as those employed in CSMs; namely, Eqs.\,\eqref{eq:alpha_s DSE ghost--gluon}-\eqref{eq:alpha_s 4g vertex},  defining $\alpha_s^{\rm gh}$, $\alpha_s^{\rm qg}$, $\alpha_s^{\rm 3g}$,  $\alpha_s^{\rm 4g}$. 
Nevertheless, there are other possibilities.
For instance, the strong running coupling is accessible via Wilson loops, which provide the long distance static quark-quark potential; hence, in turn $\alpha_V$, with recent results available elsewhere \cite{Takaura:2018vcy, Karbstein:2018mzo}.
Wilson-loop calculations of $\alpha_{s}$ are gauge invariant.

The basics of LGT are summarized in Sec.\,\ref{LGT determinations}. 
Since fermion fields are difficult to handle using a lattice regularization,%
\footnote{How to include dynamical quarks in LGT calculations is well studied.  
Today, they are included in many LGT studies. 
However, it prohibitively increases the CPU costs of calculations in which large lattices are necessary, as is the case for any study of $\alpha_s$ in the IR. 
Therefore, when quarks are not crucial, \textit{e.g}., to compute $\alpha_s^{\rm gh}$, $\alpha_s^{\rm 3g,~4g}$ -- in contrast to $\alpha_s^{\rm qg}$, dynamical quark fields are often quenched out of the computation. 
It is expected that, generally, qualitative features revealed by pure gauge studies should hold if quarks were present, but this is not guaranteed \cite{Biddle:2022acd}.}
we will ignore them for now and first discuss $\alpha_{s}$ calculated in the pure gauge sector. 
Just as with CSMs (unless the pinch-technique is used), gauge-dependence affects LGT results if the studies make use of $n$-point Schwinger functions. 
Such calculations are typically performed in the Landau gauge and one of the MOM-type RSs.

For the computation of $\alpha_{s}$ using dressing functions, discretized fields must be defined. 
Such definitions, while not unique, are equivalent up to a proportionality factor \cite{Giusti:1998ur, Giusti:2001xf}. 
On a lattice, the discretized gluon field, $\mathcal{A}$, is given at the lattice site,
$x$, by the difference between a link $U_{\overrightarrow{\mu}}$ and its adjoint: 
\begin{equation}
 \mathcal{A}_{\overrightarrow{\mu}}^{c}(x) \equiv  \frac{1}{2i} \left(U_{\overrightarrow{\mu}}(x)-
U_{\overrightarrow{\mu}}^{\dagger}(x)\right)
 \xrightarrow[a \to 0]{}  2a\sqrt{\pi\alpha_{s}^{\rm bare}}A_{\mu}^{c}(x),
\end{equation}
where $c$ is the color index, $\overrightarrow{\mu}$ the link direction, $a$ the lattice spacing and
$A_{\mu}^{c}$ the physical gluon field. 

The gluon propagator in the functional-integral formalism is the 2-point Schwinger function 
\begin{equation} 
\label{GluonProp}
D_{\mu\nu}^{bc}(x-y)=\int d\mu(A,\bar\psi,\psi) \, A_{\mu}^{b}(x)A_{\nu}^{c}(y)e^{-S}
\equiv\left\langle A_{\mu}^{b}(x)A_{\nu}^{c}(y)\right\rangle,
\end{equation}
where $d\mu(A,\bar\psi,\psi)$ is the integration measure and $S$ is the Euclidean-space QCD action.
On a lattice, Eq.\,\eqref{GluonProp} becomes
\begin{equation}
\mathcal{D}_{\overrightarrow{\mu}\overrightarrow{\nu}}^{bc}\left(x-y\right)=\left\langle \mathcal{A}_{\overrightarrow{\mu}}^{b}\left(x\right)\mathcal{A}_{\overrightarrow{\nu}}^{c}\left(y\right)\right\rangle. 
\end{equation}
Its Fourier transform yields the discretized gluon propagator dressing function, 
$\mathcal{Z}(Q^{2})\equiv\delta^{bc}\delta^{\mu \nu}\mathcal{D}^{bc}_{\mu \nu}(Q^{2})Q^{2}.$
The inverse of the Faddeev--Popov operator $(-\partial+A)\partial$ defines the ghost propagator, $D^{\rm G}_{bc}$, which, after discretization and Fourier-transformation to momentum space, yields the ghost propagator dressing function:
$\mathcal{G}=\delta^{bc}Q^{2}\mathcal{D}_{bc}^{\rm G}\left(Q^{2}\right)$.
Discretized dressing functions for vertices are obtained from the 3- and 4-point correlation functions.
With these dressing functions, the couplings $\alpha_{s}^{gh}$, $\alpha_s^{\rm 3g}$ and  $\alpha_s^{\rm 4g}$ are obtained from Eqs.\,\eqref{eq:alpha_s DSE ghost--gluon}--\eqref{eq:alpha_s 4g vertex}.

As with other methods, LGT calculations are verified by comparing the UV running of the obtained $\alpha_s(Q^2)$, regardless of its definition, with the RGE expectation. 
For LGT however, the discretization limits the UV reach and computation accuracy. 
Likewise, finite size effects limit the IR reach.
This latter issue makes it difficult to verify phenomena expected in the deep-IR, such as the zero-crossing of the 3-gluon dressing function responsible for the minima of $\alpha_s^{\rm 3g}$ around $Q=0.1\,$GeV; nevertheless, as described below, recent analyses point toward confirmation \cite{Athenodorou:2016oyh, Boucaud:2017obn, Aguilar:2021lke}.
Typical lattice physical size used by current LGT studies range from 2-20~fm, with spacing from 0.05-0.2~fm.
As noted above, modern LGT calculations of $\alpha_{s}$ from $n$-point correlation functions deliver results that confirm the decoupling scenario.

Herein, we sketch only those LGT results pertaining to $\alpha_s$ that were obtained in the past quinquennium.  Earlier analyses may be traced from Ref.\,\cite{Deur:2016tte}.
Beside the results themselves, LGT outputs also play a crucial role in checking and improving the Schwinger functions obtained using other methods, such as CSMs.  
Thus, much of LGT's impact is already discussed in those sections, with the references provided therein. 
There is also a the large body of LGT analyses that compute the Schwinger functions alone, without making the step to a result for one of the vertex-based couplings.  
This body includes work on the center vortex approach to confinement, which involves simulations of the gluon propagator and/or the long-distance static quark-quark  potential \cite{Langfeld:2003ev, Bowman:2010zr, OCais:2008kqh, OMalley:2011aa, Trewartha:2015ida, Trewartha:2017ive, Biddle:2018dtc, Biddle:2022acd, Leinweber:2022ukj, Virgili:2022ybm, Biddle:2022zgw}.  
These topics are omitted from our discussion. 

As already discussed, LGT results for Schwinger functions are a crucial component in the parameter-free prediction of the PI charge -- see Sec.\,\ref{DSE:alpha_PI}.
Likewise, in Ref.\,\cite{Boucaud:2017obn}, both LGT and DSEs are used to determine the behavior of the 3-gluon vertex. 
Specifically, Refs.\,\cite{Athenodorou:2016oyh, Boucaud:2017obn} study $SU(3)$-color in Landau gauge with the aim of locating a zero-crossing in the 3-gluon vertex, using a realistic (but pure gauge sector) high-statistics LGT simulation, wherein LGT artifacts are well controlled. 
The zero-crossing is expected to occur at a scale of $\sim 0.01\,$GeV$^2$, which is difficult to reach for LGT simulations.
Nevertheless, a zero-crossing is found $Q^2/{\rm GeV}^2 = 0.026 \pm 0.08(\rm{stat})$. 
The $Q^2$ dependence of $\alpha_s^{\rm 3g}$ and the related vertex dressing function support the DSE conclusion on the origin of the zero-crossing.
Namely, the crossing marks entry into the domain whereupon loop diagrams from massless ghosts, which contribute negatively to the gluon vacuum polarization, come to dominate over the positively contributing gluon loops -- recall that the effective masses of the latter suppress their contribution in the IR. 
The consistency between LGT and DSE for the 3-gluon vertex is also checked in Ref.~\cite{Aguilar:2021lke}.

Another calculation of ({\it inter alia}) $\alpha_s^{\rm 3g}$ \cite{Athenodorou:2016gsa} has been used to test the effect of the ``Wilson flow'' smoothing technique. 
That Landau gauge calculation, using the MOM scheme, with both quenched and $n_f=(2+1+1)$-unquenched field configurations, agrees well with the older LGT calculations of $\alpha_s^{\rm 3g}$. 
An interpretation of the IR behavior of $\alpha_s^{\rm 3g}$ in term of quasi-classical instantons is offered therein.
This has some similarity in spirit to the semi-classical analysis that drives the derivation of $\alpha_{g_1}^{\rm AdS}$ -- see Sec.\,\ref{HLFQCD}.

Possible biases arising from finite volume and finite lattice spacing were studied in Refs.\,\cite{Duarte:2016iko, Boucaud:2017ksi, Duarte:2017wte} for ghost and gluon propagators and for $\alpha_s^{\rm gh}$ (Landau gauge, MOM RS, pure gauge sector). 
Finite lattice spacing effects were found to be small in the UV ($Q^2>1\,$GeV$^2$), and finite volume effects to be small everywhere, at least for the lattice sizes used in current calculations. 
For instance, a three-fold increase in the lattice spacing ($a/{\rm fm}=0.06\to 0.18$) 
increased $\alpha_s^{\rm gh}$ by about 10\% at its maximum, located on $Q^2 \simeq 1\,$GeV$^2$ (Fig.~\ref{Flo:lowq alpha. DSE}), leaving the IR values unchanged;
and decreased the $Q^2=0$ value of the gluon propagator by roughly 10\%.

The origins of the effects seem to be different for the gluon and ghost propagators \cite{Boucaud:2018xup}. 
The issue with the gluon propagator appears to arise from a non-optimal lattice spacing calibration and can thus be fixed by a careful scale-setting procedure. 
That with ghost propagator, on the other hand, seems to follow from the breaking of the O(4) symmetry (the group of rotations in 4 dimensions). 
Extrapolation methods exist to avoid the artifacts arising from such breaking. 
Applying both scale-setting and extrapolation procedures should thus yield an $\alpha_s^{\rm gh}$ largely free of the lattice discretization artifacts \cite{Boucaud:2018xup}. 

This approach was applied to the gluon propagator in Ref.~\cite{Aguilar:2021okw}, which then used it as an input, along with a LGT determination of the 3-gluon vertex, to solve the DSE for the ghost-gluon vertex and obtain the ghost propagator, $\alpha_s^{\rm gh}$ and $\alpha_s^{\rm 3g}$, all essentially free of DSE truncation and LGT artifacts. 
Both couplings vanish in the IR, as expected for LGT results on $\alpha_s^{\rm gh}$ since LGT realizes solely the decoupling scenario.
At any $Q^2$ value, $\alpha_s^{\rm gh} > \alpha_s^{\rm 3g}$.  
The difference vanishes in the UV. 
These results were obtained in the Landau gauge, MOM RS and pure gauge sector.

\section{Refined Gribov--Zwanziger formalism} 
\label{Gribov--Zwanziger}
A somewhat different approach to strong-coupling QCD may be traced to Ref.\,\cite{Gribov:1977wm} and its focus on the manner by which one might properly fix the gauge in QCD.
The issue identified therein relates to the question of how one can nonperturbatively ensure that the gauge fixing procedure selects only unique gauge-inequivalent field configurations from the function space of gauge fields.
(No gauge transformation can connect two gauge inequivalent configurations.)
It is possible, even probable, that standard schemes -- implemented by introducing Faddeev-Popov ghost fields \cite{Faddeev:1967fc} -- fail in this regard by admitting (perhaps infinitely) many equivalent gauge fields that satisfy the gauge fixing constraint; and this happens for every gauge-inequivalent trajectory in the space of gauge fields.  
Along any one trajectory, each of these fields is a member of the set of ``Gribov copies'' of the desired field configuration.
All members of this set are related, one to another, by a gauge transformation; so, the action has the same value for each and they all contribute the same amount to the functional integral.
Moreover, such a set is attached to each inequivalent gauge field trajectory, further compounding the problem.
Proper gauge fixing would select only one member from each such set to be used in computing the QCD functional integral.
Plainly, the possibility of infinitely many gauge-equivalent, hence, action equivalent, Gribov copies remaining after gauge fixing, and infinitely many sets of such copies, forces one to question whether the functional integral has any meaning at all \cite{Williams:2003du}. 

Such gauge copies may contaminate and destabilize the formulation of a wide range of gauges in which ghosts fields play a dynamical role, \textit{e.g}., linear covariant gauges and Coulomb gauge.
They do not affect ghost-free formulations, like Laplacian gauge \cite{Vink:1992ys}, which was introduced for that very purpose, the axial and LC gauges, see below Eq.\,\eqref{eq:gluon loop}, or 
the Weyl gauge, in which the Coulomb potential is set to zero.

Following Ref.\,\cite{Gribov:1977wm}, the issue of Gribov copies and potential solutions were further explored in Refs.\,\cite{Zwanziger:1981kg, Zwanziger:1982na, Zwanziger:1989mf, Zwanziger:1991gz, Zwanziger:1992qr}.
The problems were partially solved.  
An overview of the program is provided in Ref.\,\cite{Vandersickel:2012tz} and subsequent developments sketched in Ref.\,\cite{Deur:2016tte}.

Such analyses are usually performed in Landau gauge, because, {\it e.g}., it is a RGI linear covariant gauge and readily implemented in LGT.
Prominent amongst the schemes is modification of the QCD action, including a ``horizon term'':
\begin{equation}
\label{horizon}
\int d^4 x\, \gamma \, g^2 f^{abc} A_\mu^b(x) [{\mathpzc M}^{-1}]^{ad}(x,x) f^{dec} A_\mu^e(x)\,,
%
\end{equation}
%
%
where 
\begin{equation}
{\mathpzc M}^{ab}(x,y) = 
[- \partial^2 \delta^{ab} + \partial_\mu f^{abc} A_\mu^c(x)]\delta^4(x-y)
\end{equation}
is the Landau gauge Faddeev-Popov operator.  The scale $\gamma$ controls the strength of the horizon constraint and is fixed by the spacetime dimension and number of gluon fields.  
(Issues concerning BRST (a-)symmetry, renormalisability, etc., of such an action are discussed in Refs.\,\cite{Dudal:2010fq, Vandersickel:2012tz}.)

The original program \cite{Gribov:1977wm, Zwanziger:1981kg} relies on an IR divergent ghost dressing function and a gluon propagator that vanishes as $Q^2\to 0$.  
As discussed above, such behavior is not realized in QCD:
the QCD ghost propagator remains that of a standard massless excitation and the gluon is characterised by a dynamically-generated IR mass-scale, ensuring that the gluon propagator is nonzero at $Q^2=0$ \cite{Binosi:2022djx, Papavassiliou:2022wrb, Ferreira:2023fva}.
Thus, the original scheme is not viable.

A modification of the approach in Refs.\,\cite{Gribov:1977wm, Zwanziger:1981kg, Zwanziger:1982na, Zwanziger:1989mf, Zwanziger:1991gz, Zwanziger:1992qr} is canvassed in Refs.\,\cite{Dudal:2007cw, Dudal:2008sp, Dudal:2010tf, Dudal:2011gd, Capri:2015ixa, Capri:2015nzw}, delivering a framework in harmony with modern continuum and LGT Schwinger function results.
(Other procedures are described elsewhere \cite{Gracey:2010cg, Serreau:2012cg, Serreau:2015yna, Reinosa:2015gxn, Reinosa:2014ooa}.)
It admits the possibility that the ghost fields used to fully fix the gauge develop a nonzero dimension-two condensate, whose presence yields a Landau-gauge 2-point Schwinger function for the gluon with the following form \cite{Dudal:2007cw, Dudal:2008sp}:
\begin{equation}
\label{overlineD}
D_{\mu\nu}^{ab}(Q)
= \delta^{bc} \left[\delta_{\mu\nu} - \frac{Q_\mu Q_\nu}{Q^2}\right]
\frac{Q^2 + M^2}{Q^4 + Q^2 \mathpzc{m}^2 + \lambda^4}\,,
\end{equation}
where
$\lambda^4 = 6 \gamma g^2  - {\mathpzc a}^2 M^2$, 
with 
$m_\gamma^4= \gamma g^2 $ expressing the Gribov horizon scale; 
$\mathpzc{m}^2=M^2-{\mathpzc a}^2$,
with ${\mathpzc a}^2\propto \langle A_\mu^a A_\mu^a\rangle$, 
a dimension-two gluon condensate, and $M^2$ related to the ghost-field condensate, both computed within a hadronic medium \cite{Brodsky:2012ku}.  
Using this scheme, which supports a nonzero value for the gluon propagator in the far-IR, one can obtain fair agreement with LGT results for the gluon two-point function \cite{Dudal:2010tf, Cucchieri:2011ig, Dudal:2011gd, Dudal:2012zx}.  
In fact, at IR momenta ($Q^2 \lesssim 1\,$GeV$^2$), a good fit to high-statistics LGT results can be obtained using Eq.\,\eqref{overlineD}.
The refined Gribov-Zwanziger approach has been employed \cite{Mintz:2017qri} to study, at one-loop, the ghost-gluon interaction vertex using a fixed perturbative $\alpha_s$ value (chosen to lie between 0.15 and 0.42). 

Consistency between Eq.\,\eqref{overlineD} and robust LGT results \cite{Bogolubsky:2009dc, Ayala:2012pb} for the Landau-gauge gluon two-point Schwinger function is also explored in Ref.\,\cite{Gao:2017uox}.  That analysis introduced a novel constraint, insisting that the values obtained for the parameters in Eq.\,\eqref{overlineD} should ensure that $D_{\mu\nu}^{ab}(Q)$ expresses a minimal level of consistency with parton-like behaviour on the UV domain.  This condition leads to a constraint relation between the two principal mass scales in Eq.\,\eqref{overlineD}: $m_\gamma$, the Gribov horizon scale; and $m_g=\lambda^2/M$, the IR gluon mass.

In general, there are three possible outcomes, which were all canvassed in Ref.\,\cite{Gao:2017uox}.
\begin{enumerate}[(i)]
\item If fitting LGT results using Eq.\,\eqref{overlineD}, subject to UV partonic constraints, were to yield $m_\gamma \gg m_g$, then the Gribov gauge-fixing horizon is affecting UV modes of the gluon.  
However, the validity of standard perturbation theory shows this is not the case.  
Consequently, $m_\gamma \gg m_g$ is unrealistic: had it been favoured by LGT results, then it would have been necessary to discard either or both those results and the Gribov gauge-field horizon condition.

\item The converse, $m_\gamma \ll m_g$, would indicate that the gluon mass alone is sufficient to screen IR gluon modes, in which case the Gribov gauge-fixing ambiguity could have no physical impact and any horizon term is redundant.

\item The detailed numerical analysis in Ref.\,\cite{Gao:2017uox} indicates that QCD occupies a middle ground: $m_\gamma \approx m_g$, with each of a size ($\sim 0.5\,$GeV) that one associates with EHM in the gauge-sector.  In this scenario, the gluon mass and horizon scale play a practically identical role in screening long-wavelength gluon modes, thereby dynamically eliminating all gauge-fixing ambiguities.  It was further found that, together, they set a confinement scale of roughly 1\,fm.  In consequence of these findings, it was conjectured that the gluon mass and Gribov gauge-field horizon are equivalent emergent phenomena.  

\end{enumerate}

A stronger conclusion requires unquenched LGT results with better sensitivity to IR momenta.
Meanwhile, Ref.\,\cite{Gao:2017uox} worked with the quenched LGT results \cite{Bogolubsky:2009dc} and found that a high-quality fit is achievable with $m_\gamma = m_g$.  
In this realization of Scenario (iii), the horizon scale, ${\mathpzc a}^2$ and ghost-field condensate can all be absorbed into a single running gluon mass, $m_g(k^2)$, whose dynamical appearance alone is sufficient to eliminate gauge-fixing ambiguities and complete the definition of QCD. 
This leads one back to the PI charge discussed in Sec.\,\ref{DSE:alpha_PI}.

\section{Curci-Ferrari model }
\label{Curci-Ferrari model}
The Curci--Ferrari model \cite{Curci:1976bt} is a massive Yang-Mills theory in which the mass force carrier (here the gluon) is an additional parameter of the theory. 
A mass term $\frac{m_g^2}{2} A_\mu^a A^\mu_a $ is added to the QCD Lagrangian after gauge fixing, {\it viz}.\ its Faddeev-Popov expression, Eq.\,\,\eqref{QCD FP Lagrangian}.
This yields a model that contains longitudinal gluon modes, which persist in the UV. 
It has been argued that the model is renormalizable but not unitary \cite{deBoer:1995dh, Forshaw:1998tf}.
The model can be made to recover a Landau-gauge Yang-Mills theory when an appropriate zero mass gluon limit is engineered. 

The rationale for employing the model to explore the IR-behavior of $\alpha_s$ and associated Schwinger functions \cite{Tissier:2010ts}, is the hypothesis that the Faddeev-Popov quantization method becomes inapplicable in the IR and should be modified to account for the presence of Gribov copies.
The gluon mass is then interpreted as an effective mass arising from the breaking of BRST symmetry owing to the Gribov ambiguity \cite{Gribov:1977wm}.
In fact, the gluon propagator obtained in the model is the same as that arising in Landau-gauge pQCD, but for a mass term that suppresses the Landau pole thereby, potentially, allowing the use of perturbation theory. 
Ref.\,\cite{Pelaez:2021tpq} reviews the application of the Curci--Ferrari model to QCD and Yang-Mills theories.
(The motivations for this approach are potentially undermined by the existence of a Schwinger mechanism in QCD \cite{Aguilar:2021uwa, Aguilar:2022thg} and the conclusions suggested in Ref.\,\cite{Gao:2017uox}, \textit{viz}.\ that the emergence of a gluon mass scale makes moot the issue of Gribov ambiguities.)

The scheme has been used to model gluon, quark and ghost propagators, as well as the quark--gluon vertex. 
A one-loop perturbative treatment without dynamical quarks results in a ghost-gluon coupling, $\alpha_s^{\rm gh}$, that vanishes in the IR and remains small enough to validate {\it a posteriori} the use of perturbation theory -- see Ref.\,\cite[Sec.\,4.15]{Deur:2016tte}. 
Although the formalism is perturbative, nonperturbative phenomena are nevertheless modelled by inclusion of the gluon mass parameter. 
Generally, the Schwinger functions obtained agree with CSM and LGT predictions in the  decoupling scenario, although a scaling scenario can be engineered \cite{Reinosa:2017qtf}.

Calculations of $\alpha_s^{\rm qg}$, $\alpha_s^{\rm 3g}$ and $\alpha_s^{\rm 4g}$ have been carried 
out in the Curci-Ferrari model \cite{Pelaez:2017bhh, Gracey:2019xom, Barrios:2020ubx, Barrios:2021cks, Barrios:2022hzr, Figueroa:2021sjm}.
The results agree with those obtained using continuum and lattice Schwinger function methods -- see Fig.\,\ref{Flo:lowq alpha. DSE} (left panel).
However, the agreement is weakened by the inclusion of dynamical quarks.
This might be explained by the observation that although $\alpha_s^{\rm gh}$, $\alpha_s^{\rm 3g}$ and $\alpha_s^{\rm 4g}$ remain small (in Landau gauge) at all $Q^2$ -- see Fig.~\ref{Flo:lowq alpha. DSE}, the quark-gluon coupling $\alpha_s^{\rm qg}$ is found to be two- or even three-times larger in the higher energy part of the IR-domain when compared with $\alpha_s^{\rm gh}$, $\alpha_s^{\rm 3g}$, respectively.
A $\sfrac{1}{N_c}$ expansion has been used to address this issue.
Two-loop results are also available for propagators, and for $\alpha_s^{\rm qg}$ and  $\alpha_s^{\rm 3g}$ \cite{Gracey:2019xom, Barrios:2020ubx, Barrios:2021cks, Barrios:2022hzr}.

There is an appealing simplicity in the fact that fair agreement with the results of continuum and lattice Schwinger function methods in QCD is obtained at 1-loop in the Curci-Ferrari model.  
This is largely a consequence of $\alpha_s^{\rm gh}$, $\alpha_s^{\rm 3g}$ and 
$\alpha_s^{\rm 4g}$ remaining small in the model.
Notably, although somewhat better grounded in the Curci-Ferrari model, 1-loop agreement is not entirely new.  
Early DSE calculations, featuring an effective gluon mass, used a 1-loop expression of 
$\alpha_s$ and the propagators -- see, \textit{e.g}., the discussion of Eq.\,\eqref{eq:alpha_s cornwall} -- to obtain results that were qualitatively and semi-quantitatively confirmed by later LGT calculations.
We observe, too, that the effective charge approach (Sec.\,\ref{EffectiveCharge}) defines the nonperturbative coupling using the leading order expression for one or another given observable. 
This indicates that for some quantities, at least, nonperturbative effects can be encapsulated in just a few well-chosen parameters, \textit{e.g}., the effective gluon mass for $\alpha_s$ or the gluon propagator.

\section{Analytic coupling}
\label{AnalyticApproaches}
As discussed in connection with the effective charge approach, Sec.\,\ref{EffectiveCharge}, defining  $\alpha_s$ so that it remains an observable is advantageous in both the UV and IR domains. 
A different method to preserve this ``observability'' attribute of $\alpha_s$ is to demand its analyticity in the complex $Q^{2}$-plane,%
\footnote{
Except on the real timelike axis where singularities may exist, corresponding to the production of  on-shell particles.} 
that is, physically speaking, for $\alpha_s$ to be causal. 
Such a requirement removes the Landau pole that appears at $Q\simeq \Lambda_s$ in the perturbative treatment: the pole violates causality because it would represent the production of causality-violating tachyons -- see the last four paragraphs of Sec.\,\ref{introduction}.

This approach to solving the Landau pole problem has a long history. 
It was introduced in 1955 \cite{Bogolyubov:1956gh}, 
applied soon thereafter to study the QED Landau pole \cite{Redmond:1958juf, Redmond:1958pe, Bogolyubov:1959vck}, 
and then applied to QCD \cite{Shirkov:1997wi}.
Since then, the approach has been pursued in various guises by several authors -- See Refs.\,\cite{Deur:2016tte, Prosperi:2006hx} for overviews.

The procedure may be formulated \cite{Shirkov:1997wi} using a spectral density, $\rho$, 
whose form is usually taken as that expected from standard pQCD. 
Then, the theory of dispersion relations is employed to mold an analytical coupling, $\alpha_{\rm an}$, free of the Landau pole.  
This method is also, therefore, called the {\it dispersive approach}.

Specifically, the K\"{a}ll\'{e}n--Lehman spectral relation \cite{Kallen:1952zz, Lehmann:1954xi} is employed:
\begin{equation}
\alpha_{\rm an}(Q^{2})=\frac{1}{\pi}\int_{0}^{\infty}d\nu\frac{\rho\left(\nu\right)}{\nu+Q^{2}},
\label{eq:Kallen-Lehman}
\end{equation}
with $\nu$ the integration variable. 
Here, $\rho(\nu)$ is the density in momentum space of a noninteracting gluon propagator, with effective mass $\sqrt\nu$, that mediates the interaction. 
Note that the original K\"{a}ll\'{e}n--Lehman identity relates the full, {\it viz}.\ interacting,  propagator, rather than $\alpha_{\rm an}$, to the weighted sum of massive free-propagators.
The integration over $\nu$ means that $\alpha_{\rm an}$ is explicitly independent of the mass term appearing in the propagator. 
Further, $\nu$ identifies as $Q^2$, rather than a squared mass, in the spectral function -- 
see Eq.\,\eqref{eq:alpha_s spectral function} below. 
(A variation that explicitly includes a gluon mass has been developed \cite{Shirkov:1999hm, Deur:2016tte}.)
The form of $\rho(\nu)$ is chosen so that $\alpha_{\rm an}$ has the desired UV and IR properties. 

To find that $\rho(\nu)$ which would yield $\alpha_{\rm an}$ from Eq.\,\eqref{eq:Kallen-Lehman}, 
the Schwarz reflection principle \cite{Schwarzreflectionprinciple} is used.
This entails that
$\rho\left(\nu\right)= \Im\big( \alpha_{\rm an}(-\nu +i\epsilon) \big)$ \cite{Alekseev:2002zn}. 
The standard choice is \cite{Shirkov:1997wi}:
\begin{equation}
\rho\left(\nu\right)= \Im\big( \alpha_s(-\nu +i\epsilon) \big), 
\label{eq:alpha_s spectral function}
\end{equation}
with $\alpha_s$ the usual pQCD coupling.
Inserting Eq.\,\eqref{eq:alpha_s spectral function} into  Eq.\,\eqref{eq:Kallen-Lehman} yields a Kramers--Kr\"{o}nig type of relation \cite{deL.Kronig:26, Kramers27} for $\alpha_{\rm an}$, which ensures that it is analytical and causal. 

The procedure described above is applicable at any order in the $\beta$ series, with the same features as the standard perturbative solutions for $\alpha^{(l)}_s$  at $l+1$-loop -- see Secs.\,\ref{alpha_s 1-loop}-\ref{alpha_s n-loops}); namely, that analytical solutions exist for the 1-loop $\beta_0$ and 2-loop $\beta_1$, but the higher order solutions are either numerical or approximately analytic. 
At $\beta_{0}$-level, the procedure gives: 
\begin{subequations}
\begin{align}
\rho_0\left(\nu\right) & =\frac{\pi}{\beta_0[\ln^2(\nu/\Lambda_s^2)+\pi^2]},
\label{eq:sepctral function, b0} \\
\alpha_{\rm an}^{(0)}\left(Q^{2}\right) & =\frac{4\pi}{\beta_{0}}\left(\frac{1}{\mbox{ln}(Q^{2}/\Lambda_s^{2})}+\frac{\Lambda_s^{2}}{\Lambda_s^{2}-Q^{2}}\right).\label{eq:alpha_s from analytic QCD}
\end{align}
\end{subequations}
This coupling agrees with the UV expectation of pQCD and is free of the Landau pole in the IR.
In fact, its value at the would-have-been Landau pole is 
$\alpha_{\rm an}^{(0)}(\Lambda_s^{2})/\pi=2/\beta_{0} \approx 0.2$.
As $Q^2\to 0$, all dependence on $\Lambda_s$ disappears and 
$\alpha_{\rm an}(0)/\pi=4/\beta_{0}\simeq 0.41$, 
a result holding at all orders \cite{Alekseev:2002zn}. 
Albeit finite, $\alpha_{\rm an}(Q^{2})$ does not freeze in the IR, but increases monotonically until reaching the $4\pi/\beta_{0}$ value.

The last term in Eq.\,\eqref{eq:alpha_s from analytic QCD} does not depend on the higher-order terms in the $\beta$ series and does not involve any $\ln(Q^2/\Lambda_s^2)$ term. 
Thus, it is a nonperturbative contribution, which expands at high $Q^2$ as $(\Lambda_s/Q)^{2n}$, \textit{i.e}., may be identified with the nonperturbative higher-twist terms of the OPE formalism. 

The couplings calculated at $l$-loop order, $\alpha_{\rm an}^{(l)}(Q^{2})$, are numerically close to each other, {\it viz}.\ their values do not much depend on the loop order. 
This was to be expected because all $\alpha_{\rm an}^{(l)}$ share the same IR and UV limits. 

The small UV and IR values of $\alpha_{\rm an}^{(0)}$, and the fact that $\alpha_{\rm an}^{(l)}\simeq \alpha_{\rm an}^{(0)}$ suggests that the perturbative series associated with a quantity expanded in $\alpha_{\rm an}^{(l)}$ should be valid on both the UV and IR domains, providing that all nonperturbative effects are encompassed by $\alpha_{\rm an}$.
To consistently include the $\alpha^{(l)}_{\rm an}$ at each order $n$ in the pQCD series, 
an $n$-th order spectral function is defined as $\rho_n(\nu) = \Im\big( \alpha_s^n(-\nu) \big)$. 
This provides the $n$-th order coupling,  
\begin{equation}
\alpha^{(l)}_{{\rm an},(n)}=\frac{1}{\pi}\int_{0}^{\infty}d\nu\frac{\rho_n\left(\nu\right)}{\nu+Q^{2}},
\end{equation}
and a standard pQCD series,
$1+\Sigma_n a_n \big[\alpha_s^{(l)}(Q^2)\big]^n$,
is re-expressed as  $1+\Sigma_n a_n \alpha^{(l)}_{{\rm an},(n)}(Q^2)$.
Notice that the series is unusual because it lacks a unique expansion parameter $\alpha_s$.
Instead, it uses parameters $\alpha_{{\rm an},(n)}$ in which the power of $n$ is implicit rather than explicit. 
As a consequence, $\alpha_{{\rm an},(m+n)} \neq \alpha_{{\rm an},(m)}^n \neq \alpha_{{\rm an},(m)}\alpha_{{\rm an},(n)}$; and, in particular, $\alpha_{{\rm an},(n)} \neq \alpha_{{\rm an}}^n$. 
This approach is called analytical perturbation theory (APT) \cite{Shirkov:1997wi}. 
Its extension to non-integer values of $n$ is known as fractional APT (FAPT) \cite{Bakulev:2005gw}, and is motivated by the presence in QCD of operators with nonzero anomalous dimensions.

In the past quinquennium, several authors have pursued the analytical/dispersive approach. 
A method to extend the FAPT perturbative series to higher orders using an 
$1/\ln(\sfrac{Q^2}{\Lambda_s})$ expansion is proposed in Ref.~\cite{Kotikov:2022sos} and tested 
on the Bjorken sum rule, Eq.~(\ref{bjorken SR}).
The FAPT approach is used elsewhere \cite{Ayala:2016zrz, Cvetic:2020unz, Cvetic:2020naz} to compute $\alpha_s^{\rm gh}$ for $N_f = 3$ in the MiniMOM RS and Landau gauge on both the IR and UV domains.
The aim was to provide an analytical form for $\alpha_s^{\rm gh}$ that qualitatively reproduces the decoupling solution found numerically using LGT. 
To that end, $\rho$ was modeled, with ultimately two free parameters adjusted so that the correct $\tau$-decay phenomenology, at the edge of the pQCD domain, is recovered and that the maximum of $\alpha_s^{\rm gh}$ in the IR matches LGT findings. 
In Ref.\,\cite{Zheng:2017hqi}, a SU(3), pure-gauge, $\beta$-function, inspired by the 
Novikov-Shifman-Vainshtein-Zakharov $\beta$-function of supersymmetric gauge theories \cite{Novikov:1983uc, Ryttov:2007cx, Pica:2010mt}, is treated with the analytic procedure to eliminate the Landau pole. 
The resulting analytically-improved $\beta$-function provides an IR-safe $\alpha_s$ that freezes to the IR-fixed point $\alpha_s=0.31$.
Finally, in Ref.\,\cite{Brandt:2022xum}, $\alpha_{\rm an}$ is studied at the 2-loop level, with an exact analytical solution provided, albeit in an integral form that must be either numerically estimated or approximated.

\section{Screened massive expansion}
\label{Screened massive expansion}
The final approach to determining $\alpha_s$ in the IR that has seen developments in the past quinquennium is the screened massive expansion method. 
Its rationale is the smallness of vertex-based $\alpha_s(Q^2)$ forms in the 
decoupling scenario -- Fig.\,\ref{Flo:lowq alpha. DSE} (left panel), where $\alpha_s(Q^2)  \xrightarrow[Q^2 \to 0]{} 0$, irrespective of which vertex coupling is considered. 
This intimates that QCD in the IR might be treatable perturbatively, with nonperturbative effects possibly encapsulated in effective mass parameters \cite{Siringo:2014lva, Siringo:2015gia, Siringo:2015wtx, Siringo:2016jrc, Comitini:2021kxj}. 
Working from this position, \mbox{Refs.\,\cite{Siringo:2014lva, Siringo:2015gia, Siringo:2015wtx, Siringo:2016jrc, Comitini:2021kxj}} propose a double perturbative expansion for chiral QCD (massless current quarks) in the IR which, in particular, enables computation of the ghost--gluon vertex coupling $\alpha^{\rm gh}_s(Q^2)$, Eq.\,\eqref{eq:alpha_s DSE ghost--gluon}, as well as propagators. 
The coupling is first assumed to be small enough to permit the double perturbative expansion and then calculated. 
The result is a numerically small coupling, consistent with the initial assumption.

The method is discussed generally for a SU(N) Yang-Mills theory with massless fermions in Refs.\,
\cite{Siringo:2015gia, Siringo:2015wtx}.
Starting from the Faddeev--Popov Lagrangian of that theory, trial {\it massive} propagators are used for the internal gluon lines of the Feynman diagrams. 
While the mass parameter assigned to the propagators eventually cancels in the renormalization process, distinct dynamical effective masses for gluons and constituent quarks emerge from the loop diagrams. 
The disappearance of the initial mass parameters makes the method compatible with BRST symmetry, in contrast to other massive extensions of SU(N).%
\footnote{Methods displaying {\it emerging dynamical} masses, \textit{e.g}., DSE studies, also respect the BRST symmetry, as noted above.}
The coupling and propagators are obtained in Landau gauge using the $\overline{\rm MS}$ RS. 
They are consistent with other nonperturbative methods, including LGT; and the smallness of $\alpha^{\rm gh}_s(Q^2)$ in the deep-IR means $\alpha^{\rm gh}_s$ is dominated by its 1-loop solution. 
In particular, it consistently displays the expected decoupling solution.  
$\alpha^{\rm gh}_s(Q^2)$ peaks at $Q^2\simeq 0.4$~GeV$^2$, with values of 1.2--1.5, but this is {\it a priori} not inconsistent with a perturbative expansion since pQCD series are developed in the ratio $\alpha_s/\pi$.

The screened massive expansion can also be engineered \cite{Siringo:2019qwx} to produce $\alpha_s^{\rm gh}$ in line with a scaling scenario -- Fig.\,\ref{Flo:lowq alpha. DSE} (right panel).   
To achieve such an outcome, one is required to employ a ``screened MOM'' RS, \textit{i.e}., force the renormalized gluon 2-point function to match a massive free-propagator instead of the standard massless propagator at the momentum subtraction point. 
The usual MOM RS always generates the decoupling scenario. 

\section{Other approaches}
Other methods that have been used to model $\alpha_s$ include  
the Stochastic Quantization formalism, where the role of ghosts is played by longitudinal gluons;
the background field method; 
the Banks-Zaks approach;
and instanton vacuum models.
However, so far as we are aware, no new developments that are pertinent to this review have been made with these methods during the past quinquennium.  
Their current status, therefore, remains as described elsewhere \cite{Deur:2016tte}.
 Additionally, the Bern-Kosower formalism \cite{Bern:1990cu} has recently been used to compute SU(3) vertex functions, but without yet providing the corresponding couplings~\cite{Ahmadiniaz:2016qwn}.

\chapter{Summary and Perspective}
\label{epilogue}
The coupling $\alpha_s(Q^2)$ is central to QCD and to the Standard Model of particle physics. 
Ever since the dawn of the QCD era, vigorous efforts have been underway to understand it. 
As we have described herein, they can be classified into two categories.

The first category we considered is that body which aims at a precise determination of $\alpha_s$ at high energy, with the reference point set to $M_{\rm Z}$, the $Z_0$ mass. 
There, the theoretical framework defining $\alpha_s(Q^2)$ is well understood and is based on the RGE \cite{Petermann:1953wpa, GellMann:1954fq, Callan:1970yg} and perturbation theory. 
The leading behavior of $\alpha_s(Q^2)$ is a logarithmic decrease with $Q^2$ \cite{Gross:1973id, Politzer:1973fx}. 
Accordingly, 
both $\alpha_s$ and its $Q^2$-dependence vanish as $Q^2 \to \infty$, which explains Bjorken scaling \cite{Bjorken:1966jh} and enables the use of QCD perturbation theory for the prediction of high-energy phenomena -- arguably the keystone in establishing QCD as the best existing candidate for a theory of the strong force. 

Today, the main goal of high-energy studies of $\alpha_s$ is to determine this coupling with an accuracy $\Delta \alpha_s/\alpha_s \ll 1\%$ \cite{dEnterria:2022hzv}, the current global assessment being 
$\alpha_s(M_{\rm z})=0.1176(10)$ without LGT ($\Delta \alpha_s/\alpha_s=0.85 \%$) 
and 
$\alpha_s(M_{\rm z})=0.1179(09)$ with LGT \linebreak ($\Delta \alpha_s/\alpha_s=0.76\%$) \cite{Workman:2022ynf}.

High accuracy is evidently necessary for the understanding of QCD, but also for Standard Model studies and searches for physics beyond the Standard Model. 
For the former, it is because QCD corrections to Standard Model reactions often dominate the uncertainties, since $\alpha_s$ is much larger than $\alpha$ and $G_F$ and known with several orders-of-magnitude less accuracy.
An example of the latter is testing the universality of $\alpha_s(M_{\rm z})$ by measuring it at 
a scale $Q^2 \neq M^2_{\rm Z}$ and evolving it to $M^2_{\rm Z}$. 

Reaching $\Delta \alpha_s/\alpha_s \ll 1\%$ requires combining many determinations of $\alpha_s(M_{\rm Z})$ obtained from distinct observables whose approximant is well controlled. 
It also demands using either techniques minimizing renormalization scale ambiguities \cite{Brodsky:1982gc,Brodsky:2011ta} or providing calculations of pQCD approximants to at least past NLO. Several are currently known to N$^3$LO, while the $\beta$-series of QCD has 
been computed to five-loops ($\beta_4$) \cite{Kniehl:2006bg}.
Nevertheless, when the common scheme of varying the renormalization scale is employed \cite{Wang:2023ttk} 
the uncertainty coming from the truncation of pQCD series is generally a limiting factor in the extraction of $\alpha_s(M_{\rm z})$ from measurement or LGT.
In fact, LGT seems to be emerging as a leading method for high precision determinations of 
$\alpha_s(M_{\rm z})$ \cite{Aoki:2021kgd}; consequently, scrutinizing the systematics associated with LGT has become crucial. 
Even so, experimental efforts on $\alpha_s(M_{\rm z})$ remain imperative because LGT calculations do not include physics beyond the Standard Model; thus, LGT can only contribute to searches for such physics  by providing a pure-QCD baseline.

The second category we discussed is the class of studies whose goal is to understand $\alpha_s$ in the strong QCD regime. 
Like the precision efforts, this is also challenging, albeit for different reasons. 
A first stumbling block is placed by the many possible ``obvious'' definitions one can choose for $\alpha_s$ in the IR: at first glance, no one choice is plainly superior to another. 
In fact, it might be that several couplings with different values and $Q^2$-dependence are necessary 
to characterize QCD: quark-gluon, ghost-gluon, 3-gluon and 4-gluon vertices may couple distinctly. 
Whilst the Slavnov-Taylor identities -- the QCD equivalent of the Ward-Green-Takahashi identities in QED -- ensure that couplings derived from all these vertices must be equivalent in perturbation theory, practitioner-dependent choices of, \textit{e.g}., gauge, RS, and momentum flow, can obscure connections between them at IR momenta. 

Compounding the number of distinct $\alpha_s$ behaviors is the enhancement of the RS-dependence in the IR  
allied with the fact that the relation between couplings calculated in different RSs is usually known only in the pQCD  framework. 
(See, however, the HLFQCD approach, Sec.\,\ref{HLFQCD} and Ref.\,\cite{Deur:2016cxb}, for an exception.)  

Finally, and not least amongst the reasons, nonperturbative calculations are complex and difficult. 
This may lead to erroneous $\alpha_s$ behaviors owing to artifacts from approximations, such as finite size effects in LGT or truncation prescriptions for continuum Schwinger function methods.

Many  prescriptions for $\alpha_s$ in IR have been studied, resulting in $\alpha_s(Q^2 \to 0)$ values ranging from 0 to $\infty$~\cite{Deur:2016tte}. 
While it rapidly became clear that the Landau pole at $Q^2 \simeq \Lambda^2_s$ was a perturbative artifact and therefore could not represent the behavior of $\alpha_s$ nor be the cause of color confinement,
there was no consensus on how $\alpha_s$ ought to behave in the IR. 

A prominent proposal was that the coupling should possess an IR fixed point, \textit{i.e}., that it displays a plateau for $Q^2 \ll \Lambda^2_s$, a behavior called the (IR) freezing of $\alpha_s$. 
This was posited soon after the advent of QCD \cite{Eichten:1974af, Caswell:1974gg, Sanda:1979xp, Cornwall:1981zr, Banks:1981nn, Godfrey:1985xj}. 
Other proposals were that $\alpha_s$ should vanish as $Q^2 \to 0$ \cite{Dokshitzer:1995ev, Boucaud:2002fx}, or diverge as $1/Q^2$ \cite{Richardson:1978bt} or that it should increase monotonically as $1/Q^2$, but without diverging \cite{Shirkov:1997wi}.
As explained above, many causes lie behind these widely differing expectations. 


For further progress to be made in this area, a unifying definition of $\alpha_s(Q^2 \geq 0)$ must be identified and widely appreciated. 
In this connection, it is worth recalling that the effective charge prescription was long ago adopted in QED \cite{GellMann:1954fq}.  
This choice is ``optimal'' because it provides an unambiguous, readily interpretable, and widely applicable definition.  
Something equivalent has long been sought in QCD.
Namely, a single coupling, 
which has minimal sensitivity to choices of gauge or RS, 
whose origin and momentum-argument are unambiguous, 
whose value is both an expression of the interparticle interaction strength at the listed scale and directly comparable with data, 
and which is useful, in the sense that the definition should be capable of delivering predictions for measurable nonperturbative quantities. 

We have highlighted steps made in this direction during the past quinquennium.  
In particular, data and distinct non-perturbative methods have converged to a concordant global picture of $\alpha_s$ defined, like in QED, as an effective charge. 
This presents the possibility of implementing a running coupling that is both {\it consistent} and {\it fruitful}.

\textit{Consistent}, because of agreement between the experimental measurement of the effective charge $\alpha_{g_1}$ \cite{Deur:2005cf, Deur:2008rf, Deur:2022msf} 
and parameter-free predictions from 
HLFQCD \cite{Brodsky:2010ur}, 
DSE and LGT \cite{Binosi:2016nme, Cui:2019dwv}.  
In all cases, the results are gauge independent:
self-evidently for the experimental data;
and by virtue of the nonperturbative approach for HLFQCD \cite{Brodsky:2010ur} or thanks to the pinch technique for DSE and LGT \cite{Cornwall:1981zr, Binosi:2009qm}.
This coupling also agrees with an earlier determination \cite{Cornwall:1981zr} after allowance is made for the different RS \cite{Deur:2016cxb}.

\textit{Fruitful}, 
because many hadronic quantities are calculable with such a coupling and within the theoretical framework in which it is derived.
These quantities include 
hadron masses \cite{Deur:2014qfa, Chang:2011ei}, 
hadron polarized and unpolarized PDFs and GPDs \cite{deTeramond:2018ecg, deTeramond:2021lxc, Chang:2013pq, Shi:2015esa, Ding:2015rkn, Ding:2019qlr, Ding:2019lwe, Yin:2023dbw}, 
elastic and transition form factors \cite{Sufian:2016hwn, Raya:2015gva, Raya:2016yuj, Rodriguez-Quintero:2018wma},
meson decay constants \cite{Xu:2022kng},
the scale of QCD, $\Lambda_s$ \cite{Deur:2016opc}, 
and the connection between the soft and hard pomerons \cite{Dosch:2022mop}.

Thus, the availability of $\alpha_{g_1}$ at all scales (Fig.~4.1),  combined with the elimination of scale and scheme ambiguities using PMC scale setting \cite{Wang:2023ttk}, has opened a new range of predictions for physics phenomena in both the perturbative and nonperturbative domains. 
The mechanism causing the IR freezing of the coupling remains a subject of research, although progress has also been made there.

Regardless of its origin, one now has in hand a coupling that can universally characterize QCD interactions, that has the merit of being independent of the choice of gauge, vertex, and process,
and that can predict observables in the strong QCD domain.
This argues strongly for its interpretation as a canonical QCD coupling, not only in the UV region -- where it agrees with the perturbative coupling -- but in the IR as well.
%
As we have stressed, this effective charge fulfills basic criteria expected of a canonical coupling.
Other candidates for $\alpha_s$ that meet these criteria may exist, but none are presently known.  If one were found, it would provide for a valuable comparison, which could potentially deliver additional advances in our understanding of QCD and the standard model.

\paragraph{Acknowledgments.} 
This perspective is based on results obtained and insights developed through collaborations with many people, to all of whom we are greatly indebted.
%
%
This work is supported by:
U.S.\ Department of Energy, Office of Science, Office of Nuclear Physics, contract DE-AC05-06OR23177 (AD),
U.S.\ Department of Energy contract DE-AC02-76SF00515 (SJB);
and National Natural Science Foundation of China grant no.\,12135007 (CDR).

\newpage

\section*{Abbreviations}
\label{PageAbbreviations}
\addcontentsline{toc}{section}{\protect\textbf{Abbreviations}}
\noindent The following abbreviations are used in this manuscript:\\[2ex]
\noindent
\begin{longtable}{ p{.20\textwidth}  p{.80\textwidth} } 
AdS/CFT & anti-de Sitter/conformal field theory \\
AdS/QCD & anti-de Sitter/QCD \\
ALEPH & Apparatus for LEP PHysics \\
APT & analytical perturbation theory \\
ATLAS & A Toroidal LHC Apparatus \\
BLM & Brodsky-Lepage-Mackenzie \\
BRST & Becchi-Rouet-Stora-Tyutin \\
CEBAF & Continuous Electron Beam Accelerator Facility \\
CERN & European Organization for Nuclear Research\\
CIPT & contour-improved perturbation theory \\
CLAS & CEBAF Large Acceptance spectrometer \\
CMS & Compact Muon Solenoid (experiment) \\ 
CSR & commensurate scale relations \\
CSMs & continuum Schwinger function methods \\
dAFF & de Alfaro-Fubini-Furlan \\
DCSB & dynamical chiral symmetry breaking \\
DELPHI & DEtector with Lepton, Photon and Hadron Identification \\
DESY & Deutsches Elektronen-Synchrotron \\
DGLAP & Dokshitzer--Gribov--Lipatov--Altarelli--Parisi \\
DIS & deep inelastic scattering \\
DSE & Dyson-Schwinger equation \\
EIC & electron ion collider (at Brookhaven National Laboratory) \\
EicC & electron ion collider in China \\
EHM & emergent hadron mass \\
ERBL & Efremov-Radyushkin-Brodsky-Lepage \\
fAPT &  fractional APT \\
FF & (parton) fragmentation function \\
FLAG & flavor lattice averaging group \\
FOPT & fixed order perturbation theory \\
FRG & functional renormalisation group \\
GPD & generalized parton distribution \\
HERA & Hadron-Electron Ring Accelerator \\
HLFQCD & holographic light-front QCD \\
HRS & High Resolution Spectrometer (experiment at PEP, or detector system at JLab) \\ 
HT & higher-twist \\
IR & infrared (physics/phenomena) \\
JLab & Thomas Jefferson National Accelerator Facility \\
JADE & JApan, Deutschland and England (experiment at PETRA) \\
L3 & Third LEP experiment \\
LC & Light Cone \\
LEP & Large Electron-Positron collider \\
LF & Light Front \\
LL & leading logarithm \\
LGT & lattice gauge theory \\
LHC & large hadron collider (at CERN) \\
LO & leading order \\
MS (and $\overline{\rm MS}$) & Minimal Subtraction \\
NLL (and N$^i$LL) & next to ... leading logarithm \\
NLO (and N$^i$LO) & next to ... leading order \\
OPAL & Omni-Purpose Apparatus for LEP \\
(P)DF & (parton) distribution function \\
PDG & Particle Data Group and associated publications \\
PEP & Positron-Electron Project (facility at SLAC) \\
PETRA & Positron-Electron Tandem Ring Accelerator\\
PI (charge) & process-independent (charge) \\
PMC & principle of maximum conformality \\
pQCD & perturbative QCD \\
QCD & quantum chromodynamics \\
QED & quantum electrodynamics \\
QFT & quantum field theory \\
RGE & renormalization group equation \\
RGI & renormalisation group invariant \\
RHIC & relativistic heavy ion collider \\
RS & Renormalization Scheme \\
SLAC & Stanford Linear Accelerator Center \\
SLC & Stanford Linear Collider \\
SLD & SLAC Large Detector \\
SU(N) & special unitary (group of degree n) \\
TASSO & Two Arm Spectrometer SOlenoid \\
UV & ultraviolet (physics/phenomena) \\
\end{longtable}

\newpage

\addcontentsline{toc}{section}{\protect\textbf{Bibliography}}

\bibliographystyle{h-physrev4}
\bibliography{CollectBibAlpha}

\end{document}